\documentclass[a4paper,twocolumn,11pt,accepted=2019-10-29]{quantumarticle}
\pdfoutput=1
\usepackage[utf8]{inputenc}
\usepackage[english]{babel}
\usepackage[T1]{fontenc}
\usepackage{amsmath, mathtools, amssymb}
\usepackage{hyperref}
\usepackage[sort, numbers, compress]{natbib}
\usepackage{tikz}
\usepackage{simplewick}
\DeclareMathOperator{\Tr}{Tr}
\newcommand{\stkout}[1]{\ifmmode\text{\sout{\ensuremath{#1}}}\else\sout{#1}\fi}

\begin{document}

\title{Ergodicity probes: using time-fluctuations to measure the Hilbert space dimension}

\author{Charlie Nation}
\affiliation{Department of Physics and Astronomy, University of Sussex, Brighton, BN1 9QH, United Kingdom.}
\orcid{0000-0002-3770-8980}
\email{C.Nation@sussex.ac.uk}
\author{Diego Porras}
\email{D.Porras@iff.csic.es}
\orcid{0000-0003-2995-0299}
\affiliation{Institute of Fundamental Physics, IFF-CSIC, Calle Serrano 113b, 28006 Madrid, Spain}

\maketitle

\begin{abstract}
Quantum devices, such as quantum simulators, quantum annealers, and quantum computers, may be exploited to solve problems beyond what is tractable with classical computers. 
This may be achieved as the Hilbert space available to perform such `calculations' is far larger than that which may be classically simulated. 
In practice, however, quantum devices have imperfections, which may limit the accessibility to the whole Hilbert space. 
We thus determine that the dimension of the space of quantum states that are available to a quantum device is a meaningful measure of its functionality, though unfortunately this quantity cannot be directly experimentally determined. Here we outline an experimentally realisable approach to obtaining the  required Hilbert space dimension of such a device to compute its time evolution, by exploiting the thermalization dynamics of a probe qubit. This is achieved by obtaining a fluctuation-dissipation theorem for high-temperature chaotic quantum systems, which facilitates the extraction of information on the Hilbert space dimension via measurements of the decay rate, and time-fluctuations.
\end{abstract}
\section{Introduction}

The ability to control and manipulate microscopic systems at the single particle level is an essential requirement for many quantum technologies. 
Experimental setups where atoms or qubits can be arranged in ordered structures and studied in quantum non-equilibrium states include neutral atoms in optical lattices 
\cite{Bloch2012, Lewenstein2012, Aidelsburger2017}, trapped ions \cite{Porras2004,Schneider2012,Blatt2012,Clos2016a}, Rydberg atoms \cite{Kim2018, Bernien2017},  and superconducting circuits\cite{Neill2016,Roushan2018}. 
These systems can be used for the quantum simulation of many-body models, or different forms of digital or adiabatic quantum computing.
Most of these physical setups have limitations in the accessibility to
certain observables. Thus, having extra tools to characterize quantum systems in a simple and efficient way can be useful in the diagnosis and certification of quantum devices.

One of the most prominent properties of a quantum device is its size in terms of the dimension of the associated Hilbert space. The size of a quantum computer or simulator is often given in terms of number of qubits, such that the Hilbert space dimension is $2^N$ for $N$ qubits. This is a measure that ignores the effect of disorder or the possible lack of connectivity between different zones in the device. 

A more useful quantity would be the number of eigenstates of the Hamiltonian that take part in the quantum dynamics, which is bounded in this case by $2^N$, but would exclude the degrees of freedom that do not contribute to the evolution of the initial state. Thereby establishing an effective Hilbert space dimension that more accurately describes the complexity of the system, in terms of some effective fully connected Hamiltonian. This Hilbert space dimension is, however, an elusive measure in realistic experimental situations.

In this work we show that the equilibration dynamics \cite{Malabarba2014, Garcia-Pintos2017, Richter2018, DeOliveira2018, Borgonovi2019, Schiulaz2019, Alhambra2019} of a quantum system can be used to extract such information on the dimension of the Hilbert space of a quantum device, in terms of the effective number of states that contribute to the dynamics of a local observable. Indeed, advancements in quantum technologies described above have inspired a bounty of theoretical work in the field of quantum thermalization \cite{Srednicki1994, Srednicki1999, Rigol2008, Reimann2008, Yukalov2011, Rigol2012, DAlessio2016, Beugeling2014, Beugeling2015, Eisert2015, Anza2018, Deutsch2018}. In the following, we aim to help `bridge the gap' between theoretical and experimental work in this field \cite{Merali2017}.

We assume a quantum quench scenario \cite{Rigol2008, Rigol, Eisert2015, Kim2018} in which a quantum system is initialized in a fully-decohered,  infinite temperature state, except for a subsystem that acts as a sensor and is prepared in a pure state. For simplicity, we assume that this subsystem is a single qubit, which we refer to as the `probe' qubit. 
The relaxation dynamics of the probe qubit depends on the details of the underlying structure of the Hamiltonian, however, in the most generic case of non-integrable systems, an appropriate description can be given in terms of random matrix theory (RMT) \cite{Reimann2016a, DAlessio2016, Deutsch2018, Borgonovi2016, Dabelow2019, Reimann2019, Torres-Herrera2016, Hamazaki2019}. 
We show that the time-fluctuations of the probe in the long-time limit contain information about the Hilbert space dimension of the device.

Our article is structured as follows. Firstly, in Section \ref{sec:Proposal} we present the basic scheme and summarize our main result, which relies on an infinite-temperature fluctuation dissipation theorem (FDT) \cite{Kubo1966} for the dynamics of the probe qubit in order to extract information on the scaling of the effective Hilbert space of a quantum device. We continue, presenting numerical calculations that validate our predictions via exact diagonalization of a spin chain Hamiltonian in Section \ref{sec:Num_exp}. We then outline in more detail our RMT model in Section \ref{sec:Model}. In Section \ref{sec:Time_dep} we derive an expression for the time dependence of generic observables in chaotic quantum systems, and discuss how this can be exploited for our measurements. We then derive the FDT, and extensions from the RMT model, and to finite temperatures, in Section \ref{sec:FDT}. Finally, we summarise our findings in Section \ref{sec:Discussion}. Further details and proofs are included in Appendices. We note that this is arranged such that our key findings can be understood from sections \ref{sec:Proposal} and \ref{sec:Num_exp}, with the detailed calculations presented later in the text.

\section{Setup and main results}\label{sec:Proposal}
\subsection{Proposed setup}
\begin{figure}
	\includegraphics[width=0.45\textwidth]{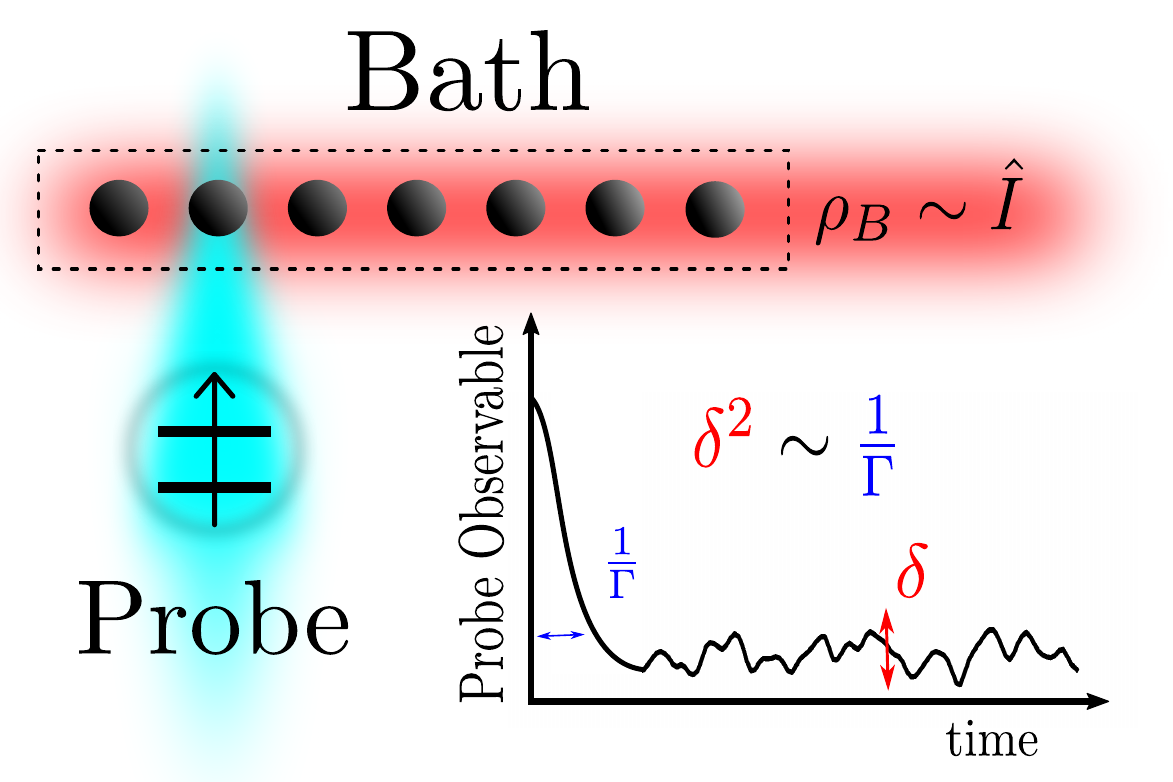}
	\caption{Illustration of our proposed setup (quantum device). A single qubit, labelled `Probe', is coupled locally to a part of a larger non-integrable bath, initialized in an infinite temperature state $\rho_B \sim \hat{I}$ (this restriction is removed below). Experimentally, for our protocol, one needs access only to an observable of the Probe qubit.}
	\label{fig:fig1} 
\end{figure}

We assume that we have a quantum system (from here on, `quantum device') made up of a single `probe' qubit, initialized in a pure state, which at $t=0$ is coupled to the rest of the quantum device (referred to as the `bath'). We initialize the bath in an infinite temperature state. This is sketched in Figure \ref{fig:fig1}.
This can be routinely achieved, for example, in a quantum computer device that has not been properly initialized.
Since quantum devices always suffer from some kind of decoherence, creating an infinite temperature (or fully decohered) state is typically a simple task.
We then let this qubit evolve in time and reach an equilibrium state.
The initial state is thus,
\begin{equation}
\rho(t = 0) = |\!\uparrow \rangle \langle \uparrow \!| \otimes \rho_B.
\label{initial.state}
\end{equation}
$\rho_B$ is the density matrix of the bath in the fully decohered state, 
\begin{equation}
\rho_B = 
\frac{1}{{\cal N}_B}
\sum_{\beta}^{{\cal N}_B} |\phi_{B,\beta} \rangle \langle \phi_{B,\beta} |,
\end{equation}
where $|\phi_{B,\alpha_B}\rangle$ is a set of orthonormal wave functions in the bath, which we take as the eigenstates of the bath Hamiltonian $H_B$.  ${\cal N}_B$ is the Hilbert space dimension of the bath, and is the parameter on which we wish to infer information via measurements of the probe qubit. Later on the  fully decohered state condition will be relaxed, allowing for finite temperatures.

The system evolves under the interacting Hamiltonian, 
\begin{equation}\label{eq:H_rm}
H = H_0 + V,
\end{equation}
where $H_0 = \hat{1} \otimes H_B$ is the Hamiltonian of the uncoupled probe + bath device, and we assume that the probe qubit does not evolve at all in the absence of coupling to the bath, which is given by the operator $V$. A prominent role in the following discussions will be played by the eigensystems of the interacting Hamiltonian,
\begin{equation}
H |\psi_\mu \rangle = E_\mu | \psi_\mu \rangle,
\label{H}
\end{equation}
and the eigensystem of the uncoupled probe-bath system
\begin{equation}
H_0 |\phi_\alpha \rangle = E^{(0)}_\alpha | \phi_\alpha \rangle.
\label{H0}
\end{equation}
All throughout this work we will use the convention that indices $\mu$, $\nu$ refer to summations over eigenstates of $H$, whereas $\alpha,\beta$ refer to eigenstates of $H_0$.

We focus on the quantum dynamics of the expectation value of a probe observable, which we take for concreteness to be $\langle \sigma_z(t) \rangle = \Tr(\rho(t) \sigma_z)$. 
Note that, however, the results should not depend on the particular choice for the probe observable, since $H_0$ has no probe energy term. The probe-bath coupling $V$, may introduce a dependence on the chosen observable, but the results below should hold in the general case of quantum chaotic or non-integrable systems.

The quantity under study, the time-averaged fluctuations of the probe observable, is defined by
\begin{equation}
\delta_{\sigma_z}^2(T) =
\frac{1}{T} \int_0^T dt \left( \langle \sigma_z(t) \rangle - \mu_{\sigma_z}(T) \right)^2, 
\end{equation}
where $\mu_{\sigma_z}(T) = \frac{1}{T} \int_0^T dt \langle \sigma_z(t) \rangle $. Assuming only that the many-body eigenenergies are non-degenerate, and that also their energy gaps are non-degenerate, we may express the time fluctuations in terms of matrix elements between eigenstates of the coupled probe-bath system \cite{DAlessio2016},
\begin{equation}\label{eq:delta_DE}
\delta_{\sigma_z}^2(\infty) = \sum_{\substack{\mu,\nu \\ \mu \neq \nu}}
|\rho_{\mu \nu}|^2 |(\sigma_z)_{\mu \nu}|^2,
\end{equation}
where $\rho_{\mu \nu} = \langle \psi_\mu |\rho| \psi_\nu\rangle$, and $(\sigma_z)_{\mu \nu} = \langle \psi_\mu |\sigma_z| \psi_\nu\rangle $.

\subsection{Summary of main results of this work}
In many practical situations the quantum device is a non-integrable, quantum chaotic system. We expect that in such cases a qualitative description can be obtained by assuming that $V$ is a random matrix. 
This assumption leads to a statistical theory for the many body wave functions in Eqs. (\ref{H}, \ref{H0}) \cite{Nation2018},
\begin{equation}\label{eq:psi_mu}
|\psi_\mu\rangle = \sum_{\alpha} c_\mu(\alpha) |\phi_\alpha \rangle,
\end{equation}
where summations are understood to be taken from $1$ to $2{\cal N}_B$, the dimension of the Hilbert space of the device. We will refer to $c_\mu(\alpha)$ as quantum chaotic or random wave functions.
We assume that $V$ belongs to the Gaussian Orthogonal Ensemble (GOE), appropriately scaled by a coupling strength $g$. 
In Ref. \cite{Nation2018} we showed that this model can be solved and it allows us to calculate matrix elements in the interacting basis.

Special care must be taken to translate the probe initial state 
defined in Eq. (\ref{initial.state}) in the random matrix formalism. 
Without loss of generality we can assume that non-interacting eigenfunctions are ordered such that odd (even) values of $\alpha$ correspond to the probe qubit in state 
$|\! \uparrow \rangle$ ($|\! \downarrow \rangle$). This leads to the initial condition
\begin{equation}\label{eq:rho_0}
\rho(t = 0) = \frac{1}{{\cal N}_B} \sum_{\alpha \in \textrm{odd} }^{2 {\cal N}_B}
|\phi_\alpha \rangle \langle\phi_\alpha |.
\end{equation}

In the following sections we prove that random matrix theory leads to a relation between the time-fluctuations and the typical decay rate of the probe qubit. This relation is the main result of this work:
\begin{equation}
\delta^2_{\sigma_z}(\infty)    = \chi(N)  \overline{\Gamma^{-1}},
\label{fdt.T}
\end{equation}
where $\overline{\Gamma^{-1}} = \frac{1}{\Delta E}\int_0^{\Delta E} dE \Gamma(E)^{-1}$ is the average inverse decay rate of the qubit. This average is an unbiased average over all initially populated initial state energies, which spans the entire energy range $\Delta E$ of the device due to the initial infinite temperature bath state. We will see that this unbiased average originates from the decay to equilibrium of each state $|\phi_\alpha\rangle$ contributing to the initial state \eqref{eq:rho_0}.
In Sections \ref{sec:Time_dep} and \ref{sec:FDT} we will see that this can be related to the decay rate obtained from a fit to the decay to equilibrium of a local probe observable in our current set up.
This quantity is thus simply an average of the decay rates experienced by the probe qubit. 

The quantity $\chi(N)$, with $N$ the total number of qubits, depends on the size of the system in the following way,
\begin{equation}
\chi(N) = C \frac{1}{{\cal N}_B \overline{D(E)}} .
\label{chi}
\end{equation}
$\overline{D(E)}$ is the average density of states (DOS) of the system, defined analogously to the average decay rate above (see also Eq. \eqref{eq:unbiased_average}). 
${\cal N}_B$ is the bath Hilbert space dimension, and finally, $C$ is a constant of order one that does not depend on the size of the system or coupling strength.

Eq. (\ref{fdt.T}) can be understood as a fluctuation-dissipation relation, which relates the time-fluctuations in the steady-state with the decay rate after a quantum quench. The ratio between fluctuations and average decay time allows us to quantify the dimension of the Hilbert space over which the ergodic quantum dynamics takes place. In a physical system if $V$ couples the probe qubit to the whole spectrum of the quantum device, we expect that the function $\chi \propto e^{c N_B}$, where $N_B$ is the number of qubits in the part of the device that acts as a bath. 

Our main result, Eq. (\ref{chi}), can be obtained by following these steps:

(i) We consider the random matrix model with an homogeneous coupling matrix $V$, and calculate $\delta^2_O(\infty)$ by extending the formalism developed in \cite{Nation2018, Nation2019} to mixed states (see Section \ref{sec:FDT_RMT}). To carry out this calculation we assume a constant DOS, $D(E) = 1/\omega_0$, with $\omega_0$ the mean energy separation. 

This calculation allows us to predict an exponential decay for a probe observable of the form:

\begin{equation}\label{eq:TDepfin}
\langle O(t) \rangle = (\langle O(t) \rangle_0 - O_{\textrm{DE}}) e^{-2\Gamma t} + O_{\textrm{DE}},
\end{equation}
where $O_{\textrm{DE}}$ is the diagonal ensemble of the observable $O$, which we find to be equal to the long-time average of $O$ as required, and $\langle O(t) \rangle_0$ refers to the free evolution of the observable $O$ under the Hamiltonian $H_0$. This result is derived in Section \ref{sec:Time_dep}. This is analogous to the result in Reference \cite{Nation2019} for pure-states. We note that the same Equation has similarly been obtained in Ref. \cite{Dabelow2019}, which allows also for the perturbation matrix $V$ to be inhomogeneous. In the specific case of interest here, Eq. \eqref{eq:TDepfin} becomes,
\begin{equation}\label{eq:TDepModel}
\langle \sigma_z(t) \rangle = e^{-2\Gamma t},
\end{equation}
as $\langle \sigma_z(t) \rangle_0 = 1$ and $\langle\sigma_z\rangle_{\textrm{DE}} = 0$. 

Using this random matrix model we also obtain a preliminary version of the infinite temperature fluctuation-dissipation theorem, Eq. (\ref{chi}), with $\chi(N) = C\omega_0/{\cal N}_B $.

(ii) In a realistic system both the DOS and the coupling strength will depend on the energy. This leads to energy-dependent qubit decay rates, $\Gamma(E)$, and DOS, $D(E)$. Due to the infinite-temperature initial condition, it is not possible a priori to approximate those quantities to any single value, since the quantum dynamics of the probe qubit result from contributions from all possible initial states. It is thus necessary to extend the RMT formalism to allow for variations in both the DOS and the decay rate in energy over the width of the initial mixed state.

It is also important to note that when applied to a realistic system ${\cal N}_B$ should be thought of as the number of eigenstates that may contribute to the dynamics of a local observable - it is thus a measure of the effective Hilbert space dimension in the sense of the number of degrees of freedom that may be explored from a given initial state. This measure accounts for effects such as locality of interactions, and disorder, which will reduce the Hilbert space available for evolution of the system. As such, ${\cal N}_B$ will in general be bounded by the total Hilbert space dimension, but is a more realistic measure of the size of the explored Hilbert space in the thermalization dynamics of the device.

In Section \ref{sec:Time_dep} we show that the decay rate observed from such an initial state, defined with an energy dependent coupling strength, is the thermal average decay rate over the energy width defined by the initial state. 

We formulate the generalization of the FDT in Section \ref{sec:FDT_real}, however a brief summary of the approach is as follows: To account for the energy variation of the DOS and decay rate we instead make the assumption that both change slowly in energy with respect to the energy width of a single eigenstate. In this case, one can reformulate the theory such that the random wave functions have a smoothly varying width. In effect this is the statement that a given random wave function contributes as if it was the eigenstate of a random matrix model with a constant DOS and decay rate, as over the width of the wave function itself these parameters do not change appreciably. 

In this case, we see that the average DOS and decay rate over the initially populated initial states contributes to the FDT, as in Eq. \eqref{fdt.T}. This is extended to finite temperature systems, where instead a thermal average can be seen to contribute, in Section \ref{sec:FDT_fin}. 

Finally, we may relate the two decay rate averages, as the thermal average observed from the decay differs slightly from that appearing in Eq. \eqref{fdt.T}. These can be seen to be related by a constant factor that does not depend on the coupling strength, or bath size, and thus scaling information of the Hilbert space dimension may be recovered. This is shown in Section \ref{sec:FDT_fin}, and Appendix \ref{App:Thermal_Averages}, where the FDT is recast in terms of thermal averages. 

In Section \ref{sec:Time_dep} we show that when an exponential decay of the form Eq. \eqref{eq:TDepfin} is observed, this indicates that $\Gamma(E)$ is approximately constant over the bulk of the initially populated states. In this case we are able to extract the numerical value of the Hilbert space dimension directly, as the averages ove $\Gamma(E)$ occurring in the FDT, and observed from the decay, are equal. 

In summary, we observe the emergence of a classical fluctuation-dissipation theorem, relating the time-fluctuations and decay rate of our probe observable $\sigma_z$. The susceptibility $\chi(N)$ in Eq. \eqref{chi} can be seen to be related to the Hilbert space dimension of the bath, ${\cal N}_B$, and thus measurements of the decay rate, $\Gamma$, and fluctuations $\delta_{\sigma_z}^2(\infty)$, which are both obtainable from the time evolution, can be exploited to obtain information on the device Hilbert space dimension, ${\cal N}_B$.

\section{Numerical Experiments}\label{sec:Num_exp}
Before going into the technical details of our derivation, we present numerical evidence that confirms the validity of the random matrix approach and our main results.
In this section, we show the application to a spin-chain system using exact diagonalization \cite{Weinberg2016, Weinberg2018}. From this, we observe the FDT numerically using a realistic experimentally observable model.
\begin{figure}
		\includegraphics[width=0.48\textwidth]{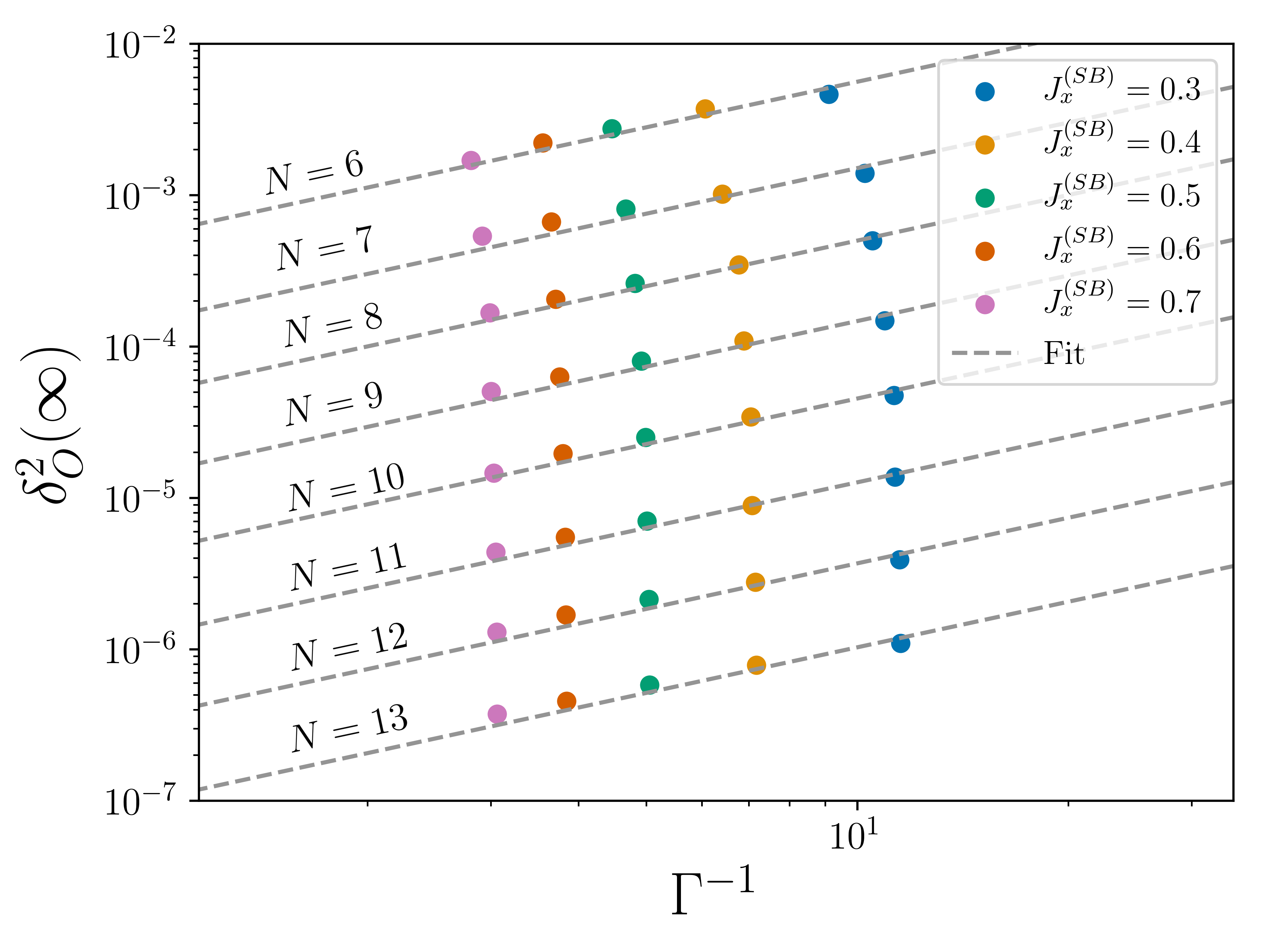}
	\caption{Observation of the FDT, Eq. \eqref{fdt.T}, for varying coupling strength, for many bath sizes $N-1$ (labelled on plot). Fits (blue dashed lines) shown for each value of $N$, are linear fits to obtain $\chi(N)$ for each individual $N$ value. We see in this case, then, that one does not require the ability to change the device length in order to observe our predictions experimentally.  Parameters: $B_Z^{(B)} = 0.1, B_x^{(B)} = 0.3, J_z = 0.1, J_x = 1, J_z^{(SB)} = 0, \beta = 0$.}
	\label{fig:fig2} 
\end{figure}

In Fig. \ref{fig:fig2}, we show  the manifestation of Eq. \eqref{fdt.T} in a spin-chain system described by the Hamiltonian $H = H_S + H_B + H_{SB}$, where $H_S = 0$ is the system Hamiltonian (acting as our probe), $H_B$ is our bath Hamiltonian, given by
\begin{equation}
\begin{split}
& H_B =  \sum_{j > 1}^N( B_{z}^{(B)}\sigma_z^{(j)}  + B_{x}^{(B)}\sigma_x^{(j)}) +\\ & \sum_{j > 1}^{N-1}[J_z\sigma_z^{(j)}\sigma_z^{(j+1)} + J_x(\sigma_+^{(j)}\sigma_-^{(j+1)} + \sigma_-^{(j)}\sigma_+^{(j+1)} )],
\end{split}
\end{equation}
which acts on sites with index $>1$, which is the probe index. The probe and bath are coupled by the interaction Hamiltonian,
\begin{equation}
\begin{split}
H_{SB} &= 
J_z^{(SB)}\sigma_z^{(1)}\sigma_z^{(N_{\rm m})} \\&+ J_x^{(SB)}(\sigma_+^{(1)}\sigma_-^{(N_{\rm m})} + \sigma_-^{(1)}\sigma_+^{(N_{\rm m})}).
\end{split}
\end{equation}
Here $N_m$ is the device site where the probe is coupled, which we set as 2 throughout. This spin-chain model may be related to the random matrix toy model, $H = H_0 + V$, via the prescription $H_0 \Leftrightarrow H_S + H_B$, and $V \Leftrightarrow H_{SB}$.
\begin{figure}
		\includegraphics[width=0.48\textwidth]{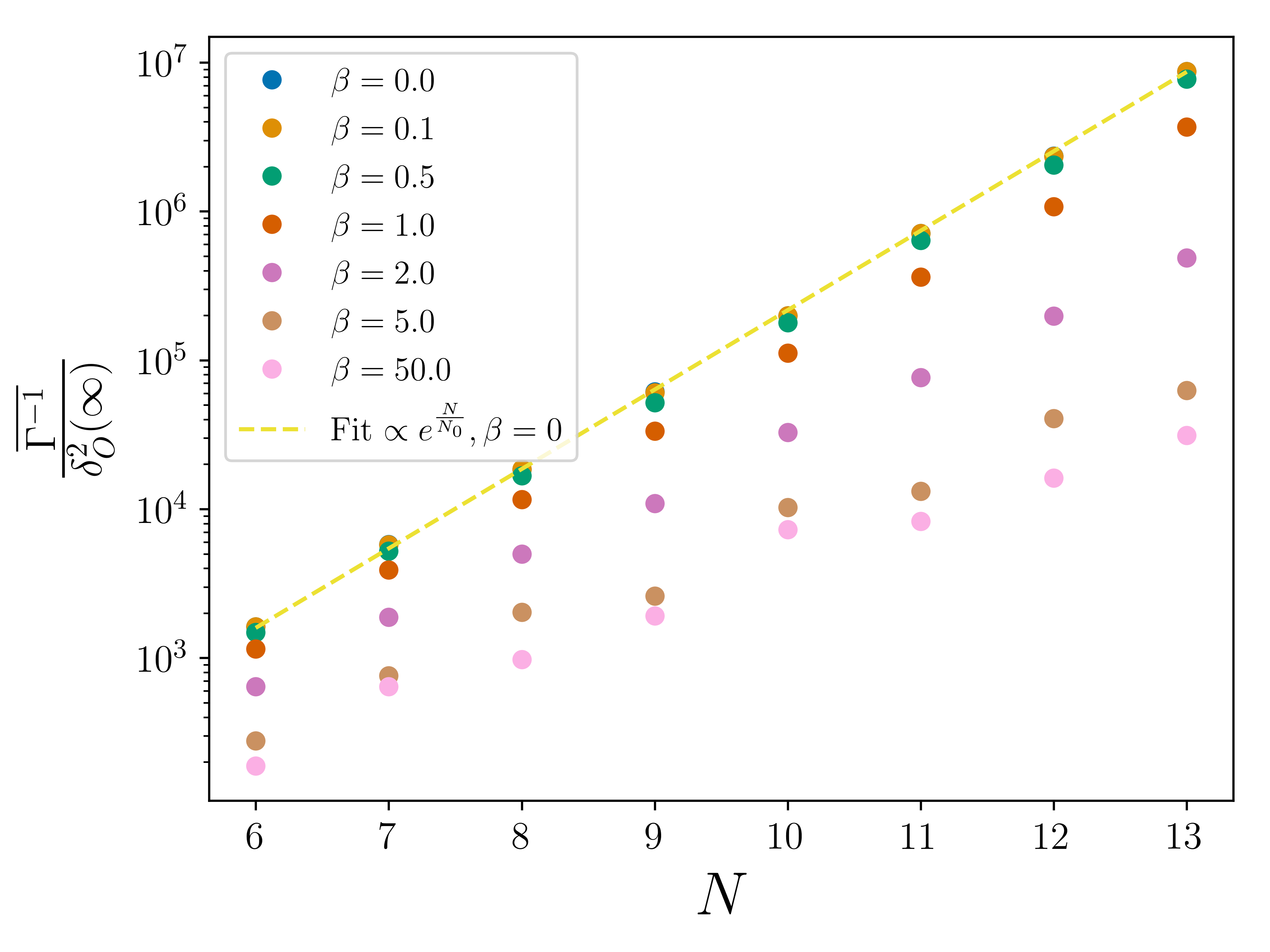}
	\caption{Observation of the FDT, Eq. \eqref{fdt.T}, for varying bath size $N-1$, for temperatures $\beta$. Fit (yellow dashed line) is performed to the function $a e^{\frac{N}{N_0}}$ for the infinite temperature case, $\beta = 0$, and thus confirms the exponential scaling of $\chi(N)$. In this case, we observe an exponential scaling for all temperatures, as the average DOS also scales exponentially with $N$. Note that here we have $\Gamma \sim 0.2$ so the high temperature limit is defined by approximately $\beta \ll 5$. Parameters: as Fig. \ref{fig:fig2} with $J_x^{(SB)} = 0.5$.}
	\label{fig:fig3} 
\end{figure}
In particular, we see that, as $\chi(N) = \frac{\delta_O^2(\infty)}{\overline{\Gamma^{-1}}} \propto {\cal N}_B^{-1}$, we expect that if all of the available Hilbert space is being utilized in the unitary dynamics we will observe the following scaling:
\begin{equation}\label{eq:exp_scaling}
\chi(N) \propto e^{-cN_B}.
\end{equation}
This is the relation that we test in Fig. \ref{fig:fig3}. 

\begin{figure}
		\includegraphics[width=0.48\textwidth]{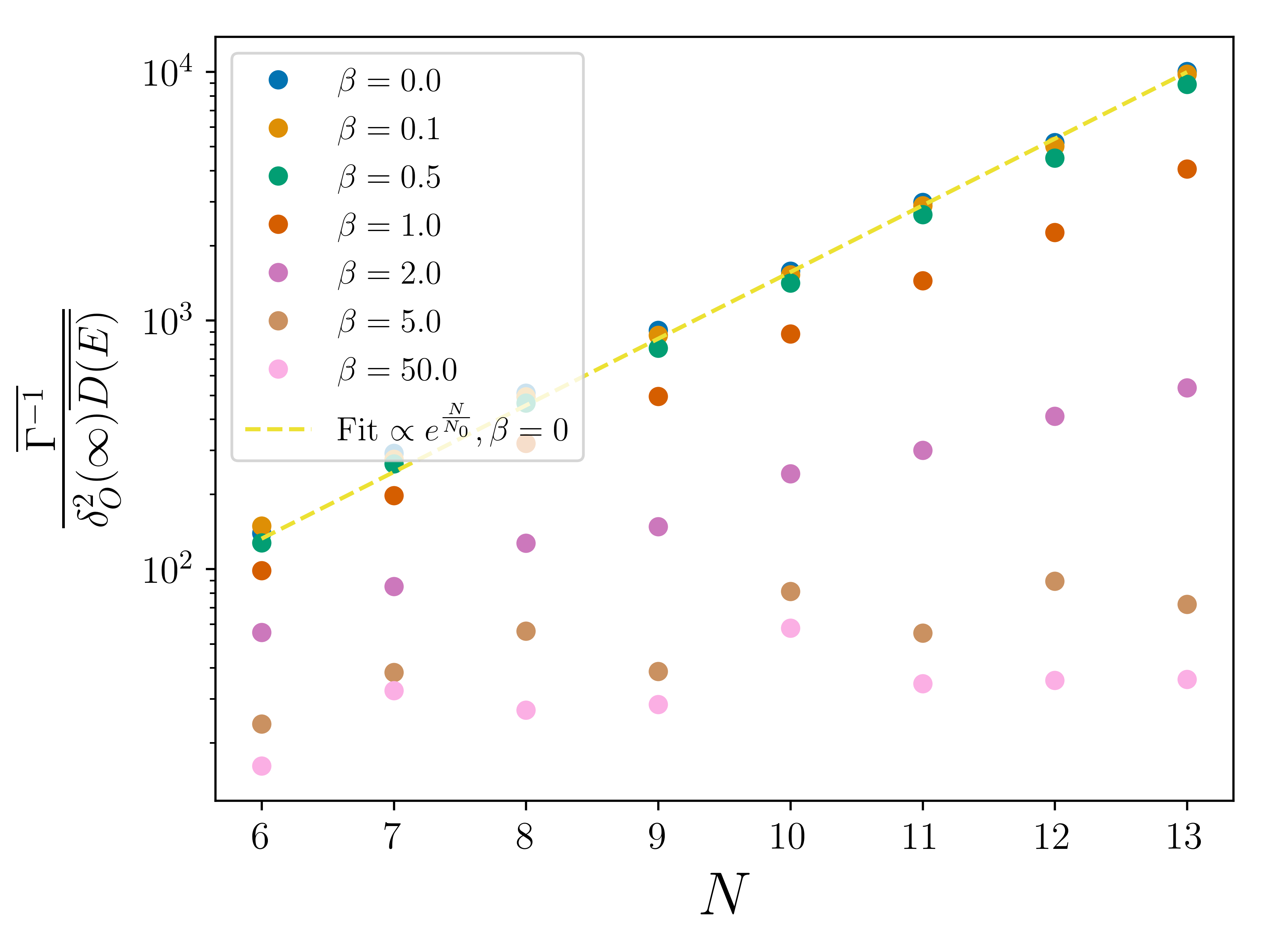}
	\caption{As in Fig. \ref{fig:fig3}, however accounting for the exponential scaling of the average DOS with device size. Fit (yellow dashed line) is performed to the function $a e^{\frac{N}{N_0}}$ for the infinite temperature case, $\beta = 0$, and thus confirms the exponential scaling of $\chi(N)\overline{D(E)} \sim \frac{1}{{\cal N}_B}$. In this case, we observe an exponential scaling for only high temperatures, and confirm that for low temperatures $\chi(N)\overline{D(E)}$ is independent of $N$. Note that here we have $\Gamma \sim 0.2$ so the high temperature limit is defined by approximately $\beta \ll 5$. Parameters: as Fig. \ref{fig:fig2} with $J_x^{(SB)} = 0.5$.}
	\label{fig:fig4} 
\end{figure}

It is important to note that this exponential scaling of $\chi(N)$, Eq. \eqref{eq:exp_scaling}, is expected from not only the contribution of ${\cal N}_B$, but also from the average DOS, $\overline{D(E)}$. This average is often trivially obtained, as for example, for an ensemble of $N$ two-level systems $\overline{D(E)} = \frac{1}{\Delta E}\int_0^{\Delta E} dE D(E) = \frac{2^N}{\Delta E}$, where $\Delta E$ is the range of energies available $E_{max} - E_{min}$ (which may itself change with $N$), regardless of the microscopic properties of the DOS. We thus also study the quantity $\chi(N) \overline{D(E)}$, as this quantity has no dependence on the DOS, and an observation of the exponential scaling in system size is confirmation that, indeed, ${\cal N}_B \propto e^{c N}$. This is shown in Fig. \ref{fig:fig4}, where we observe an exponential scaling of the Hilbert space dimension, with $c \approx 0.62$, compared to $\ln(2) \sim 0.69$ if the entire Hilbert space were explored in the dynamics. 

We further observe in Figs. \ref{fig:fig3} and \ref{fig:fig4}, that the FDT similarly applies at finite temperatures $\beta = \frac{1}{k_BT} > 0$. The extension of our theoretical approach to this case is discussed below, with additional details given in Appendix \ref{App:Thermal_Averages}. Indeed, we can show that for high temperatures, such that $\beta^{-1} \gg \Gamma$, we obtain an FDT of the same form as Eq. \eqref{fdt.T}, by employing a high energy cut-off $(\rho_B)_{\alpha\alpha} \sim e^{-\beta E_{\alpha}}$ to the bath state occupation.

For finite temperatures we show below that the FDT depends on the partition function $Z_\beta$ itself, rather than the Hilbert space dimension. Indeed, one can see that in the infinite temperature limit $Z_0 = \lim_{\beta \to 0} \sum_\alpha e^{-\beta E_\alpha}\delta_{\alpha, \textrm{odd}} = {\cal N}_B$.

\begin{figure}
		\includegraphics[width=0.48\textwidth]{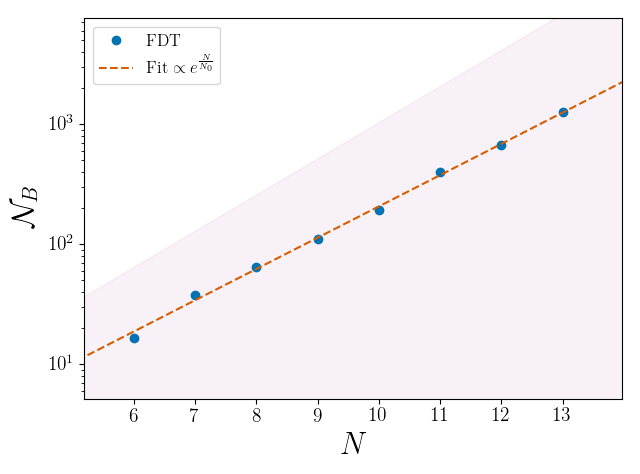}
	\caption{Hilbert space dimension of the bath ${\cal N}_B$ for varying device size $N$ inferred from the FDT, Eq. \eqref{fdt.T} (blue dots), $C = \frac{3}{8 \pi}$, which can be easily obtained from the observable (see Section \ref{sec:FDT}). For a fully connected device the maximum possible ${\cal N}_B$ is ${\cal N}_B = 2^{N-1}$, the region below this limit is shaded. We observe that the measured Hilbert space dimension indeed scales exponentially with system size (dashed line is exponential fit), though is somewhat smaller than the maximum dimension of a fully connected device. Parameters as in Fig. \ref{fig:fig2}, with $J_x^{(SB)} = 0.5$ and $\beta = 0$. }
	\label{fig:H_dim} 
\end{figure}

Finally, we see that when the observable is found to decay exponentially to its equilibrium value, this indicates that the decay rate is approximately constant over the bulk of the initially occupied states. This is shown in Section \ref{sec:Time_dep}. Exploiting this observation, we are able to directly obtain the Hilbert space dimension, as for an infinite temperature initial bath state the average $\overline{\Gamma^{-1}}$ is equal to the measured decay rate.
The bath Hilbert space dimension ${\cal N}_B$, as calculated from Eq. \eqref{fdt.T}, is plotted for varying device sizes in Fig. \ref{fig:H_dim}. Here we observe that ${\cal N}_B$ indeed increases exponentially with systems size, yet is somewhat smaller than its maximum possible value $2^{N-1}$, which is expected to the locality of interactions within the chain.

We again note that in Fig. \ref{fig:H_dim} the measurement of ${\cal N}_B$ is a measurement of the explored Hilbert space, or the total number of eigenstates that contribute to the evolution of the initial state. Thus, for a maximally connected device this would be $2^{N-1}$, whereas locality of interactions in this case restricts some areas of the Hilbert space.

We note that for models where $\Gamma(E)$ is not approximately constant in energy, which would be marked by a deviation from the exponential decay in Eq. \eqref{eq:TDepfin}, one would still have access to Figs. \ref{fig:fig2}-\ref{fig:fig4}, and thus scaling information of the Hilbert space dimension, yet the numerical value ${\cal N}_B$ would be obscured. This is explained in more detail in Section \ref{sec:Time_dep}, where we see how information on $\Gamma(E)$ is extracted from the decay, and in Section \ref{sec:FDT} and Appendix \ref{App:Thermal_Averages}, where we see how this relates to the observed FDT. The key detail is that when the FDT is expressed in terms of the same value measured from the decay it differs by a constant, thereby leaving the scaling of ${\cal N}_B$ with system size or decay rate accessible, yet obscuring the numerical value. 

We note that in Ref. \cite{Nation2019}, the current authors obtained a FDT for pure states, which can be seen to be recovered in the low temperature limit, $\beta^{-1} \ll \Gamma$, for which $\chi(N)$ does not depend explicitly on the Hilbert space dimension ${\cal N}_B$. This can also be analytically seen to be the same as the low temperature limit of our treatment below, which indicates that there is a smooth transition between these two cases. This is indeed observed in the numerics of Figs \ref{fig:fig3} and \ref{fig:fig4}.

\section{Model}\label{sec:Model}
\subsection{RMT Approach}

Our approach relies on the calculation of correlation functions from a statistical theory of random wave functions $c_\mu(\alpha)$. 
Here we summarize the essential ingredients to our model, and give details on the calculations in the sections below. 

Our theory, developed in Ref. \cite{Nation2018} by extending Deutch's RMT model \cite{Deutsch1991, Deutscha, Reimann2015}, can be used to obtain arbitrary correlation functions 
$\langle c_\mu(\alpha)c_\nu(\alpha)\cdots \rangle_V$, where 
$\langle \cdots \rangle_V$ denotes the ensemble average
over an ensemble of random matrix perturbations, $V$, for a ${\cal N} \times {\cal N}$ Hamiltonian of the form \eqref{eq:H_rm}, with $(H_0)_{\alpha\beta} = \alpha \omega_0 \delta_{\alpha\beta}$, with $\omega_0 = \frac{1}{{\cal N}}$ and $V$ a random matrix selected from the GOE, with $\langle V_{\alpha\beta}^2 \rangle_V  = \frac{(1 + \delta_{\alpha\beta})g^2}{{\cal N}}$. In practice, those correlations allow us to calculate any dynamical quantity of interest within the RMT formalism. 

To illustrate the use of such correlation functions we briefly consider the simple example of the diagonal observable matrix elements $O_{\mu\mu}$. These can be written as 
\begin{equation}
O_{\mu\mu} = \sum_{\alpha\beta}c_\mu(\alpha)c_\mu(\beta)O_{\alpha\beta}.
\end{equation}
Now, using the self-averaging property of random matrices, which we prove for this model in Appendix \ref{App:SelfAveraging}, we see that
\begin{equation}\label{eq:diag_example}
O_{\mu\mu} = \sum_{\alpha\beta}\langle c_\mu(\alpha)c_\mu(\beta)\rangle_V O_{\alpha\beta}.
\end{equation}
Note that only the random wave functions $c_\mu(\alpha)$ remain inside the ensemble average, as all other factors are in the non-interacting basis and thus do not depend on $V$. We thus observe that the diagonal observable matrix elements depend on the correlation function $\langle c_\mu(\alpha)c_\mu(\beta)\rangle_V$ (note that we will see how to deal with the summation and non-interacting observable elements in Section \ref{sec:obs_assumptions} below).

More generally, when calculating more complicated quantities, we have that there is also a non-trivial contribution of a four-point correlation function of the form 
$\langle c_\mu(\alpha) c_\nu(\beta) c_\mu(\alpha') c_\nu(\beta') \rangle$ 
which is given by (for $\mu \neq \nu$),
\begin{widetext}
\begin{equation}\label{eq:4pt_off_diag}
\begin{split}
\langle  c_\mu(\alpha) c_\nu(\beta) c_\mu(\alpha^\prime) c_\nu(\beta^\prime) \rangle_V = \Lambda(\mu, \alpha)&\Lambda(\nu, \beta)\delta_{\alpha\alpha^\prime}\delta_{\beta\beta^\prime} \\&
- \frac{\Lambda(\mu, \alpha)\Lambda(\nu, \beta)\Lambda(\mu, \alpha^\prime)\Lambda(\nu, \beta^\prime)}{\Lambda^{(2)}(\mu, \nu)} (\delta_{\alpha\beta}\delta_{\alpha^\prime\beta^\prime} +\delta_{\alpha\beta^\prime}\delta_{\beta\alpha^\prime} ),
\end{split}
\end{equation}
\end{widetext}
where $\Lambda(\mu, \alpha)$ is defined as
\begin{subequations}\label{eq:Lambda}
\begin{eqnarray}
 \langle c_\mu(\alpha)c_\nu(\beta)\rangle_V  =& \Lambda(\mu, \alpha)\delta_{\alpha\beta}\delta_{\mu\nu} \\
 \Lambda(\mu, \alpha) :=& \frac{\omega_0 \Gamma / \pi}{(E_\alpha - E_\mu)^2 + \Gamma^2}
\end{eqnarray}
\end{subequations}
with $\Gamma = \frac{\pi g^2}{{\cal N}\omega_0}$. $\Lambda^{(2)}(\mu, \nu)$ is defined similarly to Eq. (\ref{eq:Lambda}b), with $\Gamma \to 2\Gamma$. The Lorentzian form of Eq. \eqref{eq:Lambda} is found for a homogeneous perturbation $V$ \cite{Deutscha, Nation2018}, selected from the GOE. We obtain that the four-point correlation function, Eq. \eqref{eq:4pt_off_diag}, can be described in terms of product of two-point correlators $\Lambda(\mu, \alpha)$, as if the random wave function distribution was purely Gaussian, plus a correction term originating from the effective interaction due to the mutual orthogonality of random wave functions. We will see below that an approach in terms of Gaussian and non-Gaussian contractions can be formulated to describe more general correlation functions.

We note that our theory can be extended to account for different forms of the quantum chaotic wave function $\Lambda(\mu, \alpha)$. This may appear, for example, for non-homogeneous $V$. In this case, the form of the function $\Lambda$ would change, however the algebraic structure of our theory would remain.

In order to evaluate Eq. \eqref{eq:delta_DE}, we thus use Eq. \eqref{eq:psi_mu} to write $\delta_O^2(\infty)$ in terms of the random wave functions $c_\mu(\alpha)$, and non-interacting matrix elements $\rho_{\alpha\beta}$ and $(\sigma_z^2)_{\alpha\beta}$. We then use the self-averaging property of random matrices (which we discuss in \ref{App:Summary_RMT}, and prove in Appendix \ref{App:SelfAveraging}), and obtain the relevant correlation functions $\langle c_\mu(\alpha)c_\nu(\beta)\cdots \rangle_V$.

\subsection{Computing Correlation Functions}\label{sec:corr_funcs}

As we have seen, it is important to have a systematic approach to obtaining correlation functions for this model. This is a non-trivial task as the random wave functions of our theory are not Gaussian independent variables, but include an effective interaction due to the orthogonality condition $\langle \psi_\mu|\psi_\nu\rangle = \delta_{\mu\nu}$ \cite{Nation2018}. Our approach to the statistical theory of random wave functions is summarized in Appendix \ref{App:Summary_RMT}.

Below, we present such a systematic approach to obtaining arbitrary correlation functions in terms of contractions representing the Gaussian and non-Gaussian terms in the four-point correlator (which is the largest non-factorizable correlation function of our theory).

The four-point correlation function of Eq. \eqref{eq:4pt_off_diag} may be understood in terms of the contractions of non-interacting indices, indeed it can be seen to be the sum of a Gaussian contraction $\acontraction{\langle c_\mu(}{\alpha}{)c_\nu(\beta)c_\mu (}{\alpha^\prime} \bcontraction{\langle c_\mu(\alpha) c_\nu(}{\beta}{)c_\mu(\alpha^\prime)c_\nu(}{\beta^\prime} \langle c_\mu(\alpha) c_\nu(\beta) c_\mu(\alpha^\prime) c_\nu(\beta^\prime)\rangle_V \Rightarrow \langle c_\mu^2(\alpha)\rangle_V\langle c_\nu^2(\beta)\rangle_V\delta_{\alpha\alpha^\prime}\delta_{\beta\beta^\prime} = \Lambda(\mu, \alpha) \Lambda(\nu, \beta) \delta_{\alpha\alpha^\prime}\delta_{\beta\beta^\prime}$ and non-Gaussian contractions, given by
\begin{equation}\label{eq:4legab}
\begin{split}
\acontraction{\langle c_\mu(}{\alpha}{)c_\nu(}{\beta} \acontraction[1.25ex]{\langle c_\mu(}{\alpha}{)c_\nu(}{\beta} \bcontraction{\langle c_\mu(\alpha) c_\nu(\beta) c_\mu(}{\alpha^\prime}{)c_\nu((}{\beta} \bcontraction[1.25ex]{\langle c_\mu(\alpha) c_\nu(\beta) c_\mu(}{\alpha^\prime}{)c_\nu((}{\beta} & \langle c_\mu(\alpha) c_\nu(\beta) c_\mu(\alpha^\prime) c_\nu(\beta^\prime)\rangle_V \\& \qquad \qquad \qquad \qquad \Rightarrow L_{\mu\nu}^{\alpha \beta \alpha^\prime \beta^\prime}\delta_{\alpha\beta}\delta_{\alpha^\prime\beta^\prime} \\ &
\bcontraction{\langle c_\mu(}{\alpha}{)c_\nu(\beta)c_\mu (\alpha^\prime)c_\nu(}{\beta}  \bcontraction[1.25ex]{\langle c_\mu(}{\alpha}{)c_\nu(\beta)c_\mu (\alpha^\prime)c_\nu(}{\beta}\acontraction{\langle c_\mu(\alpha) c_\nu(}{\beta}{)c_\mu(}{\beta} \acontraction[1.25ex]{\langle c_\mu(\alpha) c_\nu(}{\beta}{)c_\mu(}{\beta}\langle c_\mu(\alpha) c_\nu(\beta) c_\mu(\alpha^\prime) c_\nu(\beta^\prime)\rangle_V \\& \qquad \qquad \qquad \qquad \Rightarrow L_{\mu\nu}^{\alpha \beta \alpha^\prime \beta^\prime} \delta_{\alpha\beta^\prime}\delta_{\alpha^\prime\beta},
\end{split}
\end{equation}
where
\begin{equation}\label{eq:L}
L_{\mu\nu}^{\alpha \beta \alpha^\prime \beta^\prime} := \frac{\Lambda(\mu, \alpha)\Lambda(\nu, \beta)\Lambda(\mu, \alpha^\prime)\Lambda(\nu, \beta^\prime)}{\Lambda^{(2)}(\mu,\nu)}.
\end{equation}
We reserve the double line contraction notation of Eq. \eqref{eq:4legab} for the non-Gaussian case. Note that these must occur in pairs of contractions between different interacting indices $\mu\neq\nu$. 

Now, we can see from Eq. \eqref{eq:4legab} that each contracted pair of indices contributes a Kronecker-$\delta$ symbol, and thus, when the correlation function is summed over its non-interacting indices, the number of summations is reduced. We see that as each $\Lambda$ contributes a factor on the order ${\cal O}(\frac{\omega_0}{\Gamma})$, and a summation on the order $ {\cal O}(\frac{\Gamma}{\omega_0})$, a reduced summation will act to render a term negligible in comparison to a term with no such reduction. Further, we see that the contribution of the non-Gaussian term Eq. \eqref{eq:L} is of order ${\cal O}(\frac{\omega_0^3}{\Gamma^3})$, whereas that of the Gaussian term is $\sim \Lambda^2$, and thus ${\cal O}(\frac{\omega_0^2}{\Gamma^2})$, and as such, one can see that for the non-Gaussian contractions to contribute, they must be acted on my an extra summation. Indeed, one can see that this occurs for one of the two non-Gaussian terms when one has repeated summations, i.e. $\alpha^\prime, \beta^\prime \to \alpha, \beta$ in Eq. \eqref{eq:4legab}. 

For further details we refer the reader to Ref. \cite{Nation2019}, and the calculations in Appendices \ref{App:Contractions} and \ref{App:SelfAveraging}. Here we have seen the key intuition, however: that repeated indices in correlation functions leads to the dominant contribution of contractions that would otherwise have contracted the pair of equal indices.

\subsection{Assumptions on Observables}\label{sec:obs_assumptions}
After obtaining the relevant correlation function, one needs to perform the summations over remaining indices. See, for example, the simple case of Eq. \eqref{eq:diag_example} above. To perform the summations, certain assumptions on the form of observable matrix elements in the non-interacting basis must be made. We note that in this basis, local system observables are usually of a known form. 

Our theory relies on assumptions that we expect to be satisfied for such local observables. The key assumption is related to the behaviour of matrix elements, which must have a well defined average, that does not vary pathologically with energy. 
A more formal definition of our assumption can be written in terms of the function $\Lambda$, as,
\begin{equation}\label{eq:def_MCaverages1}
\sum_\alpha \Lambda(\mu, \alpha) \Lambda(\nu, \alpha) O_{\alpha\alpha} = \overline{[O_{\alpha\alpha}]}_{\overline{\mu}}\Lambda^{(2)}(\mu, \nu),
\end{equation}
with $E_{\overline{\mu}} := \frac{E_\mu + E_\nu}{2}$, and 
\begin{equation}\label{eq:micro_ave}
\overline{[O_{\alpha\alpha}]}_{\overline{\mu}} = \sum_\alpha \Lambda(\overline{\mu}, \alpha) O_{\alpha\alpha}.
\end{equation}
We will see that this assumption
will be necessary in order to compute summations over the non-interacting indices. In this section we explain in more detail the requirements on the form of $O_{\alpha\alpha}$ for Eq. \eqref{eq:def_MCaverages1} to be valid, as well as the physical interpretation of the assumption.

The essential assumption here, which we label \emph{smoothness} of $O_{\alpha\alpha}$, as in Ref. \cite{Nation2019}, is that the microcanonical average $\overline{[O_{\alpha\alpha}]}_{\overline{\mu}}$ changes slowly over the width $\Gamma$ of the function $\Lambda(\mu, \alpha) \Lambda(\nu, \alpha)$. We showed in Ref. \cite{Nation2019} that this is the case under the assumptions,
\begin{eqnarray}
&& \frac{\Gamma}{\omega_0} \gg 1, \nonumber \\
&& \Gamma^2 
\frac{d^2}{d E_\mu^2} \overline{[O_{\alpha,\alpha}]}_{\mu} \ll 1,
\label{smoothness}
\end{eqnarray}
which thus leads us to two reasonable conditions,
\begin{enumerate}
\item There are many states in the energy width $\Gamma$
\item The microcanonical average changes slowly over the width $\Gamma$. 
\end{enumerate}
We note that the latter condition, combined with the fact that the microcanonical average and time average are equal (which is shown below), is equivalent to the statement that the time-average of the observable is not sensitive to the particular initial state (microstate), rather, it's macroscopic energy. In fact, one can see that the conditions \eqref{smoothness} are precisely those required in order to define a microcanonical average that does not vary pathologically with small changes in the energy window. In this sense, this assumption is the minimal assumption one would expect to require for thermalization to a microcanonical average that changes smoothly with initial state energy to occur.

We further note that in the consideration of time evolution below, we will consider more general observables that are not necessarily diagonal in the non-interacting basis, but fulfil a \emph{sparsity} condition. This can be written as $\sum_{\alpha\beta}O_{\alpha\beta} = \sum_\alpha \sum_n^{N_O} O_{\alpha, \alpha+n}\delta_{\beta, \alpha+n}$, where for a given observable there is a non-extensive number $N_O$ of groups of non-zero matrix elements at given energy widths, such that after the course graining procedure the observable matrix elements are non-zero for energy gaps $E_\alpha - E_\beta$ that are the possible energy gaps of $H_S$. This form can be seen \cite{Nation2019} to be reasonable for local observables. We note that it is of course possible to find observables that do not fulfil this assumption, although it is easily seen to be true for e.g. local Pauli operator observables. We will see below that our treatment of time evolution could potentially also capture a wider range of observables as well, if the form in the non-interacting basis is known. In the following we refer to observables fulfilling the above assumptions as `generic' observables.

\section{Equilibration Dynamics}\label{sec:Time_dep}

In this section we present a description of the time dependence of `generic' observables as defined above, from an arbitrary initial condition
\begin{equation}\label{eq:rho0_full}
\rho(0) = \sum_{\alpha\beta}w_{\alpha\beta}|\phi_\alpha\rangle\langle\phi_\beta|.
\end{equation}
This calculation may be performed by exploiting the methods outlined in Section \ref{sec:Model}. The general approach may be summarized in three steps: i) Writing the observable time dependence in terms of parameters in the non-interacting basis, ii) computing the relevant correlation functions (see Section \ref{sec:corr_funcs}), and iii) performing summations using the assumptions on observables (see Section \ref{sec:obs_assumptions}). 

Proceeding as such, we write the time dependent density operator in the form, 
\begin{equation}\label{eq:DM_t}
\rho(t) = \sum_{\alpha\beta} \sum_{\mu\nu} w_{\alpha\beta} c_\mu(\alpha) c_\nu(\beta) e^{-i(E_\mu - E_\nu)t} |\psi_\mu\rangle \langle \psi_\nu|, 
\end{equation}
which may be used to obtain the time evolved observable expectation value by $\langle O(t) \rangle = \Tr (\rho(t) O)$. By taking the trace over the interacting basis $\{|\psi_\mu\rangle\}$, we thus obtain
\begin{equation}
\begin{split}
\langle O(t) \rangle &= \sum_{\mu^\prime} \langle \psi_{\mu^\prime}|\sum_{\alpha\beta\mu\nu}w_{\alpha\beta}  c_\mu(\alpha)c_\nu(\beta) \\& \times e^{-i(E_\mu - E_\nu)t} |\psi_\mu\rangle \langle \psi_\nu| O |\psi_{\mu^\prime}\rangle.
\end{split}
\end{equation}
Noting the so-called diagonal ensemble contribution is defined by,
\begin{equation}
O_{\textrm{DE}} = \sum_{\alpha\mu}w_{\alpha\alpha} c^2_\mu(\alpha)O_{\mu\mu},
\end{equation}
which can be seen to be equal to the long-time average value of the observable
\begin{equation}\label{eq:time_ave_DE}
\langle \langle O(t) \rangle \rangle_\tau := \lim_{\tau\to\infty}\frac{1}{\tau}\int_0^\tau dt \langle O(t) \rangle = O_{\textrm{DE}},
\end{equation}
assuming no degenerate energy levels, 
we thus define
\begin{equation}\label{eq:DeltaO_def}
\begin{split}
\Delta O(t) :&= \langle O(t)\rangle  - O_{\textrm{DE}} \\&
= \sum_{\alpha\beta}\sum_{\substack{\mu\nu \\ \mu \neq \nu}} w_{\alpha\beta} e^{-i(E_\mu - E_\nu)t} O_{\mu\nu} \\& \qquad \times c_\mu(\alpha)c_\nu(\beta).
\end{split}
\end{equation}
Using that $O_{\mu\nu} = \sum_{\alpha\beta}c_\mu(\alpha)c_\nu(\beta)O_{\alpha\beta}$, we have
\begin{equation}
\begin{split}
\Delta O(t) &= \sum_{\alpha\beta\alpha^\prime\beta^\prime}\sum_{\substack{\mu\nu \\ \mu \neq \nu}} w_{\alpha\beta} e^{-i(E_\mu - E_\nu)t}O_{\alpha^\prime\beta^\prime} \\& \qquad \times c_\mu(\alpha)c_\nu(\beta) c_\mu(\alpha^\prime)c_\nu(\beta^\prime) .
\end{split}
\end{equation}
We see that, using the self-averaging property, this depends on the four-point correlation function given by Eq. \eqref{eq:4pt_off_diag}, such that
\begin{equation}\label{eq:three_terms}
\begin{split}
\Delta & O(t) = \sum_{\substack{\mu\nu \\ \mu \neq \nu}}e^{-i(E_\mu - E_\nu)t} \\& \times \Bigg[ \sum_{\alpha\beta} w_{\alpha\beta}O_{\alpha\beta} \Lambda(\mu, \alpha) \Lambda(\nu, \beta)\\&
 - \sum_{\alpha \beta} w_{\alpha\alpha}O_{\beta\beta}  \frac{\Lambda(\mu, \alpha)\Lambda(\mu, \beta)\Lambda(\nu, \alpha)\Lambda(\nu, \beta)}{\Lambda^{(2)}(\mu, \nu)} \\& 
 - \sum_{\alpha \beta} w_{\alpha\beta}O_{\alpha\beta}  \frac{\Lambda(\mu, \alpha)\Lambda(\mu, \beta)\Lambda(\nu, \alpha)\Lambda(\nu, \beta)}{\Lambda^{(2)}(\mu, \nu)}\Bigg]  .
\end{split}
\end{equation}
The third term can be shown to be negligible, proof of which is given in Appendix \ref{App:Bound}. We note that it is this bound that requires the sparsity assumption above. We now use the smoothness assumption, exploiting Eqs. \eqref{eq:def_MCaverages1} and \eqref{eq:micro_ave}, we obtain,
\begin{equation}\label{eq:Delta_O_1}
\begin{split}
\Delta & O(t) = \langle O(t) \rangle_0 e^{-2\Gamma t} \\ &- \sum_{\substack{\mu\nu \\ \mu \neq \nu}} \sum_{\alpha} \overline{[O_{\alpha\alpha}]}_{\overline{\mu}}w_{\alpha\alpha} e^{-i(E_{\mu} - E_{\nu})t} \Lambda(\mu, \alpha) \Lambda(\nu, \alpha),
\end{split}
\end{equation}
where to obtain the first term one may note that $\Lambda(\mu, \alpha) = \Lambda(\mu - \alpha) = \Lambda(\alpha - \mu)$, and make the change of variables $\mu(\nu) \to \mu(\nu) - \alpha(\beta)$ to perform the integrals over the new variables. Here $\langle O(t) \rangle_0 := \sum_{\alpha\beta}w_{\alpha\beta}O_{\alpha\beta}e^{-i(E_\alpha - E_\beta)t}$ is the free evolution of the observable under the Hamiltonian $H_0$. The second term may be re-expressed by defining $\tilde{\mu} = \mu - \alpha$, to obtain,
\begin{equation}
\begin{split}
\Delta & O(t) = \langle O(t) \rangle_0 e^{-2\Gamma t} \\ &- \sum_{\substack{\tilde{\mu} \tilde{\nu} \\ \tilde{\mu} \neq \tilde{\nu}}} \sum_{\alpha} \overline{[O_{\alpha\alpha}]}_{m}w_{\alpha\alpha} e^{-i(E_{\tilde{\mu}} - E_{\tilde{\mu}})t} \Lambda(\tilde{\mu}) \Lambda(\tilde{\nu}),
\end{split}
\end{equation}
where $E_m := \frac{E_{\tilde{\mu}} + E_{\tilde{\nu}}}{2} + E_\alpha$. Noting, then, that as $\Lambda(\tilde{\mu})$ is peaked around $E_{\tilde{\mu}} = 0$, and that $\overline{[O_{\alpha\alpha}]}_{\alpha}$ changes slowly over the width $\Gamma$ of the function $\Lambda$, we can make the replacement $\overline{[O_{\alpha\alpha}]}_{m} \to \overline{[O_{\alpha\alpha}]}_{\alpha}$. This allows the summations over $\tilde{\mu}, \tilde{\nu}$ to be performed, which become Fourier transforms of the Lorentzian functions $\Lambda$ in the continuum limit $\sum_\mu \to \int \frac{dE_\mu}{\omega_0}$. We thus find,
\begin{equation}\label{eq:Delta_O_2}
\begin{split}
\Delta O(t)= \langle O(t) \rangle_0 & e^{-2\Gamma t} \\& - \sum_{\alpha} \overline{[O_{\alpha\alpha}]}_{\alpha}w_{\alpha\alpha} e^{-2\Gamma t}.
\end{split}
\end{equation}

Noting that at $t = 0$ we by definition have $\Delta O(0) := \langle O(0)\rangle - O_{\textrm{DE}} =  \langle O(0)\rangle_0 - \sum_{\alpha} \overline{[O_{\alpha\alpha}]}_{\alpha}w_{\alpha\alpha} $, we obtain that  $\sum_{\alpha} \overline{[O_{\alpha\alpha}]}_{\alpha}w_{\alpha\alpha}  = O_{\textrm{DE}}$. Noting Eq. \eqref{eq:time_ave_DE}, we see that the equality of the time and microcanonical averages is derived from our RMT approach. Thus, using the definition in Eq. \eqref{eq:DeltaO_def}, we obtain
\begin{equation}\label{eq:TDepfin_S}
O(t) = (\langle O(t) \rangle_0 - O_{\textrm{DE}}) e^{-2\Gamma t} + O_{\textrm{DE}}.
\end{equation}
This is the same as that obtained in Reference \cite{Nation2019} for pure-states.

The approach outlined above is valid assuming that the decay rate $\Gamma$ is constant in energy. In fact, for the system under consideration, we have to allow $\Gamma$ to change with the initial state energy, such that $\Gamma \to \Gamma_\alpha$ (note that here the change in DOS does not affect the calculation). Accounting for this, rather than Eq. \eqref{eq:Delta_O_1}, we obtain
\begin{equation}\label{eq:Delta_O_Gamma_alpha}
\begin{split}
\Delta  O(t) = \sum_{\alpha\beta}&w_{\alpha\beta}O_{\alpha\beta}e^{-i(E_\alpha - E_\beta)t} e^{-(\Gamma_\alpha + \Gamma_\beta) t} \\ &-  \sum_{\alpha} \overline{[O_{\alpha\alpha}]}_{\alpha}w_{\alpha\alpha} e^{-2\Gamma_\alpha t}.
\end{split}
\end{equation}
Now, for our system we have that, as $H_S = 0$, the microcanonical average $ \overline{[O_{\alpha\alpha}]}_{\mu} = 0$ for all $\mu$ in the bulk of the spectrum. Also, using that for our proposed experimental protocol, both the initial state and observable are diagonal in the non-interacting basis, we have
\begin{equation}\label{eq:Delta_O_Gamma_alpha2}
\Delta O(t) = \sum_{\alpha}w_{\alpha}O_{\alpha\alpha} e^{-2\Gamma_\alpha t}.
\end{equation}
We have that, as the initial state $\langle O(0) \rangle = \sum_{\alpha}w_{\alpha}O_{\alpha\alpha} = 1$ for all initial device states, $O_{\alpha\alpha} = 1$ for all non-zero $w_\alpha$, and thus
\begin{equation}\label{eq:Delta_O_Gamma_alpha3}
\Delta O(t) = \sum_{\alpha}w_{\alpha} e^{-2\Gamma_\alpha t}.
\end{equation}
We thus wish to obtain the value $\Gamma$ that will be obtained when measuring the decay of an observable. To find this, one may simply consider the time integration of the evolution obtained above from the initial state,
\begin{equation}
\rho_\alpha = \frac{1}{Z_\beta}e^{-\beta E_\alpha} \delta_{\alpha, \textrm{odd}},
\end{equation}
describing our probe-bath model, with an initial finite temperature bath state at inverse temperature $\beta$. The time integration is then, 
\begin{equation}
\begin{split}
\lim_{\tau\to\infty} & \int_0^\tau dt \Delta O(t) \\ & 
= \lim_{\tau\to\infty} \int_0^\tau dt \sum_\alpha w_\alpha e^{-2\Gamma_\alpha t} \\&
= -\frac{1}{2}\sum_\alpha \Gamma_\alpha^{-1} w_\alpha \\&
= -\frac{1}{2 Z_\beta}\sum_\alpha \Gamma_\alpha^{-1} e^{-\beta E_\alpha} \delta_{\alpha, \textrm{odd}} \\&
:= -\frac{1}{2}\langle\langle  \Gamma(E)^{-1} \rangle\rangle_\beta,
\end{split}
\end{equation}
where we have used in the second line that $O_{\textrm{DE}} = 0$, and defined the thermal average $\langle \langle \cdots \rangle\rangle_\beta$ at inverse temperature $\beta$, and we have defined $\Gamma_\alpha$ as the decay rate of the initial state $|\phi_\alpha\rangle$. We thus see that it is the thermal average of the inverse decay rate that is measured by a fit to the time dependence of an observable. 

The integral form of the thermal average of a function $A(E)$, is given by,
\begin{equation}
\langle\langle A(E) \rangle\rangle_{\beta} := \frac{1}{Z_\beta^\prime}\int_0^{\Delta E} dE D(E) e^{-\beta (E - E_0)} A(E),
\end{equation}
with $Z_\beta^\prime := \int_0^{\Delta E} dE D(E) e^{-\beta (E-E_0)}$. We will see below that the FDT will be initially expressed in terms of a different average over the values $\Gamma(E)$. This difference is resolved in Appendix \ref{App:Thermal_Averages}, where we re-express our FDT in terms of the thermal average above. We show that the form differs only by a constant that is independent of $N$ and the coupling strength, and thus the scaling with Hilbert space dimension remains the same, and this difference is not important for our application.

Finally, we note that when $\Gamma_\alpha \approx \Gamma$ is approximately constant across the bulk of the initially populated initial states, the thermal average above is approximately equal to the unbiased average appearing in \eqref{fdt.T}. We also see that in this case, from Eq. \eqref{eq:Delta_O_Gamma_alpha3}, one expects to observe an exponential decay at the rate $\Gamma$, as in Eq. \eqref{eq:TDepModel}. Indeed, this is what we observe in our numerical example in Section \ref{sec:Num_exp} above, and thus we are able to recover the Hilbert space dimension directly in Fig. \ref{fig:H_dim}. If a non-exponential decay is observed, then the average $\langle \langle \Gamma(E)^{-1} \rangle \rangle_\beta$ is obtainable via integration as above, and the scaling of the Hilbert space dimension is still obtainable as in Fig. \ref{fig:fig4}.

\section{Fluctuation-Dissipation Theorem}\label{sec:FDT}

\subsection{Derivation from RMT}\label{sec:FDT_RMT}
Here we perform the full derivation of the FDT for the random matrix model described above. We initially focus on the case of a diagonal initial bath state $\rho_{\alpha\beta} = w_\alpha \delta_{\alpha\beta}$. We then restrict the treatment to the specific protocol outlined in Section \ref{sec:Proposal}, where the initial state is the product of a single probe qubit in a pure state, and a bath in an infinite temperature state, see Eq. \eqref{eq:rho_0}. We will follow a very similar steps as those outlined in the previous section, however we will see here that the correlation function calculation is somewhat more complicated.

The RMT model here is limited to the case of constant decay rate and DOS, we will thus extend the treatment to more realistic cases in the next section.

We are interested in the calculation of the long-time fluctuations, defined by the diagonal ensemble result,
\begin{equation}\label{eq:delta_DE_S}
\begin{split}
\delta_{\sigma_z}^2(\infty) &= \sum_{\substack{\mu,\nu \\ \mu \neq \nu}}
|\rho_{\mu \nu}|^2 |(\sigma_z)_{\mu \nu}|^2.
\end{split}
\end{equation} 
We begin be writing the initial density operator matrix elements as,
\begin{equation}
\rho_{\mu\nu} = \sum_\alpha w_{\alpha} c_\mu(\alpha)c_\nu(\alpha),
\end{equation}
then using Eqs. \eqref{eq:delta_DE_S}, and that $|\psi_\mu\rangle = \sum_\alpha c_\mu(\alpha)|\phi_\alpha\rangle$, we may write the time fluctuations as,
\begin{equation}
\begin{split}
\delta_O^2 & (\infty) = \sum_{\substack{\mu\nu \\ \mu\neq\nu}}\sum_{\alpha\beta}w_\alpha w_\beta c_\mu(\alpha)c_\nu(\alpha)c_\mu(\beta)c_\nu(\beta) \\& 
\times\sum_{\alpha^\prime \beta^\prime}O_{\alpha^\prime\alpha^\prime}O_{\beta^\prime\beta^\prime} c_\mu(\alpha^\prime)c_\nu(\alpha^\prime)c_\mu(\beta^\prime)c_\nu(\beta^\prime),
\end{split}
\end{equation}
where coefficients of the initial state are labelled as unprimed indices $\alpha, \beta$, and coefficients of the observable are labelled by primed indices.

Using the self-averaging property of random matrices (see Appendices \ref{App:Summary_RMT} and \ref{App:SelfAveraging}), we may replace the product of coefficients $c_\mu(\alpha)c_\nu(\alpha)\cdots$ by their ensemble average 
$\langle c_\mu(\alpha)c_\nu(\alpha)\cdots \rangle_V$; the above expression may then be written in terms of a sum over 8-point correlation functions, weighted by the initial state and observable coefficients $w_\alpha$ and $O_{\alpha\alpha}$:
\begin{widetext}
\begin{equation}\label{eq:flucs_full_S}
\delta_O^2(\infty)= \sum_{\substack{\mu\nu \\ \mu\neq\nu}}\sum_{\alpha\beta\alpha^\prime\beta^\prime}w_\alpha w_\beta O_{\alpha^\prime\alpha^\prime}O_{\beta^\prime\beta^\prime} \langle c_\mu(\alpha)c_\nu(\alpha)c_\mu(\beta)c_\nu(\beta) c_\mu(\alpha^\prime)c_\nu(\alpha^\prime)c_\mu(\beta^\prime)c_\nu(\beta^\prime)\rangle_V.
\end{equation}
\end{widetext}
Now, using the method of contractions outlined in Section \ref{sec:corr_funcs}, we see that this 8-point correlation function may be split up into to four-point correlation functions, each consisting of both Gaussian and non-Gaussian contractions. These are computed explicitly in Appendix \ref{App:Contractions}, in which we see that there are three dominating contributions to the fluctuations, given by,
\begin{equation}
\begin{split}
& \delta_G^2(\infty)  = \sum_{\substack{\mu\nu \\ \mu\neq\nu}} \bigg[ \overline{[w_\alpha^2]}_{\overline{\mu}}\overline{[O_{\alpha\alpha}^2]}_{\overline{\mu}} \\& \qquad \qquad \qquad + 2 \overline{[w_\alpha O_{\alpha\alpha}]}_{\overline{\mu}}^2\bigg]\Lambda^{(2)}(\mu, \nu)^2,
\end{split}
\end{equation}
\begin{equation}
\delta_{NG}^2(\infty) = 3\sum_{\substack{\mu\nu \\ \mu\neq\nu}}\overline{[w_\alpha]}_{\overline{\mu}}^2\overline{[O_{\alpha\alpha}]}_{\overline{\mu}}^2\Lambda^{(2)}(\mu, \nu)^2, 
\end{equation}
and,
\begin{equation}
\begin{split}
&\delta_{M}^2(\infty) = -\sum_{\substack{\mu\nu \\ \mu\neq\nu}}\Lambda^{(2)}(\mu, \nu)^2\bigg[\overline{[w_\alpha^2]}_{\overline{\mu}}\overline{[O_{\alpha\alpha}]}_{\overline{\mu}}^2 \\&  + 4\overline{[w_\alpha]}_{\overline{\mu}}\overline{[O_{\alpha\alpha}]}_{\overline{\mu}} \overline{[w_\alpha O_{\alpha\alpha}]}_{\overline{\mu}} + \overline{[w_\alpha]}_{\overline{\mu}}^2\overline{[O_{\alpha\alpha}^2]}_{\overline{\mu}}\bigg].
\end{split} 
\end{equation}
These three terms can be seen as the contributions to the 8-point correlation function arising due to products of Gaussian, non-Gaussian, and mixed Gaussian and non-Gaussian 4-point correlation functions respectively. In the above, in order to perform the summations over non-interacting indices in Eq. \eqref{eq:flucs_full_S} we define course grained averages of observable elements $O_{\alpha\alpha}$ as in Eq. \eqref{eq:def_MCaverages1}, as well as the mixed averages,
\begin{equation}\label{eq:def_MCaverages2}
\begin{split}
\sum_\alpha \Lambda(\mu, \alpha) \Lambda(\nu, \alpha) w_\alpha & = \overline{[w_\alpha]}_{\overline{\mu}}\Lambda^{(2)}(\mu, \nu) \\  \sum_\alpha \Lambda(\mu, \alpha) \Lambda(\nu, \alpha) w_\alpha O_{\alpha\alpha} & = \overline{[w_\alpha O_{\alpha\alpha}]}_{\overline{\mu}}\Lambda^{(2)}(\mu, \nu).
\end{split}
\end{equation}

We thus define $W_{\overline{\mu}}$ by
\begin{equation}\label{eq:W_mu_S}
\begin{split}
\delta_O^2(\infty)& = \delta_G^2(\infty) + \delta_{NG}^2(\infty) + \delta_{M}^2(\infty) \\& 
= \sum_{\substack{\mu\nu \\ \mu\neq\nu}} W_{\overline{\mu}} \Lambda^{(2)}(\mu, \nu)^2,
\end{split}
\end{equation}
with,
\begin{equation}\label{eq:W_mu2_S}
\begin{split}
W_\mu &= \overline{[w_\alpha^2]}_\mu\overline{[O_{\alpha\alpha}^2]}_\mu + 2 \overline{[w_\alpha O_{\alpha\alpha}]}_\mu^2 \\& + 3 \overline{[w_\alpha]}_\mu^2\overline{[O_{\alpha\alpha}]}_\mu^2 - \overline{[w_\alpha^2]}_\mu\overline{[O_{\alpha\alpha}]}_\mu^2 \\ & - 4\overline{[w_\alpha]}_\mu\overline{[O_{\alpha\alpha}]}_\mu \overline{[w_\alpha O_{\alpha\alpha}]}_\mu  - \overline{[w_\alpha]}_\mu^2\overline{[O_{\alpha\alpha}^2]}_\mu.
\end{split}
\end{equation}
We now take our bath to be in an initial infinite temperature state, such that $\overline{[w_\alpha]}_{\mu} = \overline{[w_\alpha]} = \frac{1}{2{\cal N}_B}$, and  $\overline{[w_\alpha^2]}_{\mu} = \overline{[w_\alpha^2]} = \frac{1}{2{\cal N}_B^2}$. As such, $W_{\overline{\mu}} = W$ is in fact energy independent, as the probe Hamiltonian $H_S = 0$, so microcanonical averages of probe observables are also energy independent. 
Now, as $\overline{[w_\alpha^2]} = 2\overline{[w_\alpha]}^2$, all terms in $W$ are $\propto \overline{[w_\alpha^2]}$, we define,
\begin{equation}
\begin{split}
W_O & = \frac{W}{\overline{[w_\alpha^2]}} \\&
= \overline{[O_{\alpha\alpha}^2]} + O_{\uparrow}^2 + \frac{3}{2}\overline{[O_{\alpha\alpha}]}^2 - \overline{[O_{\alpha\alpha}]}^2 \\& - 2\overline{[O_{\alpha\alpha}]}O_{\uparrow} - \frac{1}{2}\overline{[O_{\alpha\alpha}^2]},
\end{split}
\end{equation}
where $O_\uparrow = \langle \uparrow| O |\uparrow \rangle$, and we have used that $\overline{[w_\alpha O_{\alpha\alpha}]} = O_{\uparrow}\overline{[ w_\alpha ]}$. We see that $W_O$ is a constant of the order of unity that depends only on the observable (e.g. $W_O = 3/2$ for $O = \sigma_z$). Finally, taking the thermodynamic limit, such that $\sum_{\mu\nu} \to \int_0^{\Delta E} \int_0^{\Delta E} \frac{dE_\mu dE_\nu}{\omega_0^2}$ (not that the diagonal terms in the summation can be seen to be negligible, as they contribute to a higher order in $\frac{\omega_0}{\Gamma}$), we have
\begin{equation}\label{eq:delta_ints}
\delta_O^2(\infty) = \frac{W_O }{2{\cal N}_B^2}\int_0^{\Delta E} \int_0^{\Delta E} \frac{dE_\mu dE_\nu}{\omega_0^2} \Lambda^{(2)}(\mu, \nu)^2,
\end{equation}
which may be evaluated using,
\begin{equation}
\begin{split}
\int_0^{\Delta E} \int_0^{\Delta E} & \frac{dE_\mu  dE_\nu}{\omega_0^2}  \Lambda^{(2)}(\mu, \nu)^2 \\& = \int_0^{\Delta E} dE_\mu \frac{\rm{arccot}\left(\frac{2\Gamma}{\Delta E}\right)}{2\pi \Gamma} \\ &
\approx \frac{\Delta E}{4\pi \Gamma},
\end{split}
\end{equation}
where in the last line we have used that $\Delta E \gg \Gamma$, such that $\rm{arccot}\left(\frac{2\Gamma}{\Delta E}\right) \approx \frac{\pi}{2}$. We then obtain,
\begin{equation}\label{eq:infT_FDT_RM2}
\begin{split}
\delta_O^2(\infty) & = \frac{ W_O \omega_0}{4 \pi {\cal N}_B \Gamma},
\end{split}
\end{equation}
where we have used that $\Delta E := 2{\cal N}_B\omega_0$. 

We can see, then, that Eq. \eqref{eq:infT_FDT_RM2} is of the form of our main result, Eq. (3) of the main text, where $C = \frac{W_O}{4 \pi}$. What follows is to generalize this relation, allowing the DOS and $\Gamma$ to vary in energy, and for finite temperatures.

\subsection{Extension to Realistic Systems}\label{sec:FDT_real}
The key issue with directly applying the RMT results to realistic models is that in general the DOS, and decay rate, are energy dependent, and thus change over the width of the initial state distribution (this is especially important for the high/infinite temperatures considered here). In order to account for this, we must then go back to the evaluation of the integrals over energy, in Eq. \eqref{eq:delta_ints}, and substitute $\Gamma \to \Gamma(E)$, and $\omega_0^{-1} \to D(E)$. This is justified under the assumption that neither $\Gamma(E)$, nor $D(E)$, vary appreciably over the width $\Gamma$. i.e. $\frac{\Gamma(E) - \Gamma(E + \Gamma(E))}{\Gamma(E)} \ll 1$, and $\frac{D(E) - D(E + \Gamma(E))}{D(E)} \ll 1$.

We see then, that the integral in Eq. \eqref{eq:delta_ints} is now
\begin{equation}
\begin{split}
\delta_O^2 & (\infty) = \frac{W_O}{2{\cal N}_B^2} \int_0^{\Delta E} \int_0^{\Delta E} dE_\mu dE_\nu \\& \times \frac{D(E_\mu)D(E_\nu)}{D(E)^2} \left( \frac{2 \Gamma(E) / \pi}{(E_\mu - E_\nu)^2 + 4\Gamma(E)^2}\right)^2,
\end{split}
\end{equation}
where we have used that,
\begin{equation}
\Lambda^{(2)}(\mu, \nu) = \frac{1}{D(E)}\frac{2 \Gamma(E) / \pi}{(E_\mu - E_\nu)^2 + 4\Gamma(E)^2},
\end{equation}
with $E = \frac{E_\mu + E_\nu}{2}$. This can be seen to be valid as long as the above conditions on $D(E)$ and $\Gamma(E)$ hold, that is, as long as they change sufficiently slowly over the energy width $\Gamma$. The contribution of each eigenstate $\Lambda(\mu, \alpha)$ to the fluctuations at a given energy is then that of a local (in energy) random matrix model, with a constant DOS and decay rate. $D(E)$ and $\Gamma(E)$ can then be allowed to change over an energy much wider than a the width of $\Lambda(\mu, \alpha)$, as over such energy widths the contributions of relevant eigenstates are independent. 

Now, we further define $\omega = E_\mu - E_\nu$, and make the change of variables $E_\mu, E_\nu \to \omega, E$, and thus obtain
\begin{equation}
\begin{split}
\delta_O^2 & (\infty) = \frac{W_O}{2{\cal N}_B^2} \int_0^{\Delta E} \int_{-\Delta E}^{\Delta E} dE d\omega \\& \times \frac{D(E+\frac{\omega}{2}) D(E - \frac{\omega}{2})}{D(E)^2} \left( \frac{2 \Gamma(E) / \pi}{\omega^2 + 4\Gamma(E)^2}\right)^2 \\&
 \qquad \approx \frac{W_O}{2{\cal N}_B^2} \int_0^{\Delta E} dE \frac{1}{4\pi\Gamma(E)},
\end{split}
\end{equation}
where in the second line we have assumed that $D(E)$ and $\Gamma(E)$ is approximately constant over the width $\Gamma$. Now, we define the unbiased average of a function $A(E)$ as,
\begin{equation}\label{eq:unbiased_average}
\overline{ A(E) } = \frac{1}{\Delta E} \int_0^{\Delta E} dE A(E),
\end{equation}
(not to be confused with the average $\overline{[\cdots]}_\mu$ above) and see that, noting $\Delta E = \frac{2{\cal N}_B}{\overline{D(E)}}$,
\begin{equation}\label{eq:FDT2}
\begin{split}
\delta_O^2(\infty) & = \frac{W_O}{4\pi{\cal N}_B \overline{D(E)}} \overline{ \Gamma(E)^{-1}} \\&
= C\frac{1}{{\cal N}_B \overline{ D(E)}}\overline{ \Gamma(E)^{-1} },
\end{split}
\end{equation}
where $C = \frac{W_O}{4\pi}$ depends only on the choice of observable. We note that for the random matrix model, as the DOS and $\Gamma(E)$ are both constant in energy, the average $\overline{ \Gamma(E)^{-1}} $ is equal to the thermal average $\langle \langle \Gamma(E)^{-1} \rangle \rangle_{\beta = 0}$ obtained from a fit to the decay of an observable (see Section \ref{sec:Time_dep} above). In the case above, however, where the DOS and $\Gamma(E)$ change in energy, the unbiased average decay rate is not necessarily the same as that obtained from a fit to the decay. We may fix this problem directly, as we do in the last section, where we see that the unbiased thermal averages may be replaced by regular thermal averages weighted by the DOS at the expense of a constant that depends on the functional form of $D(E)$ and $\Gamma(E)$ (but importantly, not on $N$, or the coupling strength). We can also see, that if $\Gamma(E)$ is approximately constant over the width of the DOS, which is often the case in such systems (in fact from Eq. \eqref{eq:Delta_O_Gamma_alpha3} this can be seen to be the case if an exponential decay of the observable is observed), then the biased and unbiased thermal averages of $\Gamma(E)^{-1}$ are approximately equal for $\beta \to 0$, and Eq. \eqref{eq:FDT2} may be directly experimentally confirmed as in Fig. \ref{fig:H_dim}.

\subsection{Finite Temperature FDT}\label{sec:FDT_fin}

In this section we extend the above approach to finite temperature initial bath states, where the initial state is described by
\begin{equation}\label{eq:rho_0_S}
\rho(t = 0) = \sum_{\alpha }^{2 {\cal N}_B}
w_\alpha |\phi_\alpha \rangle \langle\phi_\alpha |,
\end{equation}
where the joint probe-bath Hamiltonian eigenbasis is built by ordering product states such that 
$| \phi_\alpha \rangle = 
|\uparrow \rangle | \phi_{B,\frac{\alpha+1}{2}} \rangle  $ 
($\alpha$ odd) , 
$| \phi_\alpha \rangle = 
|\downarrow \rangle | \phi_{B,\frac{\alpha}{2}} \rangle  $ 
($\alpha$ even).  
In this case, we have
\begin{equation}
w_\alpha =
	\begin{cases*}
      \frac{1}{Z_\beta} e^{- \beta E_\alpha} & if $\alpha \in $ odd \\
      0       & otherwise
    \end{cases*},
\end{equation}
where $Z_\beta = \sum_\alpha e^{-\beta E_\alpha}\delta_{\alpha, \textrm{odd}}$, when the bath is initially a finite temperature state at inverse temperature $\beta = \frac{1}{k_BT}$, and the probe qubit is initially in state $|\uparrow\rangle$. We thus obtain for the microcanonical averages of $w_\alpha$, assuming that $\beta^{-1} \gg \Gamma$,
\begin{equation}
\overline{[w_\alpha]}_\mu = \frac{1}{2 Z_\beta} e^{- \beta E_\mu}
\end{equation}
and
\begin{equation}
\overline{[w_\alpha^2]}_\mu = \frac{1}{2 Z_\beta^2} e^{- 2\beta E_\mu},
\end{equation}
such that $\overline{[w_\alpha^2]}_\mu = 2\overline{[w_\alpha^2]}_\mu$. Now, our most general form for the long-time fluctuations (which assumes only the ability to define the required microcanonical averages that vary smoothly over a width $\Gamma$) is
\begin{equation}\label{eq:general_flucs}
\begin{split}
\delta_O^2(\infty) = \sum_{\substack{\mu\nu \\ \mu\neq\nu}} W_{\overline{\mu}} \Lambda^{(2)}(\mu, \nu)^2,
\end{split}
\end{equation}
where $E_{\overline{\mu}} := \frac{E_\mu + E_\nu}{2}$, and $W_\mu$ is written in Eq. \eqref{eq:W_mu2_S}.
Indeed, noting that the mixed average $\overline{[w_\alpha O_{\alpha\alpha}]}_\mu = 2\overline{[w_\alpha]}_\mu O_{\uparrow}$, where $O_{\uparrow} := \langle \uparrow|O|\uparrow\rangle$, and that as each term in $W_\mu$ is $\propto \overline{[w_\alpha]}_\mu^2$ (see Eq. \eqref{eq:W_mu2_S}), we may define
\begin{equation}
W_\mu = W_O \overline{[w_\alpha^2]}_\mu = \frac{W_O}{2 Z_\beta^2}e^{-2\beta E_\mu},
\end{equation}
so
\begin{equation}
\begin{split}
\delta_O^2(\infty)& = \frac{W_O}{2Z_\beta^2}\sum_{\substack{\mu\nu \\ \mu\neq\nu}} e^{-2\beta E_{\overline{\mu}}}\Lambda^{(2)}(\mu, \nu)^2.
\end{split}
\end{equation}
This may be evaluated, including a variable DOS $D(E)$, as in the main text for the infinite temperature case, via
\begin{equation}\label{eq:delta_int_therm}
\begin{split}
\delta_O^2 & (\infty) = \frac{W_0}{2 Z_\beta^2}\int_0^{\Delta E} \int_{-\Delta E}^{\Delta E} dE d\omega e^{-2E\beta} \\& \times \frac{D(E+\frac{\omega}{2})D(E - \frac{\omega}{2})}{D(E)^2} \left( \frac{2 \Gamma(E) / \pi}{\omega^2 + 4\Gamma(E)^2}\right)^2 \\&
\qquad \approx \frac{W_O}{4 Z_\beta^2} \int_0^{\Delta E} dE e^{-2\beta E} \frac{1}{4\pi\Gamma(E)},
\end{split}
\end{equation}
where in the second line we have made the change of variables $E_\mu, E_\nu \to E, \omega$ with $E = \frac{E_\mu + E_\nu}{2}$ and $\omega = E_\mu - E_\nu$, and used that $\Delta E \gg \Gamma$ as in the main text. We now define the unbiased thermal average of the function $A(E)$ as,
\begin{equation}
\langle A(E) \rangle_{\beta} := \frac{1}{\Delta E^\prime(\beta)}\int_0^{\Delta E} dE A(E)e^{-\beta E},
\end{equation}
where $\Delta E^\prime(\beta) = \int_0^{\Delta E} dE e^{-\beta E}$. Now, we have
\begin{equation}
\delta_O^2(\infty) = \frac{W_O \Delta E^\prime(\beta)}{8\pi Z_\beta^2} \langle \Gamma(E)^{-1} \rangle_{2\beta}.
\end{equation}
Noting, then, that $\lim_{\beta\to 0} Z_{\beta} = {\cal N}_B$, and $\lim_{\beta\to 0} \Delta E^\prime(E) = \Delta E = \frac{2 {\cal N}_B}{\overline{D(E)}}$, we recover the infinite temperature case as required,
\begin{equation}
\delta_O^2(\infty) = \frac{W_O}{4\pi{\cal N}_B \overline{D(E)}} \overline{ \Gamma(E)^{-1}}.
\end{equation}
We note that, unlike in the RMT case above, the average $\langle \Gamma(E)^{-1} \rangle_\beta$ is not equal to the thermal average $\langle\langle \Gamma(E)^{-1}\rangle \rangle_\beta$, which is that obtained from a fit to the decay. In Appendix \ref{App:Thermal_Averages} we show how the FDT may be defined in terms of this thermal average. Importantly for our proposed application, we obtain that the FDT in this form is related simply by a constant $C^\prime_\beta$, defined in Appendix \ref{App:Thermal_Averages}, that does not depend on the the size of the device or coupling strength (within the weak coupling regime). For infinite temperatures we thus have
\begin{equation}
\begin{split}\label{eq:Temp_average_FDT_main}
\delta_O^2(\infty) = C^\prime_0 \frac{W_O }{4\pi {\cal N}_B \langle\langle D(E) \rangle \rangle_0} \langle\langle  \Gamma(E)^{-1} \rangle \rangle_0.
\end{split}
\end{equation}
Therefore, we can directly relate the measured inverse decay rate $\langle\langle  \Gamma(E)^{-1} \rangle \rangle_0$ to the time-averaged fluctuations, and from measurement of each for changing device size or coupling strength, as shown in the numerical experiments of Section \ref{sec:Num_exp}, yields information on the scaling of the Hilbert space dimension. We finally note that it is simply the lack of direct knowledge of the constant $C^\prime_\beta$ which prevents measurements where there is a non-negligible change in the decay rate with energy (a non-exponential decay to equilibrium) from constituting a direct measurement of the value of the Hilbert space dimension. This constant depends on the functional form $D(E)$ and $\Gamma(E)$, and thus, if these are known, inference of the Hilbert space dimension itself is thus obtainable.

Finally, we note that the finite temperature approach above can be extended to the low temperature regime, as shown in Appendix \ref{App:Thermal_Averages}, from which we can recover the pure state FDT found in Ref. \cite{Nation2019} in the low temperature limit.

\section{Discussion}\label{sec:Discussion}

The results shown above demonstrate how the chaotic dynamics of thermalization may be exploited in order to gain information on the complexity of the unitary quantum dynamics of a system. We have proposed an experimentally viable protocol, by which measurements of a local observable of a probe qubit may be exploited to measure the Hilbert space dimension of an ergodic quantum device, initialized in an infinite temperature state. We note that this measures the dimension of the states directly involved in dynamics only, and thus provides a more accurate measure of the complexity of the dynamics than a simple estimate of the Hilbert space dimension from the number of qubits. In this sense, such a measurement of a large enough quantum device, if shown to be ergodic in the sense outlined above, would be a convincing indicator of the so called `quantum supremacy' of the quantum device.

On a practical level, for a generic non-integrable Hamiltonian, our results may be observed in two ways: measurement of a probe observable for (i) changing the number of qubits/ions/... in the quantum device (as in Figs. \ref{fig:fig3} and \ref{fig:fig4}) 
, or (ii) changing the probe-bath coupling (as in Fig. \ref{fig:fig2}). The latter is perhaps the simplest experimental methodology, which we show can confirm the ergodic behaviour of a system, that is, that the unitary dynamics requires an extensive proportion of the Hilbert space, by showing a linear relationship between the long-time fluctuations and decay rate. For a model where the device size may be altered, our FDT provides even deeper insight, allowing also for the experimental observation of the scaling of the Hilbert space dimension with system size.

For cases where an exponential decay to equilibrium is observed, which we show implies that the decay rate is constant over a large range of energies, our method allows the experimenter to access the numerical value of the Hilbert space dimension itself, not simply its scaling with size or coupling strength. This can be obtained from a single time trace of the decay to equilibrium of the observable, from the measurement of the decay rate, and fluctuations around equilibrium.

\section{Acknowledgements}
We acknowledge funding from project PGC2018-094792-B-I00  (MCIU/AEI/FEDER, UE), EPSRC grant no.
EP/M508172/1, and from COST Action CA17113.

\onecolumn
\appendix

\section{Summary of RMT Approach}\label{App:Summary_RMT}
\subsection{Random Matrix Model}
The random matrix model under study may be expressed by a non-interacting part
\begin{equation}
(H_0)_{\alpha \beta} = E_\alpha \delta_{\alpha \beta}
\end{equation}
where $E_\alpha = \alpha \omega_0$, and $\omega_0 = 1/{\cal N}$ is the spacing between energy levels, and perturbation term, modelled by a random matrix,
\begin{equation}\label{eq:Hamiltonian}
V_{\alpha \beta}  = h_{\alpha\beta},
\end{equation}
where $h_{\alpha\beta}$ are independent random numbers selected from the Gaussian Orthogonal Ensemble 
(GOE), and rescaled by a coupling strength $g$, such that the matrix $h$ has the probability distribution, 
\begin{equation}
P(h) \propto \exp\left[-\frac{{\cal N}}{4 g^2} \Tr h^2\right],
\end{equation}
giving $\langle h_{\alpha\beta} \rangle = 0$, and $\langle h_{\alpha\beta}^2 \rangle = g^2 / {\cal N}$ for $\alpha \neq \beta$, and otherwise $\langle h_{\alpha\alpha}^2 \rangle = 2g^2 / {\cal N}$. 

In Ref. \cite{Nation2018} the current authors developed a consistent theoretical model of random wave functions $|\psi_\mu\rangle = \sum_\alpha c_\mu(\alpha)|\phi_\alpha\rangle$, for the random matrix model above. 
We make the ansatz on the probability distribution on the $c_\mu(\alpha)$s,
\begin{equation}\label{eq:p_c}
p(c, \Lambda) = \frac{1}{Z_p}e^{-\sum_{\mu\alpha}\frac{c^2_\mu(\alpha)}{2\Lambda(\mu, \alpha)}}\prod_{\substack{\mu\nu\\ \mu>\nu}}\delta(\sum_{\alpha}c_\mu(\alpha)c_\nu(\alpha)),
\end{equation}
 where the $\delta$-function term explicitly accounts for the orthogonality of the many-body eigenstates (which we showed to be necessary in order to obtain a consistent form for the off-diagonal matrix elements). This distribution  $\Lambda(\mu, \alpha)$ can then be shown to be a Lorentzian of width $\Gamma = \frac{\pi g^2}{{\cal N} \omega_0}$ \cite{Deutscha, Nation2018}. From Eq. (\ref{eq:p_c}), one can calculate arbitrary correlation functions of the $c_\mu(\alpha)$ coefficient by first defining the generating function,
\begin{equation}{\label{eq:gen_func_1}}
\begin{split}
G^{(\textrm{od})}_{\mu\nu} (\vec{\xi}_{\mu},\vec{\xi}_{\nu})& = \int \int \exp\bigg[-\sum_\alpha\bigg( \frac{c_\mu^2(\alpha)}{2\Lambda(\mu, \alpha)} + \frac{c_\nu^2(\alpha)}{2\Lambda(\nu, \alpha)} + \xi_{\mu,\alpha} c_{\mu}(\alpha) + \xi_{\nu,\alpha} c_{\nu}(\alpha) \bigg)\bigg] \\& \qquad \qquad \qquad \qquad \qquad \qquad \times
\delta(\sum_{\alpha}c_\mu(\alpha)c_\nu(\alpha))\prod_{\alpha}dc_\mu(\alpha) dc_\nu(\alpha) \\ &
\propto \exp \bigg[ \frac{1}{2} \sum_\alpha \xi^2_{\mu,\alpha} \Lambda(\mu,\alpha) + \frac{1}{2} \sum_\alpha \xi^2_{\nu,\alpha} \Lambda(\nu,\alpha) - 
\\& \qquad \qquad \qquad \qquad \qquad \qquad \frac{1}{2} \sum_{\alpha,\beta} \xi_{\mu,\alpha} \xi_{\mu,\beta} \xi_{\nu,\alpha} \xi_{\nu,\beta}  \frac{\Lambda(\mu,\alpha) \Lambda(\mu,\beta) \Lambda(\nu,\alpha) \Lambda(\nu,\beta)}{\Lambda^{(2)}(\mu, \nu)}\bigg],
\end{split}
\end{equation}
where in the second line we have re-expressed the $\delta$-functions in their Fourier form. The superscript $(\textrm{od})$ indicates that this is the `off-diagonal' generating function, requiring $\mu \neq \nu$. The diagonal case is discussed below.
The correlation functions may then be calculated by performing successive derivatives with respect to the force terms $\xi$ via
\begin{equation}\label{eq:Derivs}
\begin{split}
\langle c_\mu(\alpha)  c_\nu(\beta) \cdots c_\mu(\alpha_1^\prime) c_\nu(\beta_1^\prime) \rangle_V = \frac{1}{G_{\mu\nu}} \partial_{\xi_{\mu,\alpha}} \partial_{\xi_{\nu,\beta}} \cdots \partial_{\xi_{\mu,\alpha_1^\prime}} \partial_{\xi_{\nu,\beta_1^\prime}} G_{\mu\nu} {\bigg |}_{\xi_{\mu,\alpha}=0,\xi_{\nu,\alpha}=0}.
\end{split}
\end{equation}
In particular, the correlation function $\langle c_\mu(\alpha_0)c_\nu(\beta_0)c_\mu(\alpha)c_\nu(\beta)\rangle_V$ was found in \cite{Nation2018} for $\mu\neq \nu$ to be equal to
\begin{equation}\label{eq:Corr_Func}
\begin{split}
\langle c_\mu(\alpha_0)c_\nu(\beta_0) c_\mu(\alpha) c_\nu(\beta) \rangle_{V} & = \Lambda(\mu, \alpha_0)\Lambda(\nu, \beta_0)\delta_{\alpha_0\alpha}\delta_{\beta_0\beta} \\& - \frac{\Lambda(\mu, \alpha_0)\Lambda(\nu, \alpha_0)\Lambda(\mu, \alpha)\Lambda(\nu, \alpha)\delta_{\alpha_0\beta_0}\delta_{\alpha\beta}}{\Lambda^{(2)}(\mu, \nu)}  \\ & - \frac{\Lambda(\mu, \alpha_0)\Lambda(\nu, \alpha_0)\Lambda(\mu, \beta_0)\Lambda(\nu, \beta_0)\delta_{\alpha_0\beta}\delta_{\beta_0\alpha}}{\Lambda^{(2)}(\mu, \nu)},
\end{split}
\end{equation}
with
\begin{equation}
\Lambda^{(n)}(\mu, \nu) := \frac{\omega_0 n\Gamma / \pi}{(E_\mu - E_\nu)^2 + (n\Gamma)^2},
\end{equation}
where the superscript $(n)$ is left out for $n=1$. The latter two terms in Eq. (\ref{eq:Corr_Func}) arise as an explicit result of the orthogonality factor in Eq. (\ref{eq:p_c}). 

We stress here that the generating function Eq. \eqref{eq:gen_func_1} explicitly requires $\mu \neq \nu$, as it models the interactions due to mutual orthogonality of two random wave functions. For the diagonal part, we have the much simpler generating function,
\begin{equation}\label{eq:gen_func_diag}
\begin{split}
G_{\mu\mu}^{(d)} &= \int \prod_\alpha dc_\mu(\alpha) \exp\left[ - \frac{c_\mu^2(\alpha)}{2\Lambda(\mu, \alpha)} + \xi_{\mu, \alpha}c_\mu(\alpha)\right] \\&
\propto  \exp \left[ \frac{1}{2}\sum_\alpha\xi_{\mu, \alpha}^2 \Lambda(\mu, \alpha) \right]
\end{split}
\end{equation}
Thus, we have, from successive derivatives with respect to the force term $\xi_{\mu, \alpha}$ with different non-interacting indices, and taking the limit as $\xi_{\mu, \alpha} \to 0$, as in Eq. \eqref{eq:Derivs},
\begin{equation}\label{eq:CF_diag}
\begin{split}
\langle c_\mu(\alpha)&c_\mu(\beta)c_\mu(\alpha^\prime)c_\mu(\beta^\prime)\rangle_V = \Lambda(\mu, \alpha)\Lambda(\mu, \alpha^\prime)\delta_{\alpha\beta}\delta_{\alpha^\prime\beta^\prime} + \Lambda(\mu, \alpha)\Lambda(\mu, \beta)(\delta_{\alpha\alpha^\prime}\delta_{\beta\beta^\prime} + \delta_{\alpha\beta^\prime}\delta_{\alpha^\prime\beta}),
\end{split}
\end{equation}
for the diagonal case.

\subsection{Self-Averaging}

\begin{figure*}
		\includegraphics[width=0.98\textwidth]{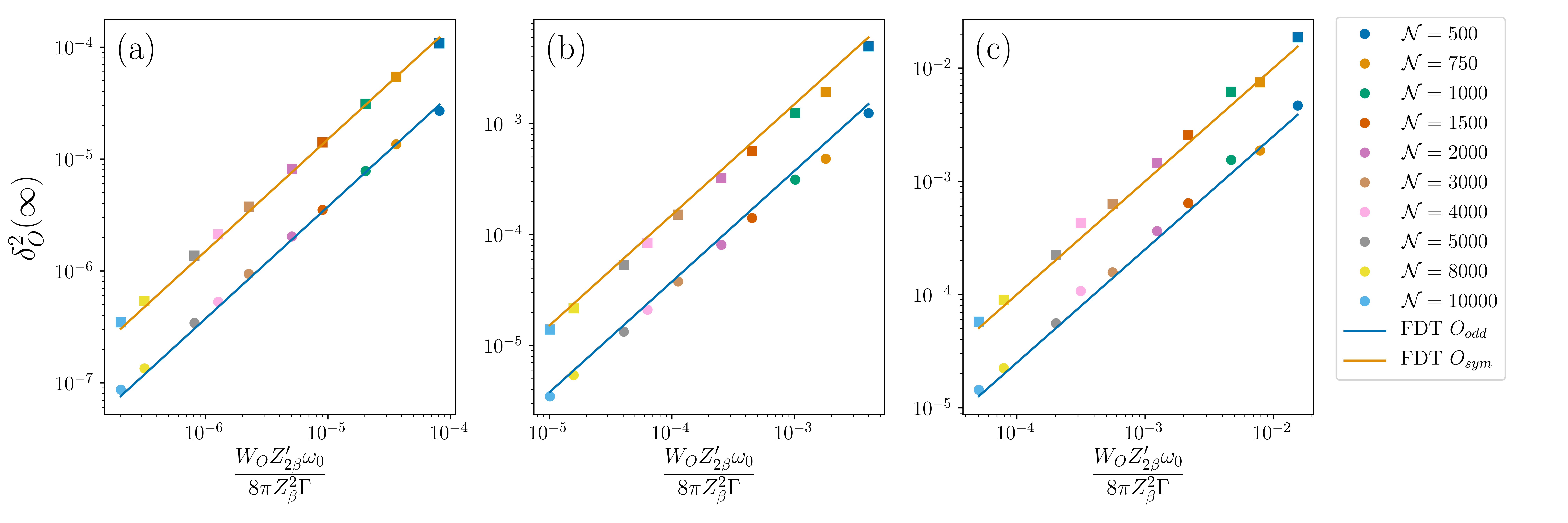}
	\caption{(a) Numerical confirmation of the random matrix FDT for an infinite temperature initial state, Eq. \eqref{eq:RM_FDT} for observables $O_{odd}$ and $O_{sym}$. (b) Shows the random matrix FDT for a high temperature initial state $\beta = 100$, and (c) for a low temperature ($\beta = 500$). Here $g =0.04$, so $\Gamma \sim 0.007$, and thus the high temperature limit $\beta \ll \Gamma^{-1}$ is approximately $\beta \ll 125$. We thus observe that the finite temperature result Eqs. \eqref{eq:FDT2} and \eqref{eq:FDT3} (which we note are equivalent, the latter is used here), is fulfilled for all temperatures. We note that the low temperature limit above uses $\rho_B \sim e^{-\beta(E - E_0)}$, with $E_0 = \frac{E_{max}}{2}$, to ensure that the initial state is not simply the ground state. For this limit we also use $W_O = \overline{[\Delta O^2]}$, as discussed in the final section. Simulations are performed with a single realization of the random matrix $V$, and thus we observe directly the self-averaging property.}
	\label{fig:RM_FDT} 
\end{figure*}

A final property we exploit is that of the self-averaging of random matrices, specifically, in summations we make the assumption that the observable time dependence and fluctuations are equal to the ensemble average over perturbations $V$, such that $\langle O(t)\rangle \approx \langle \langle O(t)\rangle \rangle_V$. This has been previously checked for this model numerically in Refs. \cite{Nation2018, Nation2019}, and can also be seen to be valid numerically in this case from Figs. \ref{fig:RM_FDT} and \ref{fig:RM_time}, which compare our analytical calculations to single realizations of the random matrix Hamiltonian. Self-averaging is a common tool in RMT \cite{Muller2015}, and whilst examples of behaviour that violates a self-averaging assumption indeed exist \cite{Schiulaz2019b, Nation2019b}, these are generally due to additional constraints preventing such statistical behaviour, for global observables such as the survival probability, or in more exotic regimes e.g. close to critical points \cite{Aharony1996, Parisi2002}. 

A recent and timely study \cite{Schiulaz2019b} has looked in detail at the self-averaging behaviour of multiple quantities for both a full GOE, and a realistic spin chain, finding that in general one should be wary of simply applying self-averaging. It is shown, for example, that the survival probability is not self-averaging at any timescale. Indeed, the current authors have recently shown that this quantity cannot be expected to behave ergodically, and thus RMT should not apply to the survival probability \cite{Nation2019b}. In the light of such studies, as well as such counterexamples to self-averaging outlined above, it is important to further justify the use of this property in more detail. 

The definition for self-averaging used in \cite{Schiulaz2019b} is that the vanishing of the quantity
\begin{equation}\label{eq:self_averaging}
{\cal R}_O(t) = \frac{\langle\langle O(t)\rangle^2\rangle_V - \langle\langle O(t)\rangle\rangle_V ^2}{\langle\langle O(t)\rangle\rangle_V ^2},
\end{equation}
as the system size increases. Indeed, a very similar quantity was bounded for this RMT model in Ref. \cite{Dabelow2019}. In Appendix \ref{App:SelfAveraging}, we will show that self-averaging in the sense of the vanishing of Eq. \eqref{eq:self_averaging} indeed occurs for our model using the assumptions on observables outlined above.

We further note that as the realistic Hamiltonians of interest are not necessarily disordered, in the sense that they themselves may have no random component, we do not require that they be self-averaging themselves (indeed this has no real meaning in this case). Rather, the assumption in our case is that when expressed in that non-interacting basis, a chaotic interaction Hamiltonian is suitably `random', such that it resembles a single realization of the ensemble we describe above, which itself is self-averaging. In fact, it is not important that the \emph{entire} interaction Hamiltonian resembles an element of this ensemble, we require instead that locally, for any energy $E$ within the bulk, the interaction Hamiltonian resembles an element of the random matrix ensemble within a width $\sim \Gamma$ from $E$.

\subsection{RMT Numerics}
 Here we confirm our analytical results with numerical calculations with the random matrix Hamiltonian.
 In particular, we show in Fig. \ref{fig:RM_FDT}a, that the infinite temperature fluctuation-dissipation theorem (FDT),
\begin{equation}\label{eq:infT_FDT_RM_S}
\begin{split}
\delta_O^2(\infty) & = \frac{ W_O \omega_0}{4 \pi {\cal N}_B \Gamma},
\end{split}
\end{equation} 
 is satisfied in this model. This is shown for two `observables' of the RMT model, $O_{odd}$ and $O_{sym}$, which are chosen to be diagonal in the non-interacting basis, with diagonal elements given by,
\begin{equation}\label{eq:O_odd}
    (O_{odd})_{\alpha\alpha} =
    \begin{cases*}
      1 & if $\alpha = \textrm{odd}$ \\
      0 & otherwise,
    \end{cases*}
\end{equation}
for $O_{odd}$, and
\begin{equation}\label{eq:O_sym}
    (O_{sym})_{\alpha\alpha} =
    \begin{cases*}
      1 & if $\alpha = \textrm{odd}$ \\
      -1 & otherwise,
    \end{cases*}
\end{equation}
for $O_{sym}$. These observables are chosen, as in Refs. \cite{Nation2018, Nation2019}, as they resemble realistic observables, such as local Pauli operators, in the sense that they are well defined, sparse, and highly degenerate \cite{Anza2018} in the non-interacting basis. For our RMT numerical calculations, we define the initial state as
\begin{equation}
\rho_{\alpha\beta} = e^{-\beta (E_\alpha - E_0)}\delta_{\alpha\beta}\delta_{\alpha, \textrm{odd}},
\end{equation}
such that 
$\langle O(0)\rangle = 1$. The energy shift $E_0$ is simply to avoid edge effects at lower temperatures, where a large fraction of the initial state population would otherwise be in the ground state. 

\begin{figure}
\begin{center}
\includegraphics[width=0.85\textwidth]{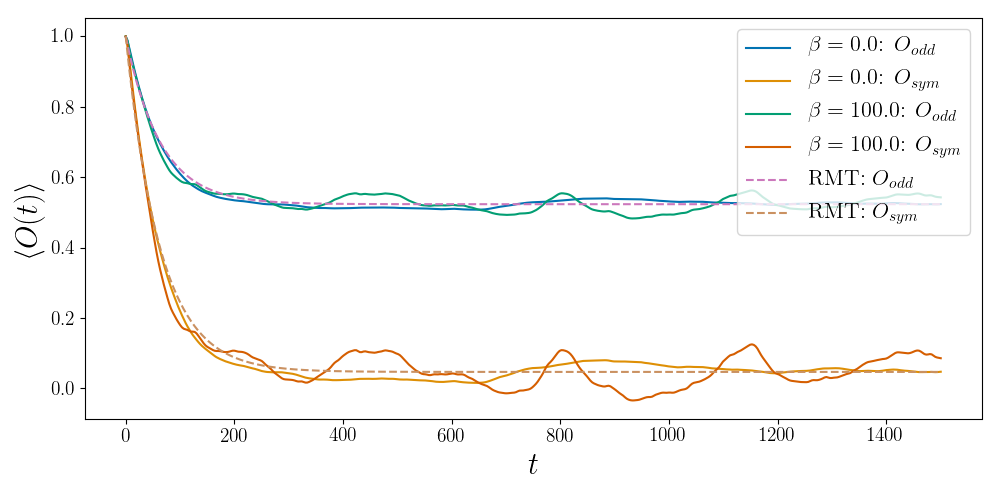}
\end{center}
	\caption{Time dependence of random matrix observables $O_{odd}$ and $O_{sym}$. Exact diagonalization numerics (solid lines) show time evolutions for a single realization of the random matrix perturbation $V$. RMT calculation (dashed lines), show Eq. \eqref{eq:TDepfin_S}, with $\langle O(t)\rangle_0 = 1$, and $\langle O\rangle_{MC} = \overline{[O_{\alpha\alpha}]} = 0 (0.5)$ for $O = O_{sym}(O_{odd})$. Parameters used are $N = 500, g = 0.05, \beta = 0, 100$.}
	\label{fig:RM_time} 
\end{figure}

In Fig. \ref{fig:RM_time} we plot the time dependence of the above observables for an infinite temperature initial state, and compare these to the observable time dependence of Eq. \eqref{eq:TDepfin_S}, derived below.

The high temperature limit, in which our FDT is derived, is defined by $\beta^{-1} \gg \Gamma$. In the numerics, we use $g = 0.05$, and thus $\Gamma \sim 0.007$, so the high temperature limit requires $\beta \lesssim 125$. We show plots for $\beta = 100$ and $\beta = 500$ in Figs. \ref{fig:RM_FDT}b and \ref{fig:RM_FDT}c, respectively. For the parameters used these correspond to a high temperature, near the edge of the expected limit, and a low temperature initial state. We observe that the finite temperature form in fact works well for \emph{all} $\beta$ values. This is further discussed analytically in the final section below, where we see that the low temperature limit of our current approach is equal to the pure state result previously obtained in Ref. \cite{Nation2019}.

\section{Gaussian, Non-Gaussian, and Mixed Contractions}\label{App:Contractions}
Here we calculate the contributions to the long-time fluctuations of Eq. \eqref{eq:flucs_full_S}. We see that the relevant correlation function is an 8-point correlation function, which we will show below may be split into three groups: products of Gaussian contractions, products of non-Gaussian contractions, and mixed products of two Gaussian and one non-Gaussian contraction. 

An example of the first form, Gaussian contractions only, is
\begin{equation}
\acontraction[1.0ex]{\langle c_\mu(}{\alpha}{)c_\nu(\alpha)c_\mu(}{\beta}
  \bcontraction[1.ex]{\langle c_\mu(\alpha)c_\nu(}{\alpha}{)c_\mu(\beta)c_\nu(}{\beta}
  \acontraction[1.0ex]{\langle c_\mu(\alpha)c_\nu(\alpha)c_\mu(\beta)c_\nu(\beta) c_\mu(}{\alpha^\prime}{)c_\nu(\alpha^\prime)c_\mu(}{\beta}
  \bcontraction[1.ex]{ \langle c_\mu(\alpha)c_\nu(\alpha)c_\mu(\beta)c_\nu(\beta) c_\mu(\alpha^\prime)c_\nu(}{\alpha^\prime}{)c_\mu(\beta^\prime)c_\nu(}{\beta}
  \langle c_\mu(\alpha) c_\nu(\alpha)c_\mu(\beta)c_\nu(\beta) c_\mu(\alpha^\prime)c_\nu(\alpha^\prime)c_\mu(\beta^\prime)c_\nu(\beta^\prime)\rangle_V = \Lambda(\mu, \alpha)\Lambda(\nu, \alpha)\Lambda(\mu, \alpha^\prime)\Lambda(\nu, \alpha^\prime)\delta_{\alpha\beta}\delta_{\alpha^\prime\beta^\prime}.
\end{equation}
Indeed, there are three such contractions, occurring each time between pairs of indices, that only contribute two kronecker-$\delta$ factors - $\sim \delta_{\alpha\beta}\delta_{\alpha^\prime\beta^\prime}, \delta_{\alpha\alpha^\prime}\delta_{\beta\beta^\prime}, \delta_{\alpha\beta^\prime}\delta_{\beta\alpha^\prime} $. Other Gaussian contractions may be defined, but may be ignored due to a reduction in the number of summations. 

In a similar manner, we may define non-Gaussian contractions that do not reduce the number of summations at all, such that they contribute on the same order as the Gaussian contractions above. An example is, 
\begin{equation}
 \acontraction[1.25ex]{ \langle c_\mu(}{\beta}{)c_\nu(}{\beta}  
 \acontraction[1.25ex]{ \langle c_\mu(}{\beta}{)c_\nu(\beta)c_\mu(\alpha)c_\nu(}{\alpha} 
 \acontraction{\langle c_\mu(}{\beta}{)c_\nu(}{\beta}  
 \acontraction{\langle c_\mu(\alpha)c_\nu(\alpha)c_\mu(}{\beta}{)c_\nu(}{\beta} 
 \acontraction[1.25ex]{ \langle c_\mu(\alpha)c_\nu(\alpha)c_\mu(}{\beta}{)c_\nu(}{\beta}  \acontraction{\langle c_\mu(\alpha)c_\nu(\alpha)c_\mu(\beta)c_\nu(\beta) c_\mu(}{\alpha^\prime}{)c_\nu(}{\alpha^\prime} 
 \acontraction[1.25ex]{\langle c_\mu(\alpha)c_\nu(\alpha)c_\mu(\beta)c_\nu(\beta) c_\mu(}{\alpha^\prime}{)c_\nu(}{\alpha^\prime} 
  \acontraction[1.25ex]{\langle c_\mu(\alpha)c_\nu(\alpha)c_\mu(\beta)c_\nu(\beta) c_\mu(}{\alpha^\prime}{)c_\nu(\alpha^\prime)c_\nu(}{\beta^\prime} 
  \acontraction{\langle c_\mu(\alpha)c_\nu(\alpha)c_\mu(\beta)c_\nu(\beta)c_\mu(\alpha^\prime)c_\nu(\alpha^\prime)c_\mu(}{\beta^\prime}{)c_\nu(}{\beta^\prime}
   \acontraction[1.25ex]{ \langle c_\mu(\alpha)c_\nu(\alpha)c_\mu(\beta)c_\nu(\beta) c_\mu(\alpha^\prime)c_\nu(\alpha^\prime)c_\mu(}{\beta^\prime}{)c_\nu(}{\beta^\prime} 
   \langle c_\mu(\alpha) c_\nu(\alpha)c_\mu(\beta) c_\nu(\beta)c_\mu(\alpha^\prime)c_\nu(\alpha^\prime)c_\mu(\beta^\prime)c_\nu(\beta^\prime)\rangle_V =  L_{\mu\nu}^{\alpha\alpha\beta\beta}L_{\mu\nu}^{\alpha^\prime\alpha^\prime\beta^\prime\beta^\prime}.
\end{equation}
It can be easily seen that there are three non-Gaussian contractions of this form, with the other two being defined by swapping pairs of primed and unprimed indices in turn.

Finally, we see that mixed contractions may also contribute, for example,
\begin{equation}
 \acontraction[1.0ex]{\langle c_\mu(}{\alpha}{)c_\nu(\alpha)c_\mu(}{\beta}
  \bcontraction[1.ex]{\langle c_\mu(\alpha)c_\nu(}{\alpha}{)c_\mu(\beta)c_\nu(}{\beta} \acontraction{\langle c_\mu(\alpha)c_\nu(\alpha)c_\mu(\beta)c_\nu(\beta) c_\mu(}{\alpha^\prime}{)c_\nu(}{\alpha^\prime} 
 \acontraction[1.25ex]{\langle c_\mu(\alpha)c_\nu(\alpha)c_\mu(\beta)c_\nu(\beta) c_\mu(}{\alpha^\prime}{)c_\nu(}{\alpha^\prime} 
  \acontraction[1.25ex]{\langle c_\mu(\alpha)c_\nu(\alpha)c_\mu(\beta)c_\nu(\beta) c_\mu(}{\alpha^\prime}{)c_\nu(\alpha^\prime)c_\nu(}{\beta^\prime} 
  \acontraction{\langle c_\mu(\alpha)c_\nu(\alpha)c_\mu(\beta)c_\nu(\beta)c_\mu(\alpha^\prime)c_\nu(\alpha^\prime)c_\mu(}{\beta^\prime}{)c_\nu(}{\beta^\prime}
   \acontraction[1.25ex]{ \langle c_\mu(\alpha)c_\nu(\alpha)c_\mu(\beta)c_\nu(\beta) c_\mu(\alpha^\prime)c_\nu(\alpha^\prime)c_\mu(}{\beta^\prime}{)c_\nu(}{\beta^\prime} 
  \langle c_\mu(\alpha)c_\nu(\alpha) c_\mu(\beta)c_\nu(\beta)c_\mu(\alpha^\prime)c_\nu(\alpha^\prime)c_\mu(\beta^\prime)c_\nu(\beta^\prime)\rangle_V = \Lambda(\mu, \alpha)\Lambda(\nu, \alpha) L_{\mu\nu}^{\alpha^\prime\alpha^\prime\beta^\prime\beta^\prime}\delta_{\alpha\beta}.
\end{equation}
We can see that there are six terms of this form that contribute only one $\delta$ factor. These are $\delta_{\alpha\beta}, \delta_{\alpha\alpha^\prime}, \delta_{\alpha\beta^\prime}, \delta_{\beta\alpha^\prime}, \delta_{\beta\beta^\prime}, \delta_{\alpha^\prime\beta^\prime}$.

We thus obtain that for the contribution from Gaussian contractions, $\delta_G^2(\infty)$, 
\begin{equation}\label{eq:G}
\begin{split}
\delta_G^2(\infty)  & = \sum_{\substack{\mu\nu \\ \mu\neq\nu}}\sum_{\alpha\beta\alpha^\prime\beta^\prime}w_\alpha w_\beta O_{\alpha^\prime\alpha^\prime}O_{\beta^\prime\beta^\prime} \Lambda(\mu, \alpha)\Lambda(\nu, \beta) \Lambda(\mu, \alpha^\prime)\Lambda(\nu, \beta^\prime)(\delta_{\alpha\beta}\delta_{\alpha^\prime\beta^\prime} + \delta_{\alpha\alpha^\prime}\delta_{\beta\beta^\prime} + \delta_{\alpha\beta^\prime}\delta_{\alpha^\prime\beta}) \\ &
= \sum_{\substack{\mu\nu \\ \mu\neq\nu}}\sum_{\alpha\beta}\Lambda(\mu, \alpha)\Lambda(\nu, \alpha)\Lambda(\mu, \beta)\Lambda(\nu, \beta) [w_\alpha^2 O_{\beta\beta}^2 + 2 w_\alpha w_\beta O_{\alpha\alpha}O_{\beta\beta}]
\end{split} 
\end{equation}
Similarly, for the non-Gaussian ($\delta_{NG}^2(\infty)$), and mixed ($\delta_{M}^2(\infty)$) contractions we have
\begin{equation}\label{eq:NG}
\begin{split}
\delta_{NG}^2(\infty)  & = 3\sum_{\substack{\mu\nu \\ \mu\neq\nu}}\sum_{\alpha\beta\alpha^\prime\beta^\prime}w_\alpha w_\beta O_{\alpha^\prime\alpha^\prime}O_{\beta^\prime\beta^\prime}L_{\mu\nu}^{\alpha\beta\alpha\beta}L_{\mu\nu}^{\alpha^\prime\beta^\prime\alpha^\prime\beta^\prime},
\end{split} 
\end{equation}
and
\begin{equation}\label{eq:M}
\begin{split}
\delta_{M}^2(\infty) = -\sum_{\substack{\mu\nu \\ \mu\neq\nu}}\sum_{\alpha\alpha^\prime\beta^\prime} \Lambda(\mu, \alpha)\Lambda(\nu, \alpha)L_{\mu\nu}^{\alpha^\prime\beta^\prime\alpha^\prime\beta^\prime}  \bigg[w_\alpha^2 O_{\alpha^\prime\alpha^\prime}  O_{\beta^\prime\beta^\prime} + 4 w_\alpha O_{\alpha\alpha} w_{\alpha^\prime} O_{\beta^\prime\beta^\prime} + O_{\alpha\alpha}^2 w_{\alpha^\prime}w_{\beta^\prime}\bigg],
\end{split} \end{equation}
respectively.
In order to perform the summations over non-interacting indices in Eq. \eqref{eq:flucs_full_S} we define course grained averages of observable elements $O_{\alpha\alpha}$ as in Eq. \eqref{eq:def_MCaverages1}. The key assumption in writing \eqref{eq:def_MCaverages1} is that the average,
\begin{equation}
\overline{[O_{\alpha\alpha}]}_\mu := \sum_\alpha \Lambda(\mu, \alpha)O_{\alpha\alpha},
\end{equation}
changes slowly in energy $E_{\overline{\mu}}$ with respect to the width $\Gamma$. Similar averages over the initial state, or mixed averages must also be defined, such as
\begin{equation}\label{eq:def_MCaverages2_S}
\begin{split}
\sum_\alpha \Lambda(\mu, \alpha) \Lambda(\nu, \alpha) w_\alpha & = \overline{[w_\alpha]}_{\overline{\mu}}\Lambda^{(2)}(\mu, \nu) \\  \sum_\alpha \Lambda(\mu, \alpha) \Lambda(\nu, \alpha) w_\alpha O_{\alpha\alpha} & = \overline{[w_\alpha O_{\alpha\alpha}]}_{\overline{\mu}}\Lambda^{(2)}(\mu, \nu).
\end{split}
\end{equation}
Note that, $\overline{[O_{\alpha\alpha}]}_{\overline{\mu}}$ can be interpreted as a microcanonical average of the observable $O$.

Now, using Eqs. \eqref{eq:def_MCaverages1} and \eqref{eq:def_MCaverages2_S}, we obtain,
\begin{equation}\label{eq:3deltas_App}
\begin{split}
& \delta_G^2(\infty)  = \sum_{\substack{\mu\nu \\ \mu\neq\nu}} \bigg[ \overline{[w_\alpha^2]}_{\overline{\mu}}\overline{[O_{\alpha\alpha}^2]}_{\overline{\mu}} + 2 \overline{[w_\alpha O_{\alpha\alpha}]}_{\overline{\mu}}^2\bigg]\Lambda^{(2)}(\mu, \nu)^2, \\ &
\delta_{NG}^2(\infty) = 3\sum_{\substack{\mu\nu \\ \mu\neq\nu}}\overline{[w_\alpha]}_{\overline{\mu}}^2\overline{[O_{\alpha\alpha}]}_{\overline{\mu}}^2\Lambda^{(2)}(\mu, \nu)^2, \\ &
\delta_{M}^2(\infty) = -\sum_{\substack{\mu\nu \\ \mu\neq\nu}}\Lambda^{(2)}(\mu, \nu)^2\bigg[\overline{[w_\alpha^2]}_{\overline{\mu}}\overline{[O_{\alpha\alpha}]}_{\overline{\mu}}^2  + 4\overline{[w_\alpha]}_{\overline{\mu}}\overline{[O_{\alpha\alpha}]}_{\overline{\mu}} \overline{[w_\alpha O_{\alpha\alpha}]}_{\overline{\mu}} + \overline{[w_\alpha]}_{\overline{\mu}}^2\overline{[O_{\alpha\alpha}^2]}_{\overline{\mu}}\bigg].
\end{split} 
\end{equation}

\section{Proof of Self-Averaging}\label{App:SelfAveraging}
In addition to the numerical demonstration in Figs. \ref{fig:RM_FDT} and \ref{fig:RM_time}, we are now able to analytically demonstrate that self-averaging, in the sense of Eq. \eqref{eq:self_averaging}, occurs for this model. We have already obtained $\langle \langle O(t) \rangle\rangle_V$ in Section \ref{sec:Time_dep} (note that here we assumed self-averaging, and thus simply used the notation $\langle O(t)\rangle$, in the current section we require the self-averaging step to be explicit, so use the angle brackets $\langle \cdots \rangle_V$ whenever self-averaging is performed), and thus, what remains is a calculation of $\langle \langle O(t) \rangle^2\rangle_V$. We can do this by writing,
\begin{equation}
\begin{split}
\Delta O(t) ^2 &= \sum_{\substack{\mu\nu\\ \mu\neq\nu}}\sum_{\substack{\mu^\prime\nu^\prime\\ \mu^\prime\neq\nu^\prime}} \sum_{\alpha\beta\alpha^\prime\beta^\prime}\sum_{\alpha_1\beta_1\alpha_1^\prime\beta_1^\prime}w_{\alpha\beta} w_{\alpha^\prime\beta^\prime}O_{\alpha^\prime\beta^\prime}O_{\alpha_1^\prime\beta_1^\prime} \\&
 \qquad \qquad \qquad \times c_\mu(\alpha)c_\nu(\beta)c_\mu(\alpha^\prime)c_\nu(\beta^\prime) c_{\mu^\prime}(\alpha_1)c_{\nu^\prime}(\beta_1)c_{\mu^\prime}(\alpha_1^\prime)c_{\nu^\prime}(\beta_1^\prime)  \\ &
= \sum_{\substack{\mu\nu\\ \mu\neq\nu}}\sum_{\substack{\mu^\prime\nu^\prime\\ \mu^\prime\neq\nu^\prime}} \sum_{\alpha\beta\alpha^\prime}\sum_{\alpha_1\beta_1\alpha_1^\prime} \sum_{n, n_1}^{N_O}  w_{\alpha\beta} w_{\alpha^\prime\beta^\prime} O_{\alpha^\prime, \alpha^\prime + n}O_{\alpha_1^\prime, \alpha_1^\prime + n_1} \\&
 \qquad \qquad \qquad \times c_\mu(\alpha)c_\nu(\beta)c_\mu(\alpha^\prime)c_\nu(\alpha^\prime + n) c_{\mu^\prime}(\alpha_1)c_{\nu^\prime}(\beta_1)c_{\mu^\prime}(\alpha_1^\prime)c_{\nu^\prime}(\alpha_1^\prime + n_1) 
\end{split}
\end{equation}
from which, we see that the relevant correlation function after self-averaging is $\langle c_\mu(\alpha)c_\nu(\beta)c_\mu(\alpha^\prime)c_\nu(\alpha^\prime + n) c_{\mu^\prime}(\alpha_1)c_{\nu^\prime}(\beta_1)c_{\mu^\prime}(\alpha_1^\prime)c_{\nu^\prime}(\alpha_1^\prime + n_1 )\rangle_V$. Indeed, the contribution from Gaussian contractions can easily be seen to be identical to that of $\langle \Delta O(t)\rangle_V^2$, as these correlators only contract identical interacting indices $\mu, \nu$. Furthermore, we see that when non-Gaussian contractions are defined between primed and unprimed interacting indices, $\mu, \nu$, there is a reduction in the number of summations. For example, we compare the contribution 
\begin{equation}
 \bcontraction[1.25ex]{ \sum_{\mu\nu\mu^\prime\nu^\prime}\sum_{\alpha\beta\alpha^\prime}\sum_{\alpha_1\beta_1\alpha_1^\prime} \langle c_\mu(}{\alpha}{)c_\nu(}{\beta}  
 \bcontraction[1.25ex]{ \sum_{\mu\nu\mu^\prime\nu^\prime}\sum_{\alpha\beta\alpha^\prime}\sum_{\alpha_1\beta_1\alpha_1^\prime} \langle c_\mu(}{\beta}{)c_\nu(\beta)c_\mu(\alpha)c_\nu((}{\alpha)} 
  \bcontraction{ \sum_{\mu\nu\mu^\prime\nu^\prime}\sum_{\alpha\beta\alpha^\prime}\sum_{\alpha_1\beta_1\alpha_1^\prime} \langle c_\mu(}{\alpha}{)c_\nu(}{\beta}  
  \bcontraction{\sum_{\mu\nu\mu^\prime\nu^\prime}\sum_{\alpha\beta\alpha^\prime}\sum_{\alpha_1\beta_1\alpha_1^\prime} \langle c_\mu(\alpha)c_\nu(\alpha)c_\mu(}{\beta}{)c_\nu((}{\alpha)}
  \bcontraction[1.25ex]{ \sum_{\mu\nu\mu^\prime\nu^\prime}\sum_{\alpha\beta\alpha^\prime}\sum_{\alpha_1\beta_1\alpha_1^\prime} \langle c_\mu(\alpha)c_\nu(\alpha)c_\mu(}{\beta}{)c_\nu((}{\alpha)}  
  \bcontraction{\sum_{\mu\nu\mu^\prime\nu^\prime}\sum_{\alpha\beta\alpha^\prime}\sum_{\alpha_1\beta_1\alpha_1^\prime} \langle c_\mu(\alpha)c_\nu(\alpha)c_\mu(\beta^\prime))c_\nu(\beta^\prime) \rangle_V \langle c_\mu(}{\alpha_1^\prime}{)c_\nu(}{\alpha_1^\prime} 
  \bcontraction[1.25ex]{\sum_{\mu\nu\mu^\prime\nu^\prime}\sum_{\alpha\beta\alpha^\prime}\sum_{\alpha_1\beta_1\alpha_1^\prime} \langle c_\mu(\alpha)c_\nu(\alpha)c_\mu(\beta^\prime))c_\nu(\beta^\prime) \rangle_V \langle c_\mu(}{\alpha_1^\prime}{)c_\nu(}{\alpha_1^\prime}     
  \bcontraction[1.25ex]{\sum_{\mu\nu\mu^\prime\nu^\prime}\sum_{\alpha\beta\alpha^\prime}\sum_{\alpha_1\beta_1\alpha_1^\prime} \langle c_\mu(\alpha)c_\nu(\alpha)c_\mu(\beta^\prime))c_\nu(\beta^\prime) \rangle_V \langle c_\mu(}{\alpha_1^\prime}{)c_\nu(\alpha_1^\prime)c_{\mu^\prime}(}{\beta_1^\prime} 
  \bcontraction{\sum_{\mu\nu\mu^\prime\nu^\prime}\sum_{\alpha\beta\alpha^\prime}\sum_{\alpha_1\beta_1\alpha_1^\prime} \langle c_\mu(\alpha)c_\nu(\alpha)c_\mu(\beta^\prime))c_\nu(\beta^\prime) \rangle_V \langle c_\mu(\alpha_1^\prime)c_\nu(\alpha_1^\prime)c_{\mu^\prime}(}{\beta_1^\prime}{)c_{\mu^\prime}(}{\beta_1^\prime}
   \bcontraction[1.25ex]{\sum_{\mu\nu\mu^\prime\nu^\prime}\sum_{\alpha\beta\alpha^\prime}\sum_{\alpha_1\beta_1\alpha_1^\prime} \langle c_\mu(\alpha)c_\nu(\alpha)c_\mu(\beta^\prime))c_\nu(\beta^\prime) \rangle_V \langle c_\mu(\alpha_1^\prime)c_\nu(\alpha_1^\prime)c_{\mu^\prime}(}{\beta_1^\prime}{)c_{\mu^\prime}(}{\beta_1^\prime}    
   \sum_{\mu\nu\mu^\prime\nu^\prime}\sum_{\alpha\beta\alpha^\prime}\sum_{\alpha_1\beta_1\alpha_1^\prime} \langle c_\mu(\alpha)c_\nu(\beta)c_\mu(\alpha^\prime)c_\nu(\alpha^\prime) \rangle_V \langle c_{\mu^\prime}(\alpha_1)c_{\nu^\prime}(\beta_1)c_{\mu^\prime}(\alpha_1^\prime)c_{\nu^\prime}(\alpha_1^\prime)\rangle_V,
\end{equation}
to that with mixed interacting indices,
\begin{equation}
 \bcontraction[1.25ex]{\sum_{\mu\nu\mu^\prime\nu^\prime}\sum_{\alpha\beta\alpha^\prime}\sum_{\alpha_1\beta_1\alpha_1^\prime} \langle c_\mu(}{\alpha}{)c_{\nu^\prime}(}{\beta_1}  
 \bcontraction[1.25ex]{\sum_{\mu\nu\mu^\prime\nu^\prime}\sum_{\alpha\beta\alpha^\prime}\sum_{\alpha_1\beta_1\alpha_1^\prime} \langle c_\mu(}{\alpha}{)c_{\nu^\prime}(\beta_1)c_\mu(\alpha^\prime)c_{\nu^\prime}(}{\alpha_1^\prime)} 
\bcontraction{\sum_{\mu\nu\mu^\prime\nu^\prime}\sum_{\alpha\beta\alpha^\prime}\sum_{\alpha_1\beta_1\alpha_1^\prime} \langle c_\mu(}{\alpha}{)c_{\nu^\prime}(}{\beta_1}   
\bcontraction{ \sum_{\mu\nu\mu^\prime\nu^\prime}\sum_{\alpha\beta\alpha^\prime}\sum_{\alpha_1\beta_1\alpha_1^\prime} \langle c_\mu(\alpha)c_{\nu^\prime}(\beta_1)c_\mu(}{\alpha^\prime}{)c_{\nu^\prime}(}{\alpha_1^\prime)}
\bcontraction[1.25ex]{ \sum_{\mu\nu\mu^\prime\nu^\prime}\sum_{\alpha\beta\alpha^\prime}\sum_{\alpha_1\beta_1\alpha_1^\prime} \langle c_\mu(\alpha)c_{\nu^\prime}(\beta_1)c_\mu(}{\alpha^\prime}{)c_{\nu^\prime}(}{\alpha_1^\prime)}
\bcontraction{ \sum_{\mu\nu\mu^\prime\nu^\prime}\sum_{\alpha\beta\alpha^\prime}\sum_{\alpha_1\beta_1\alpha_1^\prime} \langle c_\mu(\alpha)c_{\nu^\prime}(\beta_1)c_\mu(\alpha^\prime)c_{\nu^\prime}(\alpha_1^\prime) \rangle_V \langle c_{\mu^\prime}(}{\alpha_1}{)c_\nu(}{\beta} 
\bcontraction[1.25ex]{ \sum_{\mu\nu\mu^\prime\nu^\prime}\sum_{\alpha\beta\alpha^\prime}\sum_{\alpha_1\beta_1\alpha_1^\prime} \langle c_\mu(\alpha)c_{\nu^\prime}(\beta_1)c_\mu(\alpha^\prime)c_{\nu^\prime}(\alpha_1^\prime) \rangle_V \langle c_{\mu^\prime}(}{\alpha_1}{)c_\nu(}{\beta} 
\bcontraction[1.25ex]{ \sum_{\mu\nu\mu^\prime\nu^\prime}\sum_{\alpha\beta\alpha^\prime}\sum_{\alpha_1\beta_1\alpha_1^\prime} \langle c_\mu(\alpha)c_{\nu^\prime}(\beta_1)c_\mu(\alpha^\prime)c_{\nu^\prime}(\alpha_1^\prime) \rangle_V \langle c_{\mu^\prime}(}{\alpha_1}{)c_\nu(\beta)c_{\mu^\prime}(}{\alpha_1^\prime} 
  \bcontraction{  \sum_{\mu\nu\mu^\prime\nu^\prime}\sum_{\alpha\beta\alpha^\prime}\sum_{\alpha_1\beta_1\alpha_1^\prime} \langle c_\mu(\alpha)c_{\nu^\prime}(\beta_1)c_\mu(\alpha^\prime)c_{\nu^\prime}(\alpha_1^\prime) \rangle_V \langle c_{\mu^\prime}(\alpha_1)c_\nu(\beta)c_{\mu^\prime}(}{\alpha_1^\prime}{)c_{\nu}(}{\alpha^\prime}
   \bcontraction[1.25ex]{  \sum_{\mu\nu\mu^\prime\nu^\prime}\sum_{\alpha\beta\alpha^\prime}\sum_{\alpha_1\beta_1\alpha_1^\prime} \langle c_\mu(\alpha)c_{\nu^\prime}(\beta_1)c_\mu(\alpha^\prime)c_{\nu^\prime}(\alpha_1^\prime) \rangle_V \langle c_{\mu^\prime}(\alpha_1)c_\nu(\beta)c_{\mu^\prime}(}{\alpha_1^\prime}{)c_{\nu}(}{\alpha^\prime} 
   \sum_{\mu\nu\mu^\prime\nu^\prime}\sum_{\alpha\beta\alpha^\prime}\sum_{\alpha_1\beta_1\alpha_1^\prime} \langle c_\mu(\alpha)c_{\nu^\prime}(\beta_1)c_\mu(\alpha^\prime)c_{\nu^\prime}(\alpha_1^\prime) \rangle_V \langle c_{\mu^\prime}(\alpha_1)c_\nu(\beta)c_{\mu^\prime}(\alpha_1^\prime)c_\nu(\alpha^\prime)\rangle_V,
\end{equation}
and observe that the latter contribution is negligible, due to the extra reduced summation, as there are no repeated non-interacting indices within the non-Gaussian contraction. We note that the dominating contribution from non-Gaussian contractions is inevitably from the $n, n_1 = 0$ term shown above (as this reduced summation does not change the order of the contribution as $N_O \ll {\cal N}$). Thus we have a single dominating contribution given by the 4-point correlation functions defined with primed, or unprimed, interaction indices only:
\begin{equation}
\begin{split}
\langle \Delta O(t) ^2 \rangle_V &= \sum_{\substack{\mu\nu\\ \mu\neq\nu}}\sum_{\substack{\mu^\prime\nu^\prime\\ \mu^\prime\neq\nu^\prime}} \sum_{\alpha\beta\alpha^\prime}\sum_{\alpha_1\beta_1\alpha_1^\prime} \sum_{n, n_1}^{N_O} w_{\alpha\beta} w_{\alpha^\prime\beta^\prime} O_{\alpha^\prime,\alpha^\prime + n} O_{\alpha_1^\prime, \alpha_1^\prime + n_1} e^{-i (E_\mu - E_\nu + E_{\mu^\prime} - E_{\nu^\prime})t} \\&
 \qquad \qquad \times \langle c_\mu(\alpha)c_\nu(\beta)c_\mu(\alpha^\prime)c_\nu(\alpha^\prime + n) c_{\mu^\prime}(\alpha_1)c_{\nu^\prime}(\beta_1)c_{\mu^\prime}(\alpha_1^\prime)c_{\nu^\prime}(\alpha_1^\prime + n_1)\rangle_V\\& 
= \left( \sum_{\substack{\mu\nu\\ \mu\neq\nu}} \sum_{\alpha\beta\alpha^\prime}\sum_n^{N_O} w_{\alpha\beta}O_{\alpha^\prime,\alpha^\prime + n}e^{-i (E_\mu - E_\nu)t}  \langle c_\mu(\alpha)c_\nu(\beta)c_\mu(\alpha^\prime)c_\nu(\alpha^\prime + n) \rangle_V \right)\\& \qquad \qquad \times \left( \sum_{\substack{\mu^\prime\nu^\prime\\ \mu^\prime\neq\nu^\prime}}\sum_{\alpha_1\beta_1\alpha_1^\prime} \sum_{n_1}^{N_O} w_{\alpha^\prime\beta^\prime} O_{\alpha_1^\prime, \alpha_1^\prime + n_1} e^{-i( E_{\mu^\prime} - E_{\nu^\prime})t} \langle  c_{\mu^\prime}(\alpha_1)c_{\nu^\prime}(\beta_1)c_{\mu^\prime}(\alpha_1^\prime)c_{\nu^\prime}(\alpha_1^\prime + n_1)\rangle_V \right) \\&
= \langle  \Delta O(t) \rangle_V^2.
\end{split}
\end{equation}
Therefore, we can see that the transient component $\Delta O(t)$ of the time evolution is indeed self-averaging. Now, we can similarly see that the long-time average is self-averaging via,
\begin{equation}
\begin{split}
O_{\textrm{MC}} & = \sum_{\mu}\sum_{\alpha\beta}w_{\alpha\beta}O_{\mu\mu}c_\mu(\alpha)c_\mu(\beta) \\&
= \sum_{\mu}\sum_{\alpha\beta\alpha^\prime \beta^\prime}w_{\alpha\beta}O_{\alpha^\prime\beta^\prime}c_\mu(\alpha)c_\mu(\beta)c_\mu(\alpha^\prime)c_\mu(\beta^\prime) \\&
= \sum_{\mu}\sum_{\alpha\beta\alpha^\prime}\sum_n^{N_O}w_{\alpha\beta}O_{\alpha^\prime, \alpha^\prime + n}c_\mu(\alpha)c_\mu(\beta)c_\mu(\alpha^\prime)c_\mu(\alpha^\prime + n).
\end{split}
\end{equation}
The ensemble average of this is then, 
\begin{equation}
\begin{split}
\langle O_{\textrm{MC}}\rangle_V & = \sum_{\mu}\sum_{\alpha\beta\alpha^\prime}\sum_n^{N_O}w_{\alpha\beta}O_{\alpha^\prime, \alpha^\prime + n}\langle c_\mu(\alpha)c_\mu(\beta)c_\mu(\alpha^\prime)c_\mu(\alpha^\prime + n)\rangle_V  \\&
= \sum_{\mu}\sum_{\alpha\alpha^\prime}w_{\alpha\alpha}O_{\alpha^\prime\alpha^\prime}\Lambda(\mu, \alpha)\Lambda(\mu, \alpha^\prime),
\end{split}
\end{equation}
where one can easily see that the terms with $n \neq 0$ require an additional contraction, and thus the contribution is on a lower order, with fewer summations over the non-interacting indices. We then see that,
\begin{equation}
\begin{split}
O_{\textrm{MC}}^2 & = \sum_{\mu\nu}\sum_{\alpha\beta\alpha_1 \beta_1}w_{\alpha\beta}w_{\alpha_1\beta_1}O_{\mu\mu}O_{\nu\nu} c_\mu(\alpha)c_\mu(\beta)c_\nu(\alpha_1)c_\nu(\beta_1)\\&
= \sum_{\mu\nu}\sum_{\alpha\beta\alpha_1 \beta_1}\sum_{\alpha^\prime \beta^\prime \alpha_1^\prime \beta_1^\prime}w_{\alpha\beta}w_{\alpha_1\beta_1}O_{\alpha^\prime\beta^\prime}O_{\alpha_1^\prime\beta_1^\prime}c_\mu(\alpha)c_\mu(\beta)c_\nu(\alpha_1)c_\nu(\beta_1)c_\mu(\alpha^\prime)c_\mu(\beta^\prime)c_\nu(\alpha_1^\prime)c_\nu(\beta_1^\prime) \\&
=  \sum_{\mu\nu}\sum_{\alpha\beta\alpha_1 \beta_1}\sum_{\alpha^\prime \alpha_1^\prime} \sum_{n n^\prime}^{N_O} w_{\alpha\beta}w_{\alpha_1\beta_1}O_{\alpha^\prime, \alpha^\prime  +n}O_{\alpha_1^\prime, \alpha_1^\prime + n^\prime} \\& \qquad \qquad \qquad \qquad \times c_\mu(\alpha)c_\mu(\beta)c_\nu(\alpha_1)c_\nu(\beta_1)c_\mu(\alpha^\prime)c_\mu(\alpha^\prime + n)c_\nu(\alpha_1^\prime)c_\nu(\alpha_1^\prime + n^\prime),
\end{split}
\end{equation}
which, after self-averaging becomes,
\begin{equation}
\begin{split}
\langle O_{\textrm{MC}}^2 \rangle_V & = \sum_{\mu\nu}\sum_{\alpha\beta\alpha_1 \beta_1}\sum_{\alpha^\prime \alpha_1^\prime} \sum_{n n^\prime}^{N_O} w_{\alpha\beta}w_{\alpha_1\beta_1}O_{\alpha^\prime, \alpha^\prime  +n}O_{\alpha_1^\prime, \alpha_1^\prime + n^\prime} \\& \qquad \qquad \qquad \qquad \times  \langle c_\mu(\alpha)c_\mu(\beta)c_\nu(\alpha_1)c_\nu(\beta_1)c_\mu(\alpha^\prime)c_\mu(\alpha^\prime + n)c_\nu(\alpha_1^\prime)c_\nu(\alpha_1^\prime + n^\prime)\rangle_V \\&
=\sum_{\mu\nu}\sum_{\alpha\alpha_1}\sum_{\alpha^\prime \alpha_1^\prime} w_{\alpha\alpha}w_{\alpha_1\alpha_1}  O_{\alpha^\prime \alpha^\prime}O_{\alpha_1^\prime \alpha_1^\prime}\Lambda(\mu, \alpha) \Lambda(\nu, \alpha_1) \Lambda(\mu, \alpha^\prime) \Lambda(\nu, \alpha_1^\prime) \\&
= \langle O_{\textrm{MC}}\rangle_V^2 
\end{split}
\end{equation}
where we see once more that the $n, n^\prime = 0$ terms dominate, and that contractions between the primed and unprimed indices induce additional kronecker-delta factors, and thus are negligible. 

Thus, we have that ${\cal R} = 0$ (see Eq. \eqref{eq:self_averaging}), and we analytically observe self-averaging in this model. We note here that in the derivation of the 4-point correlator, we assume that the only `interactions' between the eigenstates themselves are those enforced due to the mutual orthogonality of eigenstates via the condition $\sum_\alpha c_\mu(\alpha)c_\nu(\alpha) = \delta_{\mu\nu}$ \cite{Nation2018}. As these interactions occur pairwise between eigenstates, correlation functions with more than two interacting indices $\mu, \nu$ contribute an additional $\delta_{\mu\nu}$. 

\section{Bound of dynamical term}\label{App:Bound}
In this section we bound third term in Eq. \eqref{eq:three_terms}, showing that it's contribution is negligible. To do so, we proceed by defining,
\begin{equation}
A(t) = \sum_{\substack{\mu\nu \\ \mu \neq \nu}}\sum_{\alpha\beta}w_{\alpha\beta}O_{\alpha\beta} \frac{\Lambda(\mu, \alpha)\Lambda(\mu, \beta)\Lambda(\nu, \alpha)\Lambda(\nu, \beta)}{\Lambda^{(2)}(\mu, \nu)} e^{-i(E_\mu - E_\nu)t}.
\end{equation}
We may now use the relation $|\sum_i a_i| \leq \sum_i |a_i|$, to write
\begin{equation}
\begin{split}
|A(t)| & \leq \sum_{\substack{\mu\nu \\ \mu \neq \nu}}\left| \sum_{\alpha\beta}w_{\alpha\beta}O_{\alpha\beta}  \frac{\Lambda(\mu, \alpha)\Lambda(\mu, \beta)\Lambda(\nu, \alpha)\Lambda(\nu, \beta)}{\Lambda^{(2)}(\mu, \nu)} \right| \left| e^{-i(E_\mu - E_\nu)t}\right|\\&
\leq \sum_{\substack{\mu\nu \\ \mu \neq \nu}} \sum_{\alpha\beta}\left|w_{\alpha\beta}O_{\alpha\beta}  \frac{\Lambda(\mu, \alpha)\Lambda(\mu, \beta)\Lambda(\nu, \alpha)\Lambda(\nu, \beta)}{\Lambda^{(2)}(\mu, \nu)} \right|\\&
= \sum_{\alpha\beta}\left|w_{\alpha\beta}O_{\alpha\beta}  \right| \sum_{\substack{\mu\nu \\ \mu \neq \nu}}  \frac{\Lambda(\mu, \alpha)\Lambda(\mu, \beta)\Lambda(\nu, \alpha)\Lambda(\nu, \beta)}{\Lambda^{(2)}(\mu, \nu)}\\&
\leq \frac{3\omega_0}{4\pi\Gamma}\sum_{\alpha}\left|w_{\alpha\beta}O_{\alpha\beta}  \right|,
\end{split}
\end{equation}
where we have used that,
\begin{equation}
\sum_{\substack{\mu\nu \\ \mu \neq \nu}}  \frac{\Lambda(\mu, \alpha)\Lambda(\mu, \beta)\Lambda(\nu, \alpha)\Lambda(\nu, \beta)}{\Lambda^{(2)}(\mu, \nu)} = \omega_0\frac{(E_\alpha - E_\beta)^2\Gamma + 12 \Gamma^3}{\pi((E_\alpha - E_\beta)^2 + 4\Gamma^2)^2} \leq \frac{3\omega_0}{4\pi\Gamma}. 
\end{equation}
Now, using $\sum_{\alpha\beta}O_{\alpha\beta} = \sum_\alpha \sum_n^{N_O} O_{\alpha, \alpha+n}\delta_{\beta, \alpha+n}$ \cite{Nation2019}, we have 
\begin{equation}
\begin{split}
\sum_{\alpha\beta}\left|w_{\alpha\beta}O_{\alpha\beta} \right| & = \sum_{\alpha, n}\left|w_{\alpha, \alpha+n}O_{\alpha \alpha+n} \right| \\&
\leq \max_{\alpha\beta}(O_{\alpha\beta}) \sum_{\alpha, n}w_{\alpha, \alpha + n}.
\end{split}
\end{equation}
Now, we see that in the case of $w_{\alpha\beta} \sim \delta_{\alpha\beta}$, the bound simply becomes
\begin{equation}
|A(t)| \leq \max_\alpha (O_{\alpha\alpha})\frac{3\omega_0}{4\pi\Gamma},
\end{equation}
similarly, this is the case for our application studied in the main text, where $O_{\alpha\beta}\sim \delta_{\alpha\beta}$. We note that the condition of diagonal $w_{\alpha\beta}$ correspond to a reasonable initial state in many experimental situations, since thermalization takes place typically by incoherent exchange of energy in the basis of eigenstates of $H_0$. For states with coherences, to bound this quantity we require that the coherences are not large on the off-diagonals defined by $\alpha,\alpha+n$, in the sense that $\sum_\alpha w_{\alpha, \alpha + n} \lesssim {\cal O}(1)$, so
\begin{equation}
|A(t)| \lesssim N_O \max_\alpha (O_{\alpha\alpha})\frac{3\omega_0}{4\pi\Gamma}.
\end{equation}
In this sense, `special' initial states may be chosen that do not satisfy this bound, but they are highly atypical. Further, we note that the stronger of the two assumptions made on the form of observables, that is that we may write $\sum_{\alpha\beta}O_{\alpha\beta} = \sum_\alpha \sum_n^{N_O} O_{\alpha, \alpha+n}\delta_{\beta, \alpha+n}$, is only used above in bounding $|A(t)|$, the other terms are more general. As noted above, whilst this form is reasonable for generic local variables \cite{Nation2019}, one may of course be able to build observables that do not match this form, in which case, one may either have an additional contribution due to $A(t)$, or be able to arrive at a similar bound.

\section{FDT in terms of Thermal Averages}\label{App:Thermal_Averages}

In this section, we show that the FDT may be recast in terms of the thermal averages $\langle \langle \cdots \rangle \rangle_\beta$, obtained from the a fit to the decay, shown in Sec \ref{sec:Time_dep}. We begin by re-expressing our finite temperature FDT in terms of the thermal averages, 
\begin{equation}
\langle\langle A(E) \rangle\rangle_{\beta} := \frac{1}{Z_\beta^\prime}\int_0^{\Delta E} dE D(E) e^{-\beta (E - E_0)} A(E),
\end{equation}
with $Z_\beta^\prime := \int_0^{\Delta E} dE D(E) e^{-\beta (E-E_0)}$, where we have introduced the low-energy cut-off $E_0$, which is the energy of our zero temperature pure state. The addition of this cut-off ensures that, for non-zero $E_0$, the initial state is not the ground state of the bath at zero temperature.

To include this thermal average, we return to Eq. \eqref{eq:delta_int_therm}, and note that,
\begin{equation}
\int_0^{\Delta E} dE\frac{ e^{-2\beta E} }{\Gamma(E)} = Z^\prime_{2\beta} \langle \langle D(E)^{-1} \Gamma(E)^{-1} \rangle \rangle_{2\beta}.
\end{equation}
We thus obtain,
\begin{equation}\label{eq:FDT3}
\begin{split}
\delta_O^2(\infty) &= \frac{W_O Z_{2 \beta}^\prime}{8\pi Z_\beta^2}\langle \langle D(E)^{-1} \Gamma(E)^{-1} \rangle\rangle_{2\beta}.
\end{split}
\end{equation}
We note that this can be put in the same form as the infinite temperature case, $\delta_O^2(\infty) \sim \overline{\Gamma^{-1}}$, by noting that the quantity
\begin{equation}
C^\prime_\beta = \frac{\langle\langle D(E)^{-1} \Gamma(E)^{-1} \rangle\rangle_{2\beta}}{\langle\langle D(E)\rangle\rangle_{2\beta}^{-1}\langle\langle \Gamma(E)^{-1}\rangle \rangle_{2\beta}}
\end{equation}
depends only on the particular forms of the functions $D(E)$ and $\Gamma(E)$, and the temperature - importantly, not on $N$, or on the system-bath coupling strengths (for weak couplings), as both the numerator and denominator share the same dependence in $N$ and the system-bath coupling in the thermodynamic limit. As such, we can write
\begin{equation}
\begin{split}
\delta_O^2(\infty) & = C^\prime_\beta \frac{W_O Z_{2 \beta}^\prime}{8\pi Z_\beta^2\langle\langle D(E)\rangle\rangle_{2\beta}}  \langle \langle\Gamma(E)^{-1}\rangle \rangle_{2\beta},
\end{split}
\end{equation}
and thus we recover the form of our main result, Eq. (3) of the main text, with $\chi(N) = C^\prime_\beta \frac{W_O Z_{2 \beta}}{16\pi Z_\beta^2\langle\langle D(E)\rangle\rangle_\beta}$.

For the random matrix case, where $D(E)^{-1} = \omega_0$, and $\Gamma(E) = \Gamma$ is constant, we also have, 
\begin{equation}\label{eq:RM_FDT}
\delta_O^2(\infty) = \frac{W_O Z_{2 \beta}^\prime \omega_0}{8\pi Z_\beta^2 \Gamma} 
\end{equation}
as in this case, the thermal average $\langle\langle \cdots \rangle \rangle_\beta$ and unbiased thermal average $\langle \cdots \rangle_\beta$, can be seen to be equal.

We can check that, as required, one obtains the infinite temperature limit derived above by sending $\beta \to 0$ by noting that for the infinite temperature case, we have
\begin{equation}
\delta_O^2(\infty) = \frac{W_O}{2 {\cal N}_B^2} \int_0^{\Delta E} dE \frac{1}{4\pi\Gamma(E)},
\end{equation}
which, in terms of the infinite temperature thermal average, may be written (noting that $\lim_{\beta \to 0}Z_\beta^\prime = 2{\cal N}_B$),
\begin{equation}
\begin{split}\label{eq:Temp_average_FDT}
\delta_O^2(\infty) & = \frac{W_O }{4\pi {\cal N}_B} \langle\langle D(E)^{-1} \Gamma(E)^{-1} \rangle \rangle_0 \\&
= C^\prime_0 \frac{W_O }{4\pi {\cal N}_B \langle\langle D(E) \rangle \rangle_0} \langle\langle  \Gamma(E)^{-1} \rangle \rangle_0 .
\end{split}
\end{equation}
We finally note here that it is the measurement of a thermal average of the decay rate, and not the unbiased average, that obscures the direct measurement of ${\cal N}_B$, such that in general a FDT of the form of Eq. \eqref{eq:Temp_average_FDT} would be observed experimentally. We thus limit our claim in cases where the energy dependence of the decay rate is important (which, as can be seen from Eq. \eqref{eq:Delta_O_Gamma_alpha3}, is implied by a non-exponential decay to equilibrium, due to multiple contributing decay rates) to the experimental verification of the finite size scaling of the Hilbert space dimension, rather than the particular value of ${\cal N}_B$, without additional assumptions or the numerical calculation of $C^\prime_\beta$. 

\subsection{Low Temperature FDT}
We now turn to the low temperature limit of Eq. \eqref{eq:general_flucs} for which we expect to obtain the same result as the pure state case of Ref. \cite{Nation2019}, given by,
\begin{equation}\label{eq:FDT_pure}
\delta_O^2(\infty) = \frac{\overline{[\Delta O_{\alpha\alpha}^2]}}{4\pi D(E_{\alpha_0}) \Gamma},
\end{equation}
where $D(E_{\alpha_0})$ is the DOS at the initial state energy $E_{\alpha_0}$, which is chosen to be in the bulk of the spectrum, and $\overline{[\Delta O_{\alpha\alpha}^2]} := \overline{[ O_{\alpha\alpha}^2]} - \overline{[ O_{\alpha\alpha}]}^2$. We have that in this limit,
\begin{equation}
\langle \langle A(E) \rangle \rangle_\infty = A(E_0),
\end{equation}
so 
\begin{equation}
C^\prime_\infty = \frac{ D(E_0)^{-1} \Gamma(E_0)^{-1}}{ D(E_0)^{-1} \Gamma(E_0)^{-1}} = 1,
\end{equation}
and thus,
\begin{equation}
\begin{split}
\delta_O^2(\infty) =  \frac{W_O }{4\pi D(E_0)} \Gamma(E_0)^{-1},
\end{split}
\end{equation}
which is the zero temperature limit, Eq. \eqref{eq:FDT_pure} when $W_0 = \overline{[\Delta O_{\alpha\alpha}^2]}$. Which can be seen to be the case for zero temperature as follows.
Recalling that $W_0$ is defined by
\begin{equation}
\begin{split}
W_O & = \frac{W_\mu}{\overline{[w_\alpha^2]}} 
\end{split}
\end{equation}
where,
\begin{equation}
\begin{split}
W_\mu = \overline{[w_\alpha^2]}_\mu\overline{[O_{\alpha\alpha}^2]}_\mu + 2 \overline{[w_\alpha O_{\alpha\alpha}]}_\mu^2 & + 3 \overline{[w_\alpha]}_\mu^2\overline{[O_{\alpha\alpha}]}_\mu^2 - \overline{[w_\alpha^2]}_\mu\overline{[O_{\alpha\alpha}]}_\mu^2 \\ & - 4\overline{[w_\alpha]}_\mu\overline{[O_{\alpha\alpha}]}_\mu \overline{[w_\alpha O_{\alpha\alpha}]}_\mu  - \overline{[w_\alpha]}_\mu^2\overline{[O_{\alpha\alpha}^2]}_\mu.
\end{split}
\end{equation}
We see that in the zero temperature limit $w_\alpha \sim \delta_{\alpha\alpha_0}$, and thus, the averages in Eq. \eqref{eq:def_MCaverages2_S} contribute to a lower order as the number of summations is reduced for terms with, e.g. $w_\alpha w_\beta$, over terms with, e.g. $w_\alpha^2$. This can  be seen to lead to $\overline{[w_\alpha]}_\mu^2 \ll \overline{[w_\alpha^2]}_\mu$. Similarly, both terms above with mixed averages $ \overline{[w_\alpha O_{\alpha\alpha}]}_\mu$ contribute on the order  $\overline{[w_\alpha]}_\mu^2$, as  $ \overline{[w_\alpha O_{\alpha\alpha}]}_\mu = \overline{[w_\alpha]}_\mu O_{\uparrow}$. 
Using only the remaining terms, we have that,
\begin{equation}\label{eq:W_mu3}
\begin{split}
W_\mu &= \overline{[w_\alpha^2]}_\mu\overline{[O_{\alpha\alpha}^2]}_\mu  - \overline{[w_\alpha^2]}_\mu\overline{[O_{\alpha\alpha}]}_\mu^2 
=   \overline{[w_\alpha^2]}_\mu \overline{[\Delta O_{\alpha\alpha}^2]},
\end{split}
\end{equation}
so,
\begin{equation}
W_O = \overline{[\Delta O^2]},
\end{equation}
as required.

We recall that until Eq. \eqref{eq:general_flucs}, no assumptions on the initial state or observable are made other than the ability to define the required microcanonical averages. As such, taking the low temperature limit at this point, as we have done above, does not contradict any assumptions made.

\bibliographystyle{unsrtnat}

\begin{thebibliography}{44}%
\makeatletter
\providecommand \@ifxundefined [1]{%
 \@ifx{#1\undefined}
}%
\providecommand \@ifnum [1]{%
 \ifnum #1\expandafter \@firstoftwo
 \else \expandafter \@secondoftwo
 \fi
}%
\providecommand \@ifx [1]{%
 \ifx #1\expandafter \@firstoftwo
 \else \expandafter \@secondoftwo
 \fi
}%
\providecommand \natexlab [1]{#1}%
\providecommand \enquote  [1]{``#1''}%
\providecommand \bibnamefont  [1]{#1}%
\providecommand \bibfnamefont [1]{#1}%
\providecommand \citenamefont [1]{#1}%
\providecommand \href@noop [0]{\@secondoftwo}%
\providecommand \href [0]{\begingroup \@sanitize@url \@href}%
\providecommand \@href[1]{\@@startlink{#1}\@@href}%
\providecommand \@@href[1]{\endgroup#1\@@endlink}%
\providecommand \@sanitize@url [0]{\catcode `\\12\catcode `\$12\catcode
  `\&12\catcode `\#12\catcode `\^12\catcode `\_12\catcode `\%12\relax}%
\providecommand \@@startlink[1]{}%
\providecommand \@@endlink[0]{}%
\providecommand \url  [0]{\begingroup\@sanitize@url \@url }%
\providecommand \@url [1]{\endgroup\@href {#1}{\urlprefix }}%
\providecommand \urlprefix  [0]{URL }%
\providecommand \Eprint [0]{\href }%
\providecommand \doibase [0]{http://dx.doi.org/}%
\providecommand \selectlanguage [0]{\@gobble}%
\providecommand \bibinfo  [0]{\@secondoftwo}%
\providecommand \bibfield  [0]{\@secondoftwo}%
\providecommand \translation [1]{[#1]}%
\providecommand \BibitemOpen [0]{}%
\providecommand \bibitemStop [0]{}%
\providecommand \bibitemNoStop [0]{.\EOS\space}%
\providecommand \EOS [0]{\spacefactor3000\relax}%
\providecommand \BibitemShut  [1]{\csname bibitem#1\endcsname}%
\let\auto@bib@innerbib\@empty
\bibitem [{\citenamefont {Bloch}\ \emph {et~al.}(2012)\citenamefont {Bloch},
  \citenamefont {Dalibard},\ and\ \citenamefont {Nascimb{\`e}ne}}]{Bloch2012}%
  \BibitemOpen
  \bibfield  {author} {\bibinfo {author} {\bibfnamefont {I.}~\bibnamefont
  {Bloch}}, \bibinfo {author} {\bibfnamefont {J.}~\bibnamefont {Dalibard}}, \
  and\ \bibinfo {author} {\bibfnamefont {S.}~\bibnamefont {Nascimb{\`e}ne}},\
  }\href {https://doi.org/10.1038/nphys2259} {\bibfield  {journal} {\bibinfo
  {journal} {Nat. Phys.}\ }\textbf {\bibinfo {volume} {8}},\ \bibinfo {pages}
  {267 EP } (\bibinfo {year} {2012})}\BibitemShut {NoStop}%
\bibitem [{\citenamefont {Lewenstein}\ \emph {et~al.}(2012)\citenamefont
  {Lewenstein}, \citenamefont {Sanpera},\ and\ \citenamefont
  {Ahufinger}}]{Lewenstein2012}%
  \BibitemOpen
  \bibfield  {author} {\bibinfo {author} {\bibfnamefont {M.}~\bibnamefont
  {Lewenstein}}, \bibinfo {author} {\bibfnamefont {A.}~\bibnamefont {Sanpera}},
  \ and\ \bibinfo {author} {\bibfnamefont {V.}~\bibnamefont {Ahufinger}},\
  }\href@noop {} {\emph {\bibinfo {title} {Ultracold Atoms in Optical Lattices:
  Simulating quantum many-body systems}}}\ (\bibinfo  {publisher} {Oxford
  University Press},\ \bibinfo {year} {2012})\BibitemShut {NoStop}%
\bibitem [{\citenamefont {Aidelsburger}\ \emph {et~al.}(2017)\citenamefont
  {Aidelsburger}, \citenamefont {Ville}, \citenamefont {Saint-Jalm},
  \citenamefont {Nascimb{\`{e}}ne}, \citenamefont {Dalibard},\ and\
  \citenamefont {Beugnon}}]{Aidelsburger2017}%
  \BibitemOpen
  \bibfield  {author} {\bibinfo {author} {\bibfnamefont {M.}~\bibnamefont
  {Aidelsburger}}, \bibinfo {author} {\bibfnamefont {J.~L.}\ \bibnamefont
  {Ville}}, \bibinfo {author} {\bibfnamefont {R.}~\bibnamefont {Saint-Jalm}},
  \bibinfo {author} {\bibfnamefont {S.}~\bibnamefont {Nascimb{\`{e}}ne}},
  \bibinfo {author} {\bibfnamefont {J.}~\bibnamefont {Dalibard}}, \ and\
  \bibinfo {author} {\bibfnamefont {J.}~\bibnamefont {Beugnon}},\ }\href
  {https://arxiv.org/pdf/1705.02650.pdf
  https://link.aps.org/doi/10.1103/PhysRevLett.119.190403} {\bibfield
  {journal} {\bibinfo  {journal} {Phys. Rev. Lett.}\ }\textbf {\bibinfo
  {volume} {119}},\ \bibinfo {pages} {190403} (\bibinfo {year}
  {2017})}\BibitemShut {NoStop}%
\bibitem [{\citenamefont {Porras}\ and\ \citenamefont
  {Cirac}(2004)}]{Porras2004}%
  \BibitemOpen
  \bibfield  {author} {\bibinfo {author} {\bibfnamefont {D.}~\bibnamefont
  {Porras}}\ and\ \bibinfo {author} {\bibfnamefont {J.}~\bibnamefont {Cirac}},\
  }\href
  {http://prl.aps.org/abstract/PRL/v92/i20/e207901$\backslash$npapers2://publication/doi/10.1103/PhysRevLett.92.207901}
  {\bibfield  {journal} {\bibinfo  {journal} {Phys. Rev. Lett.}\ }\textbf
  {\bibinfo {volume} {92}},\ \bibinfo {pages} {207901} (\bibinfo {year}
  {2004})}\BibitemShut {NoStop}%
\bibitem [{\citenamefont {Schneider}\ \emph {et~al.}(2012)\citenamefont
  {Schneider}, \citenamefont {Porras},\ and\ \citenamefont
  {Schaetz}}]{Schneider2012}%
  \BibitemOpen
  \bibfield  {author} {\bibinfo {author} {\bibfnamefont {C.}~\bibnamefont
  {Schneider}}, \bibinfo {author} {\bibfnamefont {D.}~\bibnamefont {Porras}}, \
  and\ \bibinfo {author} {\bibfnamefont {T.}~\bibnamefont {Schaetz}},\ }\href
  {http://iopscience.iop.org/article/10.1088/0034-4885/75/2/024401/pdf}
  {\bibfield  {journal} {\bibinfo  {journal} {Rep. Prog. Phys.}\ }\textbf
  {\bibinfo {volume} {75}},\ \bibinfo {pages} {24401} (\bibinfo {year}
  {2012})}\BibitemShut {NoStop}%
\bibitem [{\citenamefont {Blatt}\ and\ \citenamefont {Roos}(2012)}]{Blatt2012}%
  \BibitemOpen
  \bibfield  {author} {\bibinfo {author} {\bibfnamefont {R.}~\bibnamefont
  {Blatt}}\ and\ \bibinfo {author} {\bibfnamefont {C.~F.}\ \bibnamefont
  {Roos}},\ }\href {https://doi.org/10.1038/nphys2252} {\bibfield  {journal}
  {\bibinfo  {journal} {Nat. Phys.}\ }\textbf {\bibinfo {volume} {8}},\
  \bibinfo {pages} {277 EP } (\bibinfo {year} {2012})}\BibitemShut {NoStop}%
\bibitem [{\citenamefont {Clos}\ \emph {et~al.}(2016)\citenamefont {Clos},
  \citenamefont {Porras}, \citenamefont {Warring},\ and\ \citenamefont
  {Schaetz}}]{Clos2016a}%
  \BibitemOpen
  \bibfield  {author} {\bibinfo {author} {\bibfnamefont {G.}~\bibnamefont
  {Clos}}, \bibinfo {author} {\bibfnamefont {D.}~\bibnamefont {Porras}},
  \bibinfo {author} {\bibfnamefont {U.}~\bibnamefont {Warring}}, \ and\
  \bibinfo {author} {\bibfnamefont {T.}~\bibnamefont {Schaetz}},\ }\href
  {\doibase 10.1103/PhysRevLett.117.170401} {\bibfield  {journal} {\bibinfo
  {journal} {Phys. Rev. Lett.}\ }\textbf {\bibinfo {volume} {117}},\ \bibinfo
  {pages} {1} (\bibinfo {year} {2016})}\BibitemShut {NoStop}%
\bibitem [{\citenamefont {Kim}\ \emph {et~al.}(2018)\citenamefont {Kim},
  \citenamefont {Park}, \citenamefont {Kim}, \citenamefont {Sim},\ and\
  \citenamefont {Ahn}}]{Kim2018}%
  \BibitemOpen
  \bibfield  {author} {\bibinfo {author} {\bibfnamefont {H.}~\bibnamefont
  {Kim}}, \bibinfo {author} {\bibfnamefont {Y.}~\bibnamefont {Park}}, \bibinfo
  {author} {\bibfnamefont {K.}~\bibnamefont {Kim}}, \bibinfo {author}
  {\bibfnamefont {H.~S.}\ \bibnamefont {Sim}}, \ and\ \bibinfo {author}
  {\bibfnamefont {J.}~\bibnamefont {Ahn}},\ }\href
  {https://journals.aps.org/prl/pdf/10.1103/PhysRevLett.120.180502
  http://arxiv.org/abs/1712.02065} {\bibfield  {journal} {\bibinfo  {journal}
  {Phys. Rev. Lett.}\ }\textbf {\bibinfo {volume} {120}},\ \bibinfo {pages}
  {180502} (\bibinfo {year} {2018})}\BibitemShut {NoStop}%
\bibitem [{\citenamefont {Bernien}\ \emph {et~al.}(2017)\citenamefont
  {Bernien}, \citenamefont {Schwartz}, \citenamefont {Keesling}, \citenamefont
  {Levine}, \citenamefont {Omran}, \citenamefont {Pichler}, \citenamefont
  {Choi}, \citenamefont {Zibrov}, \citenamefont {Endres}, \citenamefont
  {Greiner}, \citenamefont {Vuletic},\ and\ \citenamefont
  {Lukin}}]{Bernien2017}%
  \BibitemOpen
  \bibfield  {author} {\bibinfo {author} {\bibfnamefont {H.}~\bibnamefont
  {Bernien}}, \bibinfo {author} {\bibfnamefont {S.}~\bibnamefont {Schwartz}},
  \bibinfo {author} {\bibfnamefont {A.}~\bibnamefont {Keesling}}, \bibinfo
  {author} {\bibfnamefont {H.}~\bibnamefont {Levine}}, \bibinfo {author}
  {\bibfnamefont {A.}~\bibnamefont {Omran}}, \bibinfo {author} {\bibfnamefont
  {H.}~\bibnamefont {Pichler}}, \bibinfo {author} {\bibfnamefont
  {S.}~\bibnamefont {Choi}}, \bibinfo {author} {\bibfnamefont {A.~S.}\
  \bibnamefont {Zibrov}}, \bibinfo {author} {\bibfnamefont {M.}~\bibnamefont
  {Endres}}, \bibinfo {author} {\bibfnamefont {M.}~\bibnamefont {Greiner}},
  \bibinfo {author} {\bibfnamefont {V.}~\bibnamefont {Vuletic}}, \ and\
  \bibinfo {author} {\bibfnamefont {M.~D.}\ \bibnamefont {Lukin}},\ }\href
  {https://www.nature.com/articles/nature24622.pdf} {\bibfield  {journal}
  {\bibinfo  {journal} {Nature}\ }\textbf {\bibinfo {volume} {551}},\ \bibinfo
  {pages} {579} (\bibinfo {year} {2017})}\BibitemShut {NoStop}%
\bibitem [{\citenamefont {Neill}\ \emph {et~al.}(2016)\citenamefont {Neill},
  \citenamefont {Roushan}, \citenamefont {Fang}, \citenamefont {Chen},
  \citenamefont {Kolodrubetz}, \citenamefont {Chen}, \citenamefont {Megrant},
  \citenamefont {Barends}, \citenamefont {Campbell}, \citenamefont {Chiaro},
  \citenamefont {Dunsworth}, \citenamefont {Jeffrey}, \citenamefont {Kelly},
  \citenamefont {Mutus}, \citenamefont {O'Malley}, \citenamefont {Quintana},
  \citenamefont {Sank}, \citenamefont {Vainsencher}, \citenamefont {Wenner},
  \citenamefont {White}, \citenamefont {Polkovnikov},\ and\ \citenamefont
  {Martinis}}]{Neill2016}%
  \BibitemOpen
  \bibfield  {author} {\bibinfo {author} {\bibfnamefont {C.}~\bibnamefont
  {Neill}}, \bibinfo {author} {\bibfnamefont {P.}~\bibnamefont {Roushan}},
  \bibinfo {author} {\bibfnamefont {M.}~\bibnamefont {Fang}}, \bibinfo {author}
  {\bibfnamefont {Y.}~\bibnamefont {Chen}}, \bibinfo {author} {\bibfnamefont
  {M.}~\bibnamefont {Kolodrubetz}}, \bibinfo {author} {\bibfnamefont
  {Z.}~\bibnamefont {Chen}}, \bibinfo {author} {\bibfnamefont {A.}~\bibnamefont
  {Megrant}}, \bibinfo {author} {\bibfnamefont {R.}~\bibnamefont {Barends}},
  \bibinfo {author} {\bibfnamefont {B.}~\bibnamefont {Campbell}}, \bibinfo
  {author} {\bibfnamefont {B.}~\bibnamefont {Chiaro}}, \bibinfo {author}
  {\bibfnamefont {A.}~\bibnamefont {Dunsworth}}, \bibinfo {author}
  {\bibfnamefont {E.}~\bibnamefont {Jeffrey}}, \bibinfo {author} {\bibfnamefont
  {J.}~\bibnamefont {Kelly}}, \bibinfo {author} {\bibfnamefont
  {J.}~\bibnamefont {Mutus}}, \bibinfo {author} {\bibfnamefont {P.~J.}\
  \bibnamefont {O'Malley}}, \bibinfo {author} {\bibfnamefont {C.}~\bibnamefont
  {Quintana}}, \bibinfo {author} {\bibfnamefont {D.}~\bibnamefont {Sank}},
  \bibinfo {author} {\bibfnamefont {A.}~\bibnamefont {Vainsencher}}, \bibinfo
  {author} {\bibfnamefont {J.}~\bibnamefont {Wenner}}, \bibinfo {author}
  {\bibfnamefont {T.~C.}\ \bibnamefont {White}}, \bibinfo {author}
  {\bibfnamefont {A.}~\bibnamefont {Polkovnikov}}, \ and\ \bibinfo {author}
  {\bibfnamefont {J.~M.}\ \bibnamefont {Martinis}},\ }\href
  {https://www.nature.com/articles/nphys3830.pdf} {\bibfield  {journal}
  {\bibinfo  {journal} {Nature Phys.}\ }\textbf {\bibinfo {volume} {12}},\
  \bibinfo {pages} {1037} (\bibinfo {year} {2016})}\BibitemShut {NoStop}%
\bibitem [{\citenamefont {Roushan}\ \emph {et~al.}(2018)\citenamefont
  {Roushan}, \citenamefont {Lucero}, \citenamefont {Martinis}, \citenamefont
  {Chiaro}, \citenamefont {Megrant}, \citenamefont {Kechedzhi}, \citenamefont
  {Dunsworth}, \citenamefont {Wenner}, \citenamefont {Klimov}, \citenamefont
  {Burkett}, \citenamefont {Arya}, \citenamefont {Vainsencher}, \citenamefont
  {Mutus}, \citenamefont {Neven}, \citenamefont {Fowler}, \citenamefont {Chen},
  \citenamefont {Chen}, \citenamefont {Barends}, \citenamefont {Isakov},
  \citenamefont {Giustina}, \citenamefont {Huang}, \citenamefont {Kelly},
  \citenamefont {Neeley}, \citenamefont {White}, \citenamefont {Boixo},
  \citenamefont {Sank}, \citenamefont {Foxen}, \citenamefont {Smelyanskiy},
  \citenamefont {Graff}, \citenamefont {Jeffrey}, \citenamefont {Quintana},\
  and\ \citenamefont {Neill}}]{Roushan2018}%
  \BibitemOpen
  \bibfield  {author} {\bibinfo {author} {\bibfnamefont {P.}~\bibnamefont
  {Roushan}}, \bibinfo {author} {\bibfnamefont {E.}~\bibnamefont {Lucero}},
  \bibinfo {author} {\bibfnamefont {J.~M.}\ \bibnamefont {Martinis}}, \bibinfo
  {author} {\bibfnamefont {B.}~\bibnamefont {Chiaro}}, \bibinfo {author}
  {\bibfnamefont {A.}~\bibnamefont {Megrant}}, \bibinfo {author} {\bibfnamefont
  {K.}~\bibnamefont {Kechedzhi}}, \bibinfo {author} {\bibfnamefont
  {A.}~\bibnamefont {Dunsworth}}, \bibinfo {author} {\bibfnamefont
  {J.}~\bibnamefont {Wenner}}, \bibinfo {author} {\bibfnamefont
  {P.}~\bibnamefont {Klimov}}, \bibinfo {author} {\bibfnamefont
  {B.}~\bibnamefont {Burkett}}, \bibinfo {author} {\bibfnamefont
  {K.}~\bibnamefont {Arya}}, \bibinfo {author} {\bibfnamefont {A.}~\bibnamefont
  {Vainsencher}}, \bibinfo {author} {\bibfnamefont {J.}~\bibnamefont {Mutus}},
  \bibinfo {author} {\bibfnamefont {H.}~\bibnamefont {Neven}}, \bibinfo
  {author} {\bibfnamefont {A.}~\bibnamefont {Fowler}}, \bibinfo {author}
  {\bibfnamefont {Z.}~\bibnamefont {Chen}}, \bibinfo {author} {\bibfnamefont
  {Y.}~\bibnamefont {Chen}}, \bibinfo {author} {\bibfnamefont {R.}~\bibnamefont
  {Barends}}, \bibinfo {author} {\bibfnamefont {S.~V.}\ \bibnamefont {Isakov}},
  \bibinfo {author} {\bibfnamefont {M.}~\bibnamefont {Giustina}}, \bibinfo
  {author} {\bibfnamefont {T.}~\bibnamefont {Huang}}, \bibinfo {author}
  {\bibfnamefont {J.}~\bibnamefont {Kelly}}, \bibinfo {author} {\bibfnamefont
  {M.}~\bibnamefont {Neeley}}, \bibinfo {author} {\bibfnamefont {T.~C.}\
  \bibnamefont {White}}, \bibinfo {author} {\bibfnamefont {S.}~\bibnamefont
  {Boixo}}, \bibinfo {author} {\bibfnamefont {D.}~\bibnamefont {Sank}},
  \bibinfo {author} {\bibfnamefont {B.}~\bibnamefont {Foxen}}, \bibinfo
  {author} {\bibfnamefont {V.}~\bibnamefont {Smelyanskiy}}, \bibinfo {author}
  {\bibfnamefont {R.}~\bibnamefont {Graff}}, \bibinfo {author} {\bibfnamefont
  {E.}~\bibnamefont {Jeffrey}}, \bibinfo {author} {\bibfnamefont
  {C.}~\bibnamefont {Quintana}}, \ and\ \bibinfo {author} {\bibfnamefont
  {C.}~\bibnamefont {Neill}},\ }\href
  {http://science.sciencemag.org/content/sci/360/6385/195.full.pdf
  http://arxiv.org/abs/1709.06678 http://science.sciencemag.org/} {\bibfield
  {journal} {\bibinfo  {journal} {Science}\ }\textbf {\bibinfo {volume}
  {360}},\ \bibinfo {pages} {195} (\bibinfo {year} {2018})}\BibitemShut
  {NoStop}%
\bibitem [{\citenamefont {Malabarba}\ \emph {et~al.}(2014)\citenamefont
  {Malabarba}, \citenamefont {Garcia-Pintos}, \citenamefont {Linden},
  \citenamefont {Farrelly},\ and\ \citenamefont {Short}}]{Malabarba2014}%
  \BibitemOpen
  \bibfield  {author} {\bibinfo {author} {\bibfnamefont {A.~S.~L.}\
  \bibnamefont {Malabarba}}, \bibinfo {author} {\bibfnamefont {L.~P.}\
  \bibnamefont {Garcia-Pintos}}, \bibinfo {author} {\bibfnamefont
  {N.}~\bibnamefont {Linden}}, \bibinfo {author} {\bibfnamefont {T.~C.}\
  \bibnamefont {Farrelly}}, \ and\ \bibinfo {author} {\bibfnamefont {A.~J.}\
  \bibnamefont {Short}},\ }\href
  {https://journals.aps.org/pre/abstract/10.1103/PhysRevE.90.012121} {\bibfield
   {journal} {\bibinfo  {journal} {Phys. Rev. E}\ }\textbf {\bibinfo {volume}
  {90}} (\bibinfo {year} {2014})}\BibitemShut {NoStop}%
\bibitem [{\citenamefont {Garc{\'{i}}a-Pintos}\ \emph
  {et~al.}(2017)\citenamefont {Garc{\'{i}}a-Pintos}, \citenamefont {Linden},
  \citenamefont {Malabarba}, \citenamefont {Short},\ and\ \citenamefont
  {Winter}}]{Garcia-Pintos2017}%
  \BibitemOpen
  \bibfield  {author} {\bibinfo {author} {\bibfnamefont {L.~P.}\ \bibnamefont
  {Garc{\'{i}}a-Pintos}}, \bibinfo {author} {\bibfnamefont {N.}~\bibnamefont
  {Linden}}, \bibinfo {author} {\bibfnamefont {A.~S.}\ \bibnamefont
  {Malabarba}}, \bibinfo {author} {\bibfnamefont {A.~J.}\ \bibnamefont
  {Short}}, \ and\ \bibinfo {author} {\bibfnamefont {A.}~\bibnamefont
  {Winter}},\ }\href
  {https://journals.aps.org/prx/pdf/10.1103/PhysRevX.7.031027} {\bibfield
  {journal} {\bibinfo  {journal} {Phys. Rev. X}\ }\textbf {\bibinfo {volume}
  {7}} (\bibinfo {year} {2017})}\BibitemShut {NoStop}%
\bibitem [{\citenamefont {Richter}\ \emph {et~al.}(2018)\citenamefont
  {Richter}, \citenamefont {Gemmer},\ and\ \citenamefont
  {Steinigeweg}}]{Richter2018a}%
  \BibitemOpen
  \bibfield  {author} {\bibinfo {author} {\bibfnamefont {J.}~\bibnamefont
  {Richter}}, \bibinfo {author} {\bibfnamefont {J.}~\bibnamefont {Gemmer}}, \
  and\ \bibinfo {author} {\bibfnamefont {R.}~\bibnamefont {Steinigeweg}},\
  }\href {https://arxiv.org/pdf/1805.11625.pdf http://arxiv.org/abs/1805.11625}
  {\  (\bibinfo {year} {2018})},\ \Eprint {http://arxiv.org/abs/1805.11625}
  {arXiv:1805.11625} \BibitemShut {NoStop}%
\bibitem [{\citenamefont {{De Oliveira}}\ \emph {et~al.}(2018)\citenamefont
  {{De Oliveira}}, \citenamefont {Charalambous}, \citenamefont {Jonathan},
  \citenamefont {Lewenstein},\ and\ \citenamefont {Riera}}]{DeOliveira2018}%
  \BibitemOpen
  \bibfield  {author} {\bibinfo {author} {\bibfnamefont {T.~R.}\ \bibnamefont
  {{De Oliveira}}}, \bibinfo {author} {\bibfnamefont {C.}~\bibnamefont
  {Charalambous}}, \bibinfo {author} {\bibfnamefont {D.}~\bibnamefont
  {Jonathan}}, \bibinfo {author} {\bibfnamefont {M.}~\bibnamefont
  {Lewenstein}}, \ and\ \bibinfo {author} {\bibfnamefont {A.}~\bibnamefont
  {Riera}},\ }\href {https://doi.org/10.1088/1367-2630/aab03b} {\bibfield
  {journal} {\bibinfo  {journal} {New J. Phys.}\ }\textbf {\bibinfo {volume}
  {20}},\ \bibinfo {pages} {33032} (\bibinfo {year} {2018})}\BibitemShut
  {NoStop}%
\bibitem [{\citenamefont {Borgonovi}\ \emph {et~al.}(2019)\citenamefont
  {Borgonovi}, \citenamefont {Izrailev},\ and\ \citenamefont
  {Santos}}]{Borgonovi2019}%
  \BibitemOpen
  \bibfield  {author} {\bibinfo {author} {\bibfnamefont {F.}~\bibnamefont
  {Borgonovi}}, \bibinfo {author} {\bibfnamefont {F.~M.}\ \bibnamefont
  {Izrailev}}, \ and\ \bibinfo {author} {\bibfnamefont {L.~F.}\ \bibnamefont
  {Santos}},\ }\href
  {https://journals.aps.org/pre/pdf/10.1103/PhysRevE.99.010101
  http://arxiv.org/abs/1802.08265} {\bibfield  {journal} {\bibinfo  {journal}
  {Phys. Rev. E}\ }\textbf {\bibinfo {volume} {99}},\ \bibinfo {pages} {010101}
  (\bibinfo {year} {2019})}\BibitemShut {NoStop}%
\bibitem [{\citenamefont {Schiulaz}\ \emph {et~al.}(2019)\citenamefont
  {Schiulaz}, \citenamefont {Torres-Herrera},\ and\ \citenamefont
  {Santos}}]{Schiulaz2019}%
  \BibitemOpen
  \bibfield  {author} {\bibinfo {author} {\bibfnamefont {M.}~\bibnamefont
  {Schiulaz}}, \bibinfo {author} {\bibfnamefont {E.~J.}\ \bibnamefont
  {Torres-Herrera}}, \ and\ \bibinfo {author} {\bibfnamefont {L.~F.}\
  \bibnamefont {Santos}},\ }\href
  {https://link.aps.org/doi/10.1103/PhysRevB.99.174313} {\bibfield  {journal}
  {\bibinfo  {journal} {Phys. Rev. B}\ }\textbf {\bibinfo {volume} {99}},\
  \bibinfo {pages} {174313} (\bibinfo {year} {2019})}\BibitemShut {NoStop}%
\bibitem [{\citenamefont {Srednicki}(1994)}]{Srednicki1994}%
  \BibitemOpen
  \bibfield  {author} {\bibinfo {author} {\bibfnamefont {M.}~\bibnamefont
  {Srednicki}},\ }\href
  {https://journals.aps.org/pre/abstract/10.1103/PhysRevE.50.888} {\bibfield
  {journal} {\bibinfo  {journal} {Phys. Rev. E}\ }\textbf {\bibinfo {volume}
  {50}},\ \bibinfo {pages} {888} (\bibinfo {year} {1994})},\ \Eprint
  {http://arxiv.org/abs/9403051} {9403051} \BibitemShut {NoStop}%
\bibitem [{\citenamefont {Srednicki}(1999)}]{Srednicki1999}%
  \BibitemOpen
  \bibfield  {author} {\bibinfo {author} {\bibfnamefont {M.}~\bibnamefont
  {Srednicki}},\ }\href {http://iopscience.iop.org/0305-4470/32/7/007}
  {\bibfield  {journal} {\bibinfo  {journal} {J. Phys. A: Math. Gen.}\ }\textbf
  {\bibinfo {volume} {32}},\ \bibinfo {pages} {1163} (\bibinfo {year}
  {1999})}\BibitemShut {NoStop}%
\bibitem [{\citenamefont {Rigol}\ \emph {et~al.}(2008)\citenamefont {Rigol},
  \citenamefont {Dunjko},\ and\ \citenamefont {Olshanii}}]{Rigol2008}%
  \BibitemOpen
  \bibfield  {author} {\bibinfo {author} {\bibfnamefont {M.}~\bibnamefont
  {Rigol}}, \bibinfo {author} {\bibfnamefont {V.}~\bibnamefont {Dunjko}}, \
  and\ \bibinfo {author} {\bibfnamefont {M.}~\bibnamefont {Olshanii}},\ }\href
  {https://www.nature.com/articles/nature06838} {\bibfield  {journal} {\bibinfo
   {journal} {Nature}\ }\textbf {\bibinfo {volume} {452}},\ \bibinfo {pages}
  {854} (\bibinfo {year} {2008})}\BibitemShut {NoStop}%
\bibitem [{\citenamefont {Reimann}(2008)}]{Reimann2008}%
  \BibitemOpen
  \bibfield  {author} {\bibinfo {author} {\bibfnamefont {P.}~\bibnamefont
  {Reimann}},\ }\href
  {https://journals.aps.org/prl/abstract/10.1103/PhysRevLett.101.190403}
  {\bibfield  {journal} {\bibinfo  {journal} {Phys. Rev. Lett.}\ }\textbf
  {\bibinfo {volume} {101}} (\bibinfo {year} {2008})}\BibitemShut {NoStop}%
\bibitem [{\citenamefont {Yukalov}(2011)}]{Yukalov2011}%
  \BibitemOpen
  \bibfield  {author} {\bibinfo {author} {\bibfnamefont {V.~I.}\ \bibnamefont
  {Yukalov}},\ }\href {www.lphys.org} {\bibfield  {journal} {\bibinfo
  {journal} {Laser Phys. Lett.}\ }\textbf {\bibinfo {volume} {8}},\ \bibinfo
  {pages} {485} (\bibinfo {year} {2011})}\BibitemShut {NoStop}%
\bibitem [{\citenamefont {Rigol}\ and\ \citenamefont
  {Srednicki}(2012)}]{Rigol2012}%
  \BibitemOpen
  \bibfield  {author} {\bibinfo {author} {\bibfnamefont {M.}~\bibnamefont
  {Rigol}}\ and\ \bibinfo {author} {\bibfnamefont {M.}~\bibnamefont
  {Srednicki}},\ }\href {\doibase 10.1103/PhysRevLett.108.110601} {\bibfield
  {journal} {\bibinfo  {journal} {Phys. Rev. Lett.}\ }\textbf {\bibinfo
  {volume} {108}},\ \bibinfo {pages} {1} (\bibinfo {year} {2012})}\BibitemShut
  {NoStop}%
\bibitem [{\citenamefont {D'Alessio}\ \emph {et~al.}(2016)\citenamefont
  {D'Alessio}, \citenamefont {Kafri}, \citenamefont {Polkovnikov},\ and\
  \citenamefont {Rigol}}]{DAlessio2016}%
  \BibitemOpen
  \bibfield  {author} {\bibinfo {author} {\bibfnamefont {L.}~\bibnamefont
  {D'Alessio}}, \bibinfo {author} {\bibfnamefont {Y.}~\bibnamefont {Kafri}},
  \bibinfo {author} {\bibfnamefont {A.}~\bibnamefont {Polkovnikov}}, \ and\
  \bibinfo {author} {\bibfnamefont {M.}~\bibnamefont {Rigol}},\ }\href
  {https://www.tandfonline.com/doi/full/10.1080/00018732.2016.1198134}
  {\bibfield  {journal} {\bibinfo  {journal} {Adv Phys.}\ }\textbf {\bibinfo
  {volume} {65}},\ \bibinfo {pages} {239} (\bibinfo {year} {2016})}\BibitemShut
  {NoStop}%
\bibitem [{\citenamefont {Beugeling}\ \emph {et~al.}(2014)\citenamefont
  {Beugeling}, \citenamefont {Moessner},\ and\ \citenamefont
  {Haque}}]{Beugeling2014}%
  \BibitemOpen
  \bibfield  {author} {\bibinfo {author} {\bibfnamefont {W.}~\bibnamefont
  {Beugeling}}, \bibinfo {author} {\bibfnamefont {R.}~\bibnamefont {Moessner}},
  \ and\ \bibinfo {author} {\bibfnamefont {M.}~\bibnamefont {Haque}},\ }\href
  {https://journals.aps.org/pre/pdf/10.1103/PhysRevE.89.042112} {\bibfield
  {journal} {\bibinfo  {journal} {Phys. Rev. E}\ }\textbf {\bibinfo {volume}
  {89}} (\bibinfo {year} {2014})}\BibitemShut {NoStop}%
\bibitem [{\citenamefont {Beugeling}\ \emph {et~al.}(2015)\citenamefont
  {Beugeling}, \citenamefont {Moessner},\ and\ \citenamefont
  {Haque}}]{Beugeling2015}%
  \BibitemOpen
  \bibfield  {author} {\bibinfo {author} {\bibfnamefont {W.}~\bibnamefont
  {Beugeling}}, \bibinfo {author} {\bibfnamefont {R.}~\bibnamefont {Moessner}},
  \ and\ \bibinfo {author} {\bibfnamefont {M.}~\bibnamefont {Haque}},\ }\href
  {https://journals.aps.org/pre/pdf/10.1103/PhysRevE.91.012144} {\bibfield
  {journal} {\bibinfo  {journal} {Phys. Rev. E}\ }\textbf {\bibinfo {volume}
  {91}} (\bibinfo {year} {2015})}\BibitemShut {NoStop}%
\bibitem [{\citenamefont {Eisert}\ \emph {et~al.}(2014)\citenamefont {Eisert},
  \citenamefont {Friesdorf},\ and\ \citenamefont {Gogolin}}]{Eisert2015}%
  \BibitemOpen
  \bibfield  {author} {\bibinfo {author} {\bibfnamefont {J.}~\bibnamefont
  {Eisert}}, \bibinfo {author} {\bibfnamefont {M.}~\bibnamefont {Friesdorf}}, \
  and\ \bibinfo {author} {\bibfnamefont {C.}~\bibnamefont {Gogolin}},\ }\href
  {http://arxiv.org/abs/1408.5148{\%}0Ahttp://dx.doi.org/10.1038/nphys3215}
  {\bibfield  {journal} {\bibinfo  {journal} {Nat. Phys.}\ }\textbf {\bibinfo
  {volume} {11}},\ \bibinfo {pages} {124} (\bibinfo {year} {2014})}\BibitemShut
  {NoStop}%
\bibitem [{\citenamefont {Anza}\ \emph {et~al.}(2018)\citenamefont {Anza},
  \citenamefont {Gogolin},\ and\ \citenamefont {Huber}}]{Anza2018}%
  \BibitemOpen
  \bibfield  {author} {\bibinfo {author} {\bibfnamefont {F.}~\bibnamefont
  {Anza}}, \bibinfo {author} {\bibfnamefont {C.}~\bibnamefont {Gogolin}}, \
  and\ \bibinfo {author} {\bibfnamefont {M.}~\bibnamefont {Huber}},\ }\href
  {https://journals.aps.org/prl/pdf/10.1103/PhysRevLett.120.150603} {\bibfield
  {journal} {\bibinfo  {journal} {Phys. Rev. Lett.}\ }\textbf {\bibinfo
  {volume} {120}} (\bibinfo {year} {2018})}\BibitemShut {NoStop}%
\bibitem [{\citenamefont {Deutsch}(2018)}]{Deutsch2018}%
  \BibitemOpen
  \bibfield  {author} {\bibinfo {author} {\bibfnamefont {J.~M.}\ \bibnamefont
  {Deutsch}},\ }\href
  {http://stacks.iop.org/0034-4885/81/i=8/a=082001?key=crossref.72e037fa24eddc6b5d4a21a7bac77e1b}
  {\bibfield  {journal} {\bibinfo  {journal} {Rep. Prog. Phys.}\ }\textbf
  {\bibinfo {volume} {81}},\ \bibinfo {pages} {082001} (\bibinfo {year}
  {2018})}\BibitemShut {NoStop}%
\bibitem [{\citenamefont {Merali}(2017)}]{Merali2017}%
  \BibitemOpen
  \bibfield  {author} {\bibinfo {author} {\bibfnamefont {Z.}~\bibnamefont
  {Merali}},\ }\href {https://arxiv.org/abs/1707.04930
  http://www.nature.com/doifinder/10.1038/551020a
  http://www.ncbi.nlm.nih.gov/pubmed/10107130} {\bibfield  {journal} {\bibinfo
  {journal} {Nature}\ }\textbf {\bibinfo {volume} {551}},\ \bibinfo {pages}
  {20} (\bibinfo {year} {2017})}\BibitemShut {NoStop}%
\bibitem [{\citenamefont {Rigol}(2009)}]{Rigol}%
  \BibitemOpen
  \bibfield  {author} {\bibinfo {author} {\bibfnamefont {M.}~\bibnamefont
  {Rigol}},\ }\href
  {https://journals.aps.org/pra/pdf/10.1103/PhysRevA.80.053607} {\bibfield
  {journal} {\bibinfo  {journal} {Phys. Rev. A.}\ }\textbf {\bibinfo {volume}
  {80}} (\bibinfo {year} {2009})}\BibitemShut {NoStop}%
\bibitem [{\citenamefont {Reimann}(2016)}]{Reimann2016a}%
  \BibitemOpen
  \bibfield  {author} {\bibinfo {author} {\bibfnamefont {P.}~\bibnamefont
  {Reimann}},\ }\href {https://www.nature.com/articles/ncomms10821.pdf}
  {\bibfield  {journal} {\bibinfo  {journal} {Nat. Comm.}\ }\textbf {\bibinfo
  {volume} {7}} (\bibinfo {year} {2016})}\BibitemShut {NoStop}%
\bibitem [{\citenamefont {Borgonovi}\ \emph {et~al.}(2016)\citenamefont
  {Borgonovi}, \citenamefont {Izrailev}, \citenamefont {Santos},\ and\
  \citenamefont {Zelevinsky}}]{Borgonovi2016}%
  \BibitemOpen
  \bibfield  {author} {\bibinfo {author} {\bibfnamefont {F.}~\bibnamefont
  {Borgonovi}}, \bibinfo {author} {\bibfnamefont {F.~M.}\ \bibnamefont
  {Izrailev}}, \bibinfo {author} {\bibfnamefont {L.~F.}\ \bibnamefont
  {Santos}}, \ and\ \bibinfo {author} {\bibfnamefont {V.~G.}\ \bibnamefont
  {Zelevinsky}},\ }\href
  {https://www.sciencedirect.com/science/article/pii/S0370157316000831}
  {\bibfield  {journal} {\bibinfo  {journal} {Phys. Rep.}\ }\textbf {\bibinfo
  {volume} {626}},\ \bibinfo {pages} {1} (\bibinfo {year} {2016})}\BibitemShut
  {NoStop}%
\bibitem [{\citenamefont {Dabelow}\ and\ \citenamefont
  {Reimann}(2019)}]{Dabelow2019}%
  \BibitemOpen
  \bibfield  {author} {\bibinfo {author} {\bibfnamefont {L.}~\bibnamefont
  {Dabelow}}\ and\ \bibinfo {author} {\bibfnamefont {P.}~\bibnamefont
  {Reimann}},\ }\href {https://arxiv.org/pdf/1903.11881.pdf
  http://arxiv.org/abs/1903.11881} {\  (\bibinfo {year} {2019})},\ \Eprint
  {http://arxiv.org/abs/1903.11881} {arXiv:1903.11881} \BibitemShut {NoStop}%
\bibitem [{\citenamefont {Reimann}\ and\ \citenamefont
  {Dabelow}(2019)}]{Reimann2019}%
  \BibitemOpen
  \bibfield  {author} {\bibinfo {author} {\bibfnamefont {P.}~\bibnamefont
  {Reimann}}\ and\ \bibinfo {author} {\bibfnamefont {L.}~\bibnamefont
  {Dabelow}},\ }\href
  {https://journals.aps.org/prl/pdf/10.1103/PhysRevLett.122.080603} {\bibfield
  {journal} {\bibinfo  {journal} {Phys. Rev. Lett.}\ }\textbf {\bibinfo
  {volume} {122}} (\bibinfo {year} {2019})}\BibitemShut {NoStop}%
\bibitem [{\citenamefont {Torres-Herrera}\ \emph {et~al.}(2016)\citenamefont
  {Torres-Herrera}, \citenamefont {Karp}, \citenamefont {Tavora},\ and\
  \citenamefont {Santos}}]{Torres-Herrera2016}%
  \BibitemOpen
  \bibfield  {author} {\bibinfo {author} {\bibfnamefont {E.~J.}\ \bibnamefont
  {Torres-Herrera}}, \bibinfo {author} {\bibfnamefont {J.}~\bibnamefont
  {Karp}}, \bibinfo {author} {\bibfnamefont {M.}~\bibnamefont {Tavora}}, \ and\
  \bibinfo {author} {\bibfnamefont {L.~F.}\ \bibnamefont {Santos}},\ }\href
  {https://arxiv.org/pdf/1608.06636.pdf} {\bibfield  {journal} {\bibinfo
  {journal} {Entropy}\ }\textbf {\bibinfo {volume} {18}} (\bibinfo {year}
  {2016})}\BibitemShut {NoStop}%
\bibitem [{\citenamefont {Kubo}(1966)}]{Kubo1966}%
  \BibitemOpen
  \bibfield  {author} {\bibinfo {author} {\bibfnamefont {R.}~\bibnamefont
  {Kubo}},\ }\href
  {http://iopscience.iop.org/article/10.1088/0034-4885/29/1/306/pdf
  http://stacks.iop.org/0034-4885/29/i=1/a=306} {\bibfield  {journal} {\bibinfo
   {journal} {Rep. Prog. Phys.}\ }\textbf {\bibinfo {volume} {29}},\ \bibinfo
  {pages} {255} (\bibinfo {year} {1966})}\BibitemShut {NoStop}%
\bibitem [{\citenamefont {Nation}\ and\ \citenamefont
  {Porras}(2018)}]{Nation2018}%
  \BibitemOpen
  \bibfield  {author} {\bibinfo {author} {\bibfnamefont {C.}~\bibnamefont
  {Nation}}\ and\ \bibinfo {author} {\bibfnamefont {D.}~\bibnamefont
  {Porras}},\ }\href
  {http://iopscience.iop.org/article/10.1088/1367-2630/aae28f} {\bibfield
  {journal} {\bibinfo  {journal} {New J. Phys.}\ }\textbf {\bibinfo {volume}
  {20}},\ \bibinfo {pages} {103003} (\bibinfo {year} {2018})}\BibitemShut
  {NoStop}%
\bibitem [{\citenamefont {Deutsch}(1991{\natexlab{a}})}]{Deutsch1991}%
  \BibitemOpen
  \bibfield  {author} {\bibinfo {author} {\bibfnamefont {J.~M.}\ \bibnamefont
  {Deutsch}},\ }\href {\doibase 10.1103/PhysRevA.43.2046} {\bibfield  {journal}
  {\bibinfo  {journal} {Phys. Rev. A}\ }\textbf {\bibinfo {volume} {43}},\
  \bibinfo {pages} {2046} (\bibinfo {year} {1991}{\natexlab{a}})}\BibitemShut
  {NoStop}%
\bibitem [{\citenamefont {Deutsch}(1991{\natexlab{b}})}]{Deutscha}%
  \BibitemOpen
  \bibfield  {author} {\bibinfo {author} {\bibfnamefont {J.~M.}\ \bibnamefont
  {Deutsch}},\ }\href {https://deutsch.physics.ucsc.edu/pdf/quantumstat.pdf}
  {\bibfield  {journal} {\bibinfo  {journal} {(unpublished)}\ } (\bibinfo
  {year} {1991}{\natexlab{b}})}\BibitemShut {NoStop}%
\bibitem [{\citenamefont {Reimann}(2015)}]{Reimann2015}%
  \BibitemOpen
  \bibfield  {author} {\bibinfo {author} {\bibfnamefont {P.}~\bibnamefont
  {Reimann}},\ }\href
  {https://iopscience.iop.org/article/10.1088/1367-2630/17/5/055025/pdf}
  {\bibfield  {journal} {\bibinfo  {journal} {New J. Phys.}\ }\textbf {\bibinfo
  {volume} {17}} (\bibinfo {year} {2015})}\BibitemShut {NoStop}%
\bibitem [{\citenamefont {Nation}\ and\ \citenamefont
  {Porras}(2019)}]{Nation2019}%
  \BibitemOpen
  \bibfield  {author} {\bibinfo {author} {\bibfnamefont {C.}~\bibnamefont
  {Nation}}\ and\ \bibinfo {author} {\bibfnamefont {D.}~\bibnamefont
  {Porras}},\ }\href
  {https://journals.aps.org/pre/pdf/10.1103/PhysRevE.99.052139} {\bibfield
  {journal} {\bibinfo  {journal} {Phys. Rev. E}\ }\textbf {\bibinfo {volume}
  {99}},\ \bibinfo {pages} {052139} (\bibinfo {year} {2019})}\BibitemShut
  {NoStop}%
\bibitem [{\citenamefont {Weinberg}\ and\ \citenamefont
  {Bukov}(2017)}]{Weinberg2016}%
  \BibitemOpen
  \bibfield  {author} {\bibinfo {author} {\bibfnamefont {P.}~\bibnamefont
  {Weinberg}}\ and\ \bibinfo {author} {\bibfnamefont {M.}~\bibnamefont
  {Bukov}},\ }\href {\doibase 10.21468/SciPostPhys.2.1.003} {\bibfield
  {journal} {\bibinfo  {journal} {SciPost Phys.}\ }\textbf {\bibinfo {volume}
  {2}},\ \bibinfo {pages} {003} (\bibinfo {year} {2017})}\BibitemShut {NoStop}%
\bibitem [{\citenamefont {Weinberg}\ and\ \citenamefont
  {Bukov}()}]{Weinberg2018}%
  \BibitemOpen
  \bibfield  {author} {\bibinfo {author} {\bibfnamefont {P.}~\bibnamefont
  {Weinberg}}\ and\ \bibinfo {author} {\bibfnamefont {M.}~\bibnamefont
  {Bukov}},\ }\href {http://arxiv.org/abs/1804.06782} {\ }\Eprint
  {http://arxiv.org/abs/1804.06782} {arXiv:1804.06782} \BibitemShut {NoStop}%
\end{thebibliography}%

\end{document}